\documentclass[12pt]{iopart}

\usepackage{graphicx}  
\usepackage{amssymb}  
\begin{document}

\title[Nuclear astrophysics]{Nuclear astrophysics: the unfinished quest for the origin of the elements}

\author{Jordi Jos\'e}
\address{Departament de F\'\i sica i Enginyeria Nuclear, EUETIB, Universitat 
Polit\`ecnica de Catalunya, E-08036 Barcelona, Spain; Institut d'Estudis 
Espacials de Catalunya, E-08034 Barcelona, Spain}
\ead{jordi.jose@upc.edu}
\author{Christian Iliadis}
\address{Department of Physics \& Astronomy, University of North Carolina, 
Chapel Hill, North Carolina, 27599, USA; Triangle Universities Nuclear 
Laboratory, Durham, North Carolina 27708, USA}
\ead{iliadis@unc.edu}

\begin{abstract}
Half a century has passed since the foundation of nuclear astrophysics. Since then, this discipline has reached its maturity.
Today, nuclear astrophysics constitutes a multidisciplinary crucible of knowledge that combines the achievements in theoretical astrophysics, observational
astronomy, cosmochemistry and nuclear physics. New tools and developments have revolutionized our understanding of the origin of the elements:
supercomputers have provided astrophysicists with the required computational capabilities to study the evolution of stars in a multidimensional 
framework; the emergence of high-energy astrophysics with space-borne
observatories has opened new windows to observe the Universe, from a novel 
panchromatic perspective; cosmochemists have isolated tiny pieces of 
stardust embedded in primitive meteorites, giving clues on the processes
operating in stars as well as on the way matter condenses to form solids; and
nuclear physicists have measured reactions near
stellar energies, through the combined efforts using
stable and radioactive ion beam facilities. 

This review provides comprehensive insight into the nuclear history of the Universe and related topics: starting from the Big Bang, when the ashes from the primordial 
explosion were transformed to hydrogen, helium, and few trace elements, to the rich variety of nucleosynthesis mechanisms and sites in the Universe.
Particular attention is paid to the hydrostatic processes governing the evolution of low-mass stars, 
red giants and asymptotic giant-branch stars, as well as to the explosive nucleosynthesis occurring in
core-collapse and thermonuclear supernovae, $\gamma$-ray bursts, 
classical novae, X-ray bursts, superbursts, and stellar mergers. 
\end{abstract}

(Some figures in this article are in color only in the electronic version)

\vspace{2pc}
\submitto{Reports on Progress in Physics}
\maketitle

\section{Introduction}
\begin{figure}
\includegraphics[scale=.70]{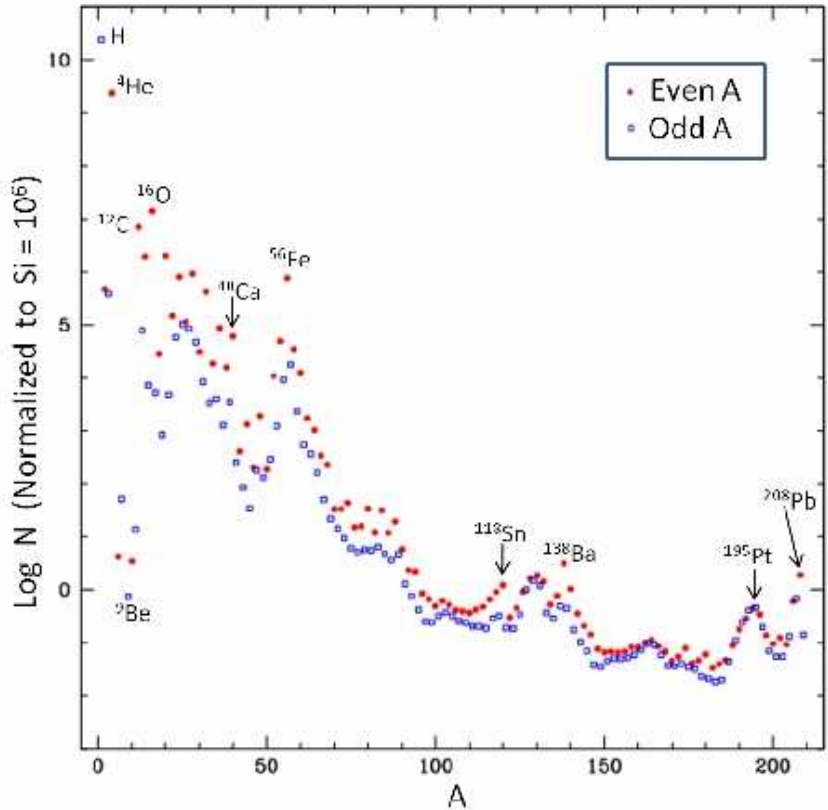}
\caption{Solar System isotopic abundances versus mass number, $A$.
Abundances are based on meteoritic samples as well as on values inferred
from the solar photosphere 
(Sect. 2). Values taken from Lodders \cite{Lod03}. 
\label{SSabun}}
\end{figure}

Most of the ordinary (visible) matter in the Universe, from a tiny terrestrial
pebble to a giant star, is composed of protons and neutrons (nucleons). These
nucleons can assemble in a suite of different nuclear configurations. 
The masses of such bound nuclei are actually smaller than the sum of 
the masses of the free individual nucleons. The difference is a measure of 
the nuclear binding energy that holds the nucleus together,
according to Einstein's famous equation $E=\Delta mc^2$, where $c$ is the speed of light and $\Delta m$ denotes the mass difference between the bound nuclei
and the free nucleons. For example, the nuclide $^{56}$Fe is among the most tightly bound species, with a binding energy near 8.8 MeV per nucleon. 
There are 82 elements that have stable isotopes, 
all the way from H to Pb (except for Tc and Pm). 
Several dozen elements have only unstable isotopes,
naturally abundant or artificially produced in nuclear physics laboratories 
(the super-heaviest nucleus 
discovered to date has 118 protons and a half-life of 0.89 ms \cite{Oga07}). 
The distribution of the nuclide abundances
in the Solar System, plotted as a function of the number of nucleons (mass number), $A$, is shown in Fig. 
\ref{SSabun}. It displays a complicated pattern, with H and He being by far the most abundant 
species. The abundances generally decrease for increasing mass number. Exceptions to this
behavior include the light elements Li, Be and B, which are extremely underabundant with respect to
their neighbors, a pronounced peak near $^{56}$Fe (the so-called {\it iron-peak}),
and several noticeable maxima in the region of $A\gtrsim 100$, for the distributions of both odd- and even-A species.

 The reasons for favoring the presence of some species and for the specific
 pattern inferred for the Solar System abundances have spawned
 large debates.
 Following earlier attempts, 
 Hoyle proposed in 1946 that all heavy elements
 are synthesized in stars \cite{Hoy46}. A major breakthrough was achieved with 
the detection of technetium, an element with no stable isotopes, in the 
spectra of several S stars by Merrill \cite{Mer52} in the early 1950s.
Since the technetium isotopes have
rather short half-lives on a Galactic time scale, the obvious conclusion was
that technetium must have been produced in the observed stars and thus
nucleosynthesis must still be going on in the Universe. 
Shortly after, two seminal papers that provided
the theoretical framework for the  origin of the {\it chemical} species 
elements, were published (almost exactly a century after Darwin's treatise on 
the origin of {\it biological} species) by Burbidge, 
 Burbidge, Fowler \& Hoyle \cite{Bur57}, and independently, by Cameron 
 \cite{Cam57}. 
 These works identified stars as Galactic factories that synthesize atomic 
 nuclei via nuclear reactions. The stars return
 this processed material to the interstellar medium through various
hydrostatic or explosive means during their evolution and thereby enrich the interstellar medium 
in {\it metals}, that is, any species heavier than He  
 according to the astronomical jargon. Out of this matter new generations of stars are born 
 and a new cycle of nuclear transmutations will be initiated.
 In fact, without concourse of the many nuclear processes 
 that take place in the stars and that are discussed in this review, the Universe would have remained 
 a chemically poor place, and probably no life form would have ever emerged.

 This review summarizes more than 50 years of progress in our understanding
 of the nuclear processes that shaped the current chemical abundance pattern
 of the Universe.
 The structure of the paper is as follows. In Section 2, we briefly address the 
 main aspects associated with the determination of the Solar System 
 abundances. 
 This is followed
 by an overview on Big Bang nucleosynthesis (Sect. 3), and by an account of
 the nuclear processes governing low-mass stars (Sect. 4). Particular
 attention is paid to the Standard Solar model and to lessons learned from 
 neutrino physics.
 In Sect. 5, we outline the basic properties of red giant and AGB stars,
 with emphasis on the {\it s-process}. The series of explosive processes
 that characterize core-collapse supernovae and $\gamma$-ray bursts 
 (e.g., {\it p-, $\nu$-, 
 $\nu$p-, $\alpha$-, and r-processes}) are reviewed in Sect. 6.
 This is supplemented with an overview of different astrophysical scenarios
 that involve binary stellar systems, such as classical and recurrent
 novae (Sect. 7), type Ia supernovae (Sect. 8), X-ray bursts and superbursts
 (Sect. 9), and stellar mergers (Sect. 10). We briefly address as well 
 non-stellar processes, such as spallation reactions (Sect. 11), and conclude
 with a number of key puzzles and future perspectives
 in this field (Sect. 12).

\section{An insight into Solar System abundances}
The determination of the chemical composition of a distant star is not a 
trivial task. It basically relies on the analysis 
of the  electromagnetic radiation emitted at different
wavelengths (so-called {\it spectra}). For the case of the 
Solar System abundances, several sources of information,
that combine radiation and matter samples, 
are used to this end: the Sun (mainly photospheric absorption lines), 
meteorites, planets and moons (mainly Earth and Moon). 
Recently, several space probes have for the first time carried material 
from the interplanetary medium to Earth.  
NASA's Genesis mission
collected  about 0.4 mg of solar wind particles for two years, 
using 250 hexagonal (10 cm wide) wafers
made of silicon, sapphire, diamond, and gold. During descent in 2007
the parachutes never opened and the capsule crashed, but fortunately the samples
could be retrieved and analyzed.
NASA's Stardust mission, in turn, flew through the tail of comet 
Wild-2 in 2004 and captured dust particles at a speed of 6.1 km s$^{-1}$ 
both in aerogel and in Al foils. Those samples were brought 
to Earth in 2006. 

Inferring the elemental  composition of the Solar System from this variety
of sources becomes a complex enterprise. First, Earth does not show a 
homogeneous chemical composition. Second, meteorites do not constitute a 
chemically unique class of objects, but different types exhibit large 
abundance differences.  The usual approach to infer the bulk of
Solar System abundances relies on a very rare and primitive type of 
meteorites:
CI carbonaceous chondrites, representing the least modified and chemically
unaltered class of meteorites. Named after the Ivuna meteorite, the
CI carbonaceous chondrite class includes only four other members:
Orgueil, Alais, Tonk and Revelstoke. 
Because of the extreme similarities of the abundances in CI chondrites 
(determined through mass spectroscopy, wet chemistry, X-ray fluorescence,
and neutron activation)
and those inferred from the solar photosphere (for non-volatile elements),
it is believed that this class of meteorites represents pristine material 
from which the Solar System formed about 4.6 Gyr 
ago. Unfortunately, noble gases and other very volatile elements do not
easily condense into solids. 
Hence, values inferred from solar photospheric spectra 
are used instead to account for the 
C, N or O  abundances, while other specific techniques are adopted for 
some special elements, such as Ar, Kr, Xe, or Hg. 
Moreover, the meteoritic abundances have to be expressed relative to 
an element other than H. Silicon is the species traditionally chosen, with a
defined number abundance of N(Si)=$10^6$. This 
requires renormalization of the meteoritic values, such that
the photospheric and meteoritic Si abundances agree. The procedure guarantees
that both abundance determinations rely on the same absolute scale.

It is worth noting that photospheric abundances face two 
major drawbacks (besides the fact that Li is depleted in the solar atmosphere).
On the one hand, the precision achieved in photospheric
abundance determinations is much lower than what can be obtained using mass 
spectroscopy measurements of meteoritic samples in laboratories.
On the other hand, photospheric abundances are not directly measured
but inferred from observations. This requires state-of-the-art models
of the solar atmosphere as well as an accurate account of  
spectrum-formation processes. As an example, the recent improvements
in atomic and molecular transition probabilities, coupled with the
use of three-dimensional (3-D) 
hydrodynamic solar atmosphere models that relax the assumption of a
local thermodynamic equilibrium in spectral line formation, translated
into a severe reduction of the abundances of C, N and O compared to
previous estimates \cite{Asp05,Asp09}. 
Furthermore, the existing 3-D models 
proved successful in reproducing  a suite of characteristic solar features 
that 1-D models cannot account for,
including the observed solar granulation topology, typical length- and 
time-scales, convective velocities, and specific
profiles of metallic lines.
One of the major sources
of uncertainty still remaining is the lack of accurate cross sections
for excitation and ionization.

Unfortunately, photospheric abundance determinations of a few elements
are not possible through spectroscopy of the quiet Sun, and other techniques
must be implemented. For instance, this is the case for the F and Cl 
abundances, inferred from infrared spectroscopy of sunspots (through 
molecular lines of HF and HCl, respectively).
With regard to noble gases, no photospheric lines are available because of their
high excitation potentials. Indirect determinations, using helioseismology
(He), X-ray and ultraviolet spectroscopy of the solar corona and solar flares
(Ne), or solar wind measurements (Ne, Ar), among others, are used to this
end.

It is remarkable that while variations in the {\it elemental} composition 
have been found within the different existing sources, that can be
attributed to a variety
of physical, chemical and geological processes, the bulk {\it isotopic} composition
of the Solar System is extremely homogeneous.
This circumstance turns out to be extremely useful since isotopic ratios of the Sun cannot
be derived easily. In fact, direct isotopic measurements in the Sun can only be
performed through the analysis of CO vibration-rotation lines in the infrared 
spectrum. Therefore, except for H and the noble gases, isotopic abundances of the Solar 
System are directly obtained from terrestrial samples. Other sources are used for 
estimating isotopic ratios of the most volatile elements (e.g.,
$^3$He/$^4$He),
such as the solar wind or lunar samples. In other cases, such as
deuterium or $^{14}$N/$^{15}$N, estimates rely on measurements of the Jovian 
atmosphere.

It is worth noting that
independent estimates of the chemical composition of the Sun can also
be obtained by comparison with solar-like stars, OB stars (hot, short-lived,
massive main sequence stars), H II regions
(large clouds of gas and plasma, characterized by the presence of ionized H,
with recent star forming episodes, such as the Orion nebula), or with the interstellar medium in the solar
neighborhood. Once the effects of diffusion in the Sun and the chemical
enrichment of the Galaxy during the past 4.6 Gyr are taken into account,
very good agreement is achieved. 
Many compilations of the Solar System abundances
have been published and the interested reader is referred to Refs.
\cite{Lod03,Asp05,Asp09,And89,Gre98}.

\begin{figure}
\includegraphics[scale=.70]{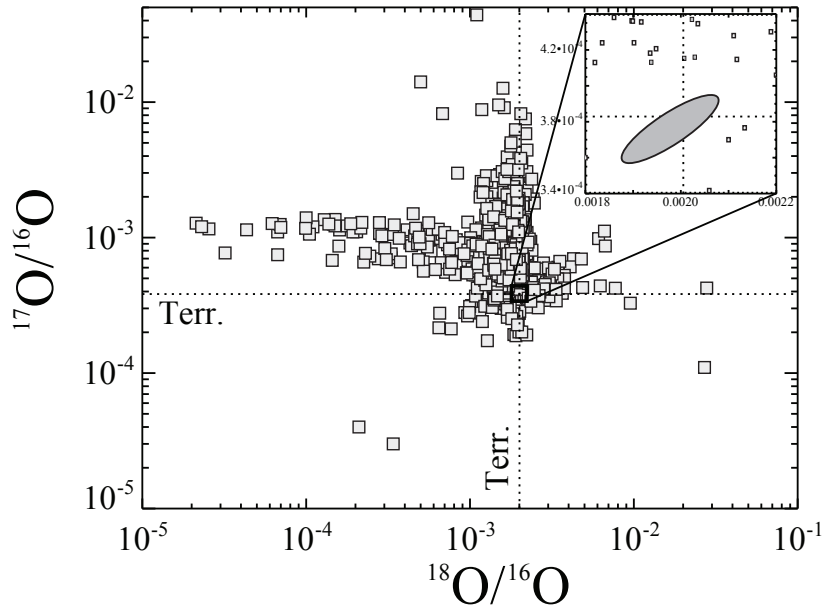}
\caption{Oxygen isotopic (number) abundance ratios measured in presolar oxide 
and silicate grains. 
The ellipse in the inset covers the range of $^{17,18}$O/$^{16}$O
ratios, on a linear scale, measured in almost all Solar System samples 
(dashed lines indicate 
terrestrial ratios). Presolar grains make up at most a few hundred parts per 
million (ppm) of the mass in the most primitive meteorites. Notice that their oxygen isotopic ratios deviate by orders of magnitude from what is found in bulk solar system matter.
Grain data are based on Nittler et al. (2008) \cite{Nit08},  
Gyngard et al. (2010) \cite{Gyn10}, and Nguyen et al. (2010) \cite{Ngu10}, 
and references therein.  The terrestrial and Solar System values are based 
on Clayton (1993) \cite{Cla93} and Lodders (2003) \cite{Lod03}. Figure
courtesy of L. Nittler.
\label{Larry}}
\end{figure}

Starting in the 1960s, studies have revealed that a
minor fraction of the material embedded in meteorites (some
micron-sized and smaller dust particles) showed huge
isotopic anomalies when compared with standard Solar System material
(see Fig. \ref{Larry}).
This has been interpreted as heterogeneities present already in the Solar 
nebula before the first solids began to condense and hence have been
coined {\it presolar grains}. Unfortunately, the tiny amount of
radioactive material present in those grains does not allow us at present to
directly confirm this conjecture through radioactive dating.
These grains are presumably produced 
by ancient stars of previous generations. 
So far, silicon carbide (SiC),
graphite (C), diamond (C), silicon nitride (Si$_3$N$_4$), silicates, and oxides
(such as corundum and spinel) have been identified as presolar
grains. Ion microprobe analyses of single presolar grains 
reveal a variety of isotopic signatures that allow, via comparison with predictions from
stellar models, an identification
of their parent stars.  
Low-mass red giants and asymptotic giant branch (AGB) stars (Sec. 5) have been 
identified as stellar sources of the majority of SiC and oxide
grains. Supernovae (most likely of the core-collapse variety; see Sect. 6) are likely
the sources of a small fraction of the SiC and Si grains,
 of low-density graphite grains, and of a few corundum grains. 
 Moreover, nanodiamonds carry a supernova signature in their Xe and Te 
 isotopic abundances. Finally, classical novae have been claimed
as the likely stellar parents of a small number of SiC, graphite
and oxide grains.
There are many excellent reviews on presolar meteoritic grains, and the
interested reader will find more information in Refs. 
\cite{Cla04,Lod05,Lug05,Mey06,Zin07}.

\section{Big Bang nucleosynthesis\label{BBN}}

\subsection{Early universe. Microwave background radiation}
Initial ideas for the Big Bang model were proposed by Gamow, Alpher, Bethe, 
Herman 
and collaborators in the 1940s in order to explain the origin of the chemical 
elements. They proposed that the Universe was initially very dense and hot, 
and that it expanded and cooled to its present state. The elements would then
be synthesized during an early time when the temperature and density 
conditions were appropriate for nuclear reactions to occur \cite{Gam46,Alp48}. 
Such a model would also predict a relic background radiation of photons with 
a present temperature of a few kelvin \cite{Alp49}. While the bulk of the 
elements are produced in stars rather than the early Universe, the idea 
proved correct for the origin of the light species $^{2}$H, $^{3}$He, $^{4}$He 
and $^{7}$Li. In addition, the observation of the cosmic microwave background 
radiation \cite{PW65}, corresponding to a blackbody spectrum with a 
temperature of 2.73 K, was of paramount importance in this regard. It 
singled out the Big Bang theory as the prime candidate for the model of our 
Universe.

Modern theories of cosmology are based on the assumption of a homogeneous and 
isotropic Universe, as implied by the Cosmological Principle. The geometry and 
evolution of the Universe can then be predicted by the theory of general 
relativity, with a number of cosmological parameters describing the spatial 
curvature and overall expansion. The Hubble parameter, $H$, provides a measure 
for the expansion rate of the Universe and its present-day value is called 
Hubble's constant, $H_0$. The mass density of baryons (more precisely, of 
nucleons), $\rho_b$, can be expressed relative to the critical density, 
$\rho_c$, by introducing the cosmological baryon density parameter, 
$\Omega_b\equiv\rho_b/\rho_c$. The critical density describes the borderline 
between a closed and an open Universe, that is, the total density at which 
the Universe is spatially flat, and is defined by the Friedmann equation as 
$\rho_c\equiv3H_0^2/8\pi G$, with $G$ the gravitational constant. The number 
density of photons is expected to have been constant since the epoch of 
electron-positron annihilation, which occurred during a time of 
$\sim$4--200~s after expansion began. The photon number density after that 
epoch can be found from the precise value for the present-day temperature of 
the cosmic microwave background radiation, $T=2.725\pm0.001$~K \cite{Mat99}, 
and is given by $n_\gamma=410.5$~cm$^{-3}$. Using the definitions $\eta 
\equiv n_b/n_\gamma$, $\eta_{10} \equiv \eta 10^{10}$ and $h \equiv 
H_0/100$~km~s$^{-1}$~Mpc$^{-1}$, one finds for the relationship of the 
parameters introduced above the expression $\Omega_b~h^2=\eta_{10}/273.9$.

The microwave background radiation carries a record of the conditions in the 
early Universe at a time of last scattering, when hydrogen and helium nuclei 
recombined with electrons to form neutral atoms. As a result, photons decoupled
from baryons and the Universe became transparent to radiation. Any oscillations
in the photon-baryon fluid around that time ($\sim$400,000 years after the 
beginning of expansion) would give rise to tiny variations of temperature (on 
the $10^{-5}$ level) in different parts of the present microwave sky. The
precise mapping of these anisotropies by NASA's Wilkinson Microwave Anisotropy 
Probe (WMAP) represented a spectacular success for cosmology. The observed 
anisotropies are usually decomposed in terms of spherical harmonics, where 
each term describes the magnitude of the anisotropy on a particular angular 
scale. The observed features in the resulting angular power spectrum are 
closely related to specific cosmological parameters. The analysis of the 
5-year WMAP data \cite{Hin09,Kom09}, when combined with other constraints 
\cite{Per07,Kow08}, yields values of $h=0.705\pm0.013$ for the dimensionless 
Hubble parameter and $\Omega_b~h^2=0.0227\pm0.0006$ for the physical baryon 
density. From these results one can derive a number of other parameters: 
$t_0=13.72\pm0.12$~ Gyr for the age of the Universe, $\rho_c=(9.3\pm0.2)\times 
10^{-30}$ g cm$^{-3}$ 
for the critical density, $\Omega_b=0.044\pm0.003$ for the 
baryon density, and $\eta=(6.2\pm0.2)\times 10^{-10}$ for the baryon-to-photon 
ratio. The WMAP data also provide estimates of $\Omega_c=0.23\pm0.01$ for the 
dark matter density and $\Omega_\Lambda=0.73\pm0.02$ for the dark energy 
density. These imply, respectively, that ordinary (baryonic) matter makes up 
only $\sim$20\% of all matter and that the expansion of the Universe is 
presently accelerating.

\subsection{Nucleosynthesis. The lithium problem}
Besides the cosmic microwave background, the other relic of the Big Bang 
is the abundance distribution of the light nuclides $^{2}$H, $^{3}$He, 
$^{4}$He and $^{7}$Li. Obviously, it is of utmost importance for any 
useful cosmological model that these two signatures provide a consistent 
description. When the Universe was less than 0.5~s old, at temperatures 
of $T\gtrsim15$ GK, the energy density was dominated by radiation (photons 
and neutrinos). The process $e^- + e^+ \leftrightarrow \nu + \bar{\nu}$ 
equilibrated the electron and neutrino gases, while the weak interactions 
$e^- + p \leftrightarrow \nu_e + n$, $e^+ + n \leftrightarrow \bar{\nu}_e + p$ 
and $n\leftrightarrow p + e^- + \bar{\nu}_e$ coupled the electron and 
neutrino gases to the baryon gas. In equilibrium the neutron-to-proton 
number ratio is determined by a Boltzmann distribution, $n_n/n_p=e^{-Q/kT}$, 
where $Q=1293.3$~keV \cite{Aud03} denotes the neutron-to-proton mass 
difference. During expansion and cooling, a temperature is eventually 
reached where the neutrino (weak) interaction processes become too slow to 
maintain equilibrium. The freeze-out of the weak interactions depends on 
their cross sections and occurs near $T\sim$15~GK, at a time of 
$\sim$0.5~s, when the neutron-to-proton number ratio is near $2/5$. 
Beyond a time of $\sim$10~s, when $T\sim$3~GK, the decay of free 
neutrons to protons, with a half-life of $T_{1/2}=614$~s \cite{PDG08}, 
becomes the dominant weak interaction. As will be seen below, further 
expansion and cooling gave rise to the onset of primordial nucleosynthesis. 
This stage is reached at a temperature and density of $T\sim$0.9~GK 
and $\rho_b \sim 2\times10^{-5}$ g cm$^{-3}$, respectively, near a time 
of $\sim$200~s, when nuclear reactions can compete with the destruction 
of nuclei by photons from the high-energy tail of the Planck distribution. 
By this time, the decay of free neutrons gave rise to a neutron-to-proton 
number ratio of about $1/7$. 

The subsequent nuclear reactions are relatively fast and, for reasons given below, 
nearly all neutrons are incorporated into the tightly bound species $^{4}$He, while 
only very small amounts of other nuclides are synthesized. Under such conditions, the 
primordial $^{4}$He abundance can be estimated by using a simple counting argument: 
for a ratio $n_n/n_p=1/7$, two out of each eight nucleons will end up bound in $^{4}$He. 
Consequently, we find for the predicted primordial helium mass fraction a value of 
$X_{\alpha}^{pred}\sim 2n_n/(n_p + n_n)=2/8=0.25$. From the above arguments it is 
apparent that the primordial $^{4}$He abundance is determined by the weak interaction 
cross sections (which are normalized by the neutron half-life), the neutron-proton mass 
difference, and the expansion rate, but is rather insensitive to either the baryon 
density or to any nuclear reaction cross sections. Observations of $^4$He in metal-poor 
clouds of ionized hydrogen (H II) in dwarf galaxies reveal that there is a small 
contribution from stellar nucleosynthesis and that it is correlated with metallicity. 
Extrapolation to zero metal abundance yields an observed primordial $^4$He mass fraction of 
$X_\alpha^{obs}=0.248\pm0.003$ \cite{Pei07}. The agreement with the predicted value 
can be regarded as a key piece of evidence for the standard cosmological model.

For a time window of $\sim$200--1000~s after expansion began, corresponding to 
temperatures and densities of $T\sim$0.9--0.4~GK and 
$\rho_b \sim 2\times10^{-5}-2\times10^{-6}$ g cm$^{-3}$, 
respectively, the early 
Universe passed through an epoch of nucleosynthesis. Note that the cosmic microwave 
background radiation and primordial nucleosynthesis probe very different eras of the 
cosmic expansion. Initially, at the higher temperature end, the strong and electromagnetic 
interactions are sufficiently fast to ensure (quasi-static) equilibrium among the abundances 
of the light nuclides $^2$H, $^3$H, $^3$He, $^4$He, $^7$Li and $^7$Be. As the Universe 
expands and cools, the nuclear reactions slow down, both because there is a decrease
 in the density and because the Coulomb barriers become harder to overcome. As a
 result, individual reactions freeze out of equilibrium at characteristic values 
of temperature. The final abundance of a particular nuclide in primordial 
nucleosynthesis is then mainly given by the rate ratio of the largest production 
and destruction reactions at the freeze-out temperature \cite{Esm91,Smi93}. Nuclear 
reactions cease completely once temperature and density are sufficiently low. 
Among the synthesized light species, $^4$He becomes by far the most abundant 
one, since it has a much higher binding energy per nucleon 
($B/A=7.074$ MeV \cite{Aud03b}) compared to all other nuclides in this mass region.

\begin{figure}
\includegraphics[scale=.60]{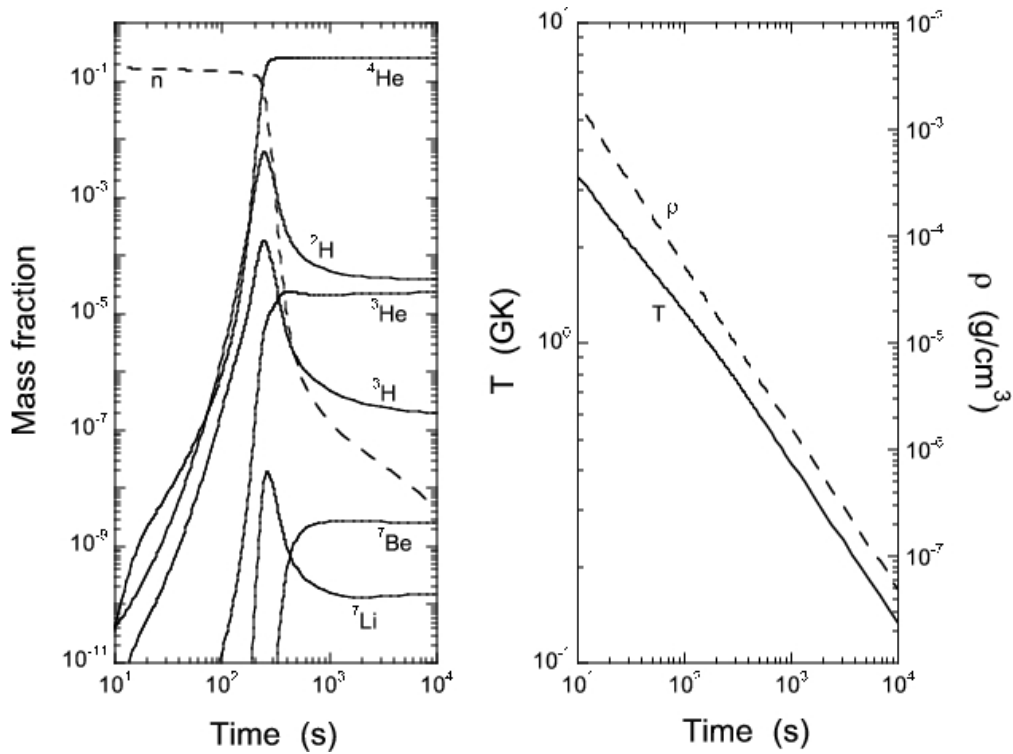}
\caption{Big Bang nucleosynthesis computed using the WMAP value of the baryon-to-photon
 ratio, $\eta_{10}^{WMAP}=6.2\pm0.2$. (Left panel) Mass fractions of the most important 
light species versus time. The hydrogen abundance (between 0.75 and 0.83 by mass) 
is almost constant on this scale and 
is not displayed. The neutron abundance (dashed line) declines even at late times 
because of radioactive decay. Note that the species $^7$Li is mainly produced as 
radioactive $^7$Be. (Right panel) Evolution of temperature and density. Significant 
nucleosynthesis starts at $t\sim$200~s, when $T\sim$0.9~GK and 
$\rho \sim 2\times 10^{-5}$ g cm$^{-3}$. For comparison, the density of air at room 
temperature amounts to $\rho_{air} \sim 1.2\times 10^{-3}$ g cm$^{-3}$. 
Data are 
courtesy of A. Coc.
\label{figBB}}
\end{figure}

Primordial nucleosynthesis starts with the p(n,$\gamma$)d reaction\footnote{The nuclear physics notation for interactions is as follows: p(n,$\gamma$)d stands for the {\it nuclear reaction} $p+n\rightarrow \gamma + d$, where $\gamma$ denotes a $\gamma$-ray; $^{13}$N($\beta^+ \nu$)$^{13}$C denotes the $\beta$ (positron)-decay of $^{13}$N to the daughter $^{13}$C, where $\nu$ denotes a neutrino. The symbols $p$, $n$, $d$, $t$ and $\alpha$ stand for protons, neutrons, deuterons ($^2$H), tritons ($^3$H) and $\alpha$-particles ($^4$He), respectively.}, resulting in a 
rapid increase of the deuterium abundance followed by destruction via d(d,n)$^3$He, 
d(d,p)t and t(d,n)$^4$He. Tritium is produced by the d(d,p)t and $^3$He(n,p)t 
reactions, and mainly destroyed via t(d,n)$^4$He. The most important production 
and destruction reactions for $^{4}$He are t(d,n)$^4$He and $^4$He(t,$\gamma$)$^7$Li, 
respectively. The species $^7$Li is mainly produced as $^7$Be via
 $^3$He($\alpha$,$\gamma$)$^7$Be, with $^7$Be(n,p)$^7$Li and $^7$Li(p,$\alpha$)$^4$He 
as the most important destruction mechanisms. For most of the reactions, direct cross
 section measurements exist at relevant energies corresponding to the individual 
freeze-out temperatures \cite{Des04}. Only for the p(n,$\gamma$)d reaction are the 
rates largely based on theory \cite{And06}, although the rate uncertainty has been 
estimated to be on the order of only $\sim$1\%. Precise primordial abundances 
are computed numerically using reaction network codes \cite{Wag67}, incorporating 
all important production and destruction reactions. The rates of nuclear reactions 
depend on the baryon density, or the baryon-to-photon ratio $\eta_{10}$, which is 
the only free parameter for Big Bang nucleosynthesis in the standard cosmological 
model. If one adopts the WMAP value, $\eta_{10}^{WMAP}=6.2\pm0.2$, in the 
nucleosynthesis calculations, then standard primordial nucleosynthesis becomes a 
parameter-free model. A comparison of computed final abundances to observed values 
can then be used to investigate the galactic chemical evolution of the light species 
or to search for possible extensions of the standard cosmological model. The results
 of such a reaction network calculation are displayed in Fig.~\ref{figBB}.

The primordial abundance of $^4$He, computed using $\eta_{10}^{WMAP}$ and 
the mean neutron lifetime from Ref. \cite{PDG08}, amounts to 
$X_\alpha^{pred}=0.2486\pm0.0002$ \cite{Cyb08}. The result agrees well 
with the observational value quoted above. For the number abundance ratio 
of deuterium to hydrogen, the predicted value using $\eta_{10}^{WMAP}$ has 
been estimated to $(D/H)^{pred}=(2.5\pm0.2)\times 10^{-5}$ \cite{Cyb08}. 
Deuterium is observed via isotope-shifted (Lyman-$\alpha$) absorption lines 
arising from low-metallicity gas clouds that fall on the line of sight 
between the observer and a high-redshift quasar. The low metallicity implies 
that the processing of deuterium by previous-generation stars, which always 
deplete deuterium during hydrogen burning \cite{Ili07}, is negligible. Averaging 
the existing seven high-quality measurements yields a value of 
$(D/H)^{obs}=(2.8\pm0.2)\times 10^{-5}$ \cite{Pet08}. Considering that the 
computed deuterium abundance depends rather strongly on $\eta_{10}$, the 
agreement between predicted and observed values of $D/H$ can be regarded as 
another key piece of evidence in favor of the standard cosmological model. 
Finally, the predicted number abundance ratio of $^7$Li to hydrogen at 
$\eta_{10}^{WMAP}$ amounts to 
$(^7Li/H)^{pred}=(5.2\pm0.7)\times 10^{-10}$ \cite{Cyb08}. It has been known 
for some time that unevolved metal-poor dwarf stars exhibit a remarkably 
constant lithium abundance \cite{Spi82}, independent of metallicity. The
 ``Spite plateau" has been interpreted as being representative for the primordial 
lithium abundance. Recent work has demonstrated that, at the lowest observed 
metallicities, there is a sizable scatter in lithium abundance and that the
 lithium abundance is correlated with metallicity \cite{Bon07}. However, the 
measurement of the primordial $^7$Li abundance remains a challenging problem.
 One source of systematic uncertainty is the determination of the effective 
temperature of the stellar atmosphere in which the lithium absorption line is 
formed. Another problem is the possible depletion of lithium, either before 
the currently observed stars were formed or within the stars we currently observe. 
Most recent studies of metal-poor field and globular cluster stars yield estimates 
for the primordial $^{7}$Li abundance in the range of 
$(^7Li/H)^{obs}\sim 1.1-2.3\times 10^{-10}$ \cite{Rya00,Asp06,Bon07}, where 
the spread in values indicates the current systematic uncertainty. Despite these
 observational difficulties, the predicted and observed values are clearly in 
disagreement (at the $4-5\sigma$ level). This ``lithium problem" represents the 
central unresolved issue in primordial nucleosynthesis. 

At present it seems unlikely that this problem is caused by either erroneous 
reaction cross sections or by wrong effective temperatures. Recent work has 
focused on two possibilities: (i) depletion of $^7$Li in stellar interiors, and 
(ii) physics beyond the standard cosmological model. The first possibility has 
been discussed in the literature for a long time and many different mechanisms 
have been proposed, such as atomic and turbulent diffusion processes, meridional 
circulation, gravity waves, or rotational mixing \cite{Mic91}. The second 
possibility is more intriguing, and may involve a time variation of the fundamental 
coupling constants \cite{Ich02,Coc07}, modification of the expansion rate during 
Big Bang nucleosynthesis \cite{Ser02,Coc06}, neutrino degeneracy \cite{Ori02},
 or negatively charged relic particles \cite{Kus07}. Clearly, the future
 resolution of the lithium problem promises to be an exciting intellectual endeavor.

\section{The Sun and low-mass stars}
Stars form in large interstellar clouds of dim, cold gas by gravitational
contraction, in timescales ranging between $\sim$0.01 -- 100 Myr \cite{Ibe65}. 
Once a critical ignition temperature of $\sim 10^6$ K is achieved, protostars
begin to fuse deuterium. Stars reach equilibrium configurations in which
the gravitational pull is balanced by the internal (thermal) pressure. 
But unlike planets, 
stars achieve equilibrium when their central temperatures are high 
enough to induce thermonuclear
fusion reactions, which become their main source of energy. 
The minimum mass required to form a star is about 0.08 
M$_\odot$.  Protostars with masses below that threshold value are called
{\it brown dwarfs}, and never reach sufficient temperatures to initiate hydrogen 
fusion. Nevertheless, brown dwarfs heavier than $\sim$13 times the mass of Jupiter do fuse deuterium during a short period.
The first thermonuclear fusion
stage that all stars undergo is central H-burning (T $\sim 10^7$ K), 
by far, the longest and most stable phase in the life of a star. 
The reason is twofold: first,
hydrogen is the most abundant element in the Universe (and hence, in stars);
and second, H has a small Coulomb 
barrier, which favors nuclear fusion even at moderately low temperatures.
Astrophysicists refer to this initial stage
as the {\it main sequence}, and its duration relies almost exclusively on the
overall stellar mass. 
A simple estimate of the time spent by a star on the main sequence
can be derived from a dimensional analysis: the energy emitted by a star, $E$, 
in a time $t$, can be expressed as $L = E/t$, where $L$ is the star's 
luminosity. Since $E$ is mainly provided
by nuclear fusion reactions, it can be expressed as $E = f M c^2$, with 
$f$ being the fraction of the star converted into energy through H-burning. 
This yields $t_{MS} = f M c^2 / L \propto M / L$. 
But the relationship between the star's luminosity and its mass is far from 
being simple.
In contrast to what is usually found in many stellar astrophysics
textbooks, no single power law for a mass-luminosity relationship fits the entire 
main sequence. Empirically, one finds $L \propto M^{\alpha}$, with $\alpha$ 
in the range of 1.9--4.8 \cite{Sho03}. 
Taking this expression into account, and using a mean
value of $\alpha = 3.5$, one finds 
$t_{MS} \propto M / L \propto M / M^{3.5} \propto M^{-2.5}$. Consequently, the larger the mass 
of a star, the faster it evolves and eventually dies.
Indeed, the duration of the main sequence phase decreases dramatically 
for massive stars. For instance, a star like the Sun spends about 
10 Gyr at this stage, fusing $\sim$600 million tons of hydrogen to helium per 
second. A star with only a tenth of a solar mass would require about 3 
trillion years to consume its hydrogen fuel, a time that is much longer than the 
current age of the Universe. In sharp contrast, a star of 100 M$_\odot$ 
would invest only 200,000 yr on the main sequence. 
Moreover, since radioactive dating yields a value of 4.6 Gyr for the age of 
the Solar System, it is expected that the Sun will spend another 
$\sim 5$ Gyr on the main sequence.

One of the natural consequences of H-burning is the fact that the 
central layers of a star become progressively enriched with He. 
Actually, once the central hydrogen content is exhausted, the He-rich 
region can again gravitationally contract 
toward a new equilibrium configuration. Depending on its mass, a star can 
either become a stable {\it He white dwarf}
or reach sufficiently high temperatures to fuse helium to carbon (Sect. 5). 
This requires temperatures near $10^8$ K because of the
larger Coulomb barrier associated with the $^4$He nucleus. 
As a result, the star is again stabilized, winning its second match against 
gravity. It is also worth noting that the overall duration of the 
He-burning phase, about 100 Myr for a 1 M$_\odot$ star \cite{SeFu07}, is shorter that the main sequence stage. This is because helium fusion produces less
energy per gram of fuel than hydrogen fusion. 

Helium burning results in a carbon- and oxygen-rich core. Central He exhaustion forces the stellar CO-rich core to contract, rising 
 once more its temperature. Stars with masses up to
8 M$_\odot$ are not able to achieve sufficient temperatures to ignite another burning episode.
For instance, carbon burning requires about $0.6 -1$ GK.
Hence, such stars end their lives as {\it CO white dwarfs}, representing
stellar remnants of planetary size that are supported by the pressure of degenerate 
electrons. 

\subsection{Hydrogen burning: proton-proton chains vs. CNO-cycles}
During H-burning, four protons are transformed into a single $^{4}$He
nucleus, with an energy release of 26 MeV per process. This is accompanied
by the emission of two positrons and two neutrinos. Two major modes
for H-burning are possible in principle. The first one is called
{\it proton-proton} (or $pp$) chains and it is the dominant mode at low
temperatures or in low metallicity environments. 
This burning regime is initiated by the reaction $p+p\rightarrow d + e^+ + \nu$, where 
$e^+$ denotes a positron. Alternatively,
two protons may fuse via the $pep$-reaction, $p+p+ e^- \rightarrow d + \nu$. The latter reaction depends on the electron density and thus can only compete with the former for densities of $> 10^4$ g cm$^{-3}$, 
which are usually not achieved during H-burning in low-mass stars. Subsequently, deuterium and hydrogen fuse via the reaction d(p,$\gamma$)$^3$He. 
This can be followed either by $^3$He($^3$He,2p)$^4$He (pp1 chain), 
$^3$He($^4$He,$\gamma$)$^7$Be(e$^-$,$\nu$)$^7$Li(p,$^4$He)$^4$He 
(pp2 chain), or $^3$He($^4$He,$\gamma$)$^7$Be(p,$\gamma$)$^8$B($\beta^+ \nu$)$^8$Be$\rightarrow ^4$He+$^4$He (pp3 chain). The
relevance of each particular chain depends critically on the temperature as 
well as on the $^4$He content. In the Sun the central temperature amounts to 
15.6 MK \cite{Bah89} 
and about 90\% of the nuclear energy is produced via the pp1 chain.

The second H-burning regime are the {\it CNO-cycles}. They require
somewhat higher temperatures than the pp-chains, as well
as the presence of suitable amounts of CNO-nuclei, which act as 
{\it catalyzers}. 
Because of the larger temperature dependence of the energy production rate
through the CNO-cycle, as compared with that of the
pp-chains, the CNO-cycle becomes the 
dominant mode above $\sim$20~MK. 
 Hence, the pp-chains dominate the energy production 
during H-burning in low-mass stars, whereas the CNO-cycles become 
the chief regime in massive stars, which are characterized by a higher central
temperature.
Since $^{12}$C is the fourth most abundant nuclide and proton captures destroy $^{12}$C much faster 
than $^{16}$O (which is the third most abundant nuclide, see Fig. 1),
the $^{12}$C(p,$\gamma$)$^{13}$N reaction often triggers the CNO-cycle.
 This is followed by $^{13}$N($\beta^+ \nu$)$^{13}$C(p,$\gamma$)$^{14}$N(p,$\gamma$)$^{15}$O($\beta^+ \nu$)$^{15}$N(p,$\alpha$)$^{12}$C (CNO1 or CN cycle). As was the case for the pp chains, four protons are
transformed into a single helium nucleus, while two positrons and two neutrinos are emitted.
The process $^{14}$N(p,$\gamma$)$^{15}$O has by far the smallest reaction rate among the reactions that form the CNO1 cycle and, therefore, 
 the major fraction of the initial $^{12}$C will be converted to $^{14}$N 
at the end of H-burning. Alternative cycles are
$^{14}$N(p,$\gamma$)$^{15}$O($\beta^+ \nu$)$^{15}$N(p,$\gamma$)$^{16}$O(p,$\gamma$)$^{17}$F($\beta^+ \nu$)$^{17}$O(p,$\alpha$)$^{14}$N (CNO2 or NO cycle),
$^{15}$N(p,$\gamma$)$^{16}$O(p,$\gamma$)$^{17}$F($\beta^+ \nu$)$^{17}$O(p,$\gamma$)$^{18}$F($\beta^+ \nu$)$^{18}$O(p,$\alpha$)$^{15}$N (CNO3 cycle),
and $^{16}$O(p,$\gamma$)$^{17}$F($\beta^+ \nu$)$^{17}$O(p,$\gamma$)$^{18}$F($\beta^+ \nu$)$^{18}$O(p,$\gamma$)$^{19}$F(p,$\alpha$)$^{16}$O (CNO4 cycle).

For a long time it was difficult to prove that energy production inside the Sun and other stars is achieved by nuclear fusion reactions. Direct evidence was obtained
by detecting the neutrinos emitted during H-burning in 
the solar interior using an underground laboratory located in the Homestake gold mine, South Dakota (USA) \cite{Bah64,Dav64}. 
Interestingly, the number of detected solar neutrinos disagreed
with the value expected from the {\it standard solar model}. A number of explanations were put forward to solve this {\it solar neutrino problem}. They included uncertainties in nuclear reaction rates,
problems with the standard solar model itself, as well as new physics
associated with neutrino properties \cite{Bah96}.
A number of experiments were conducted to solve the controversy,
including studies of the solar interior through helioseismology, 
improvements of the standard solar model, nuclear physics experiments, 
and extensive solar neutrino measurements at new 
facilities, such as GALLEX \cite{Kae10}, SAGE \cite{Abd99},
Super-Kamiokande \cite{Abe10}, SNO \cite{Aha10,Ahm02,Ahm01}, and Borexino 
\cite{Bel10}. 

\begin{figure}
\includegraphics[angle=-90,scale=.45]{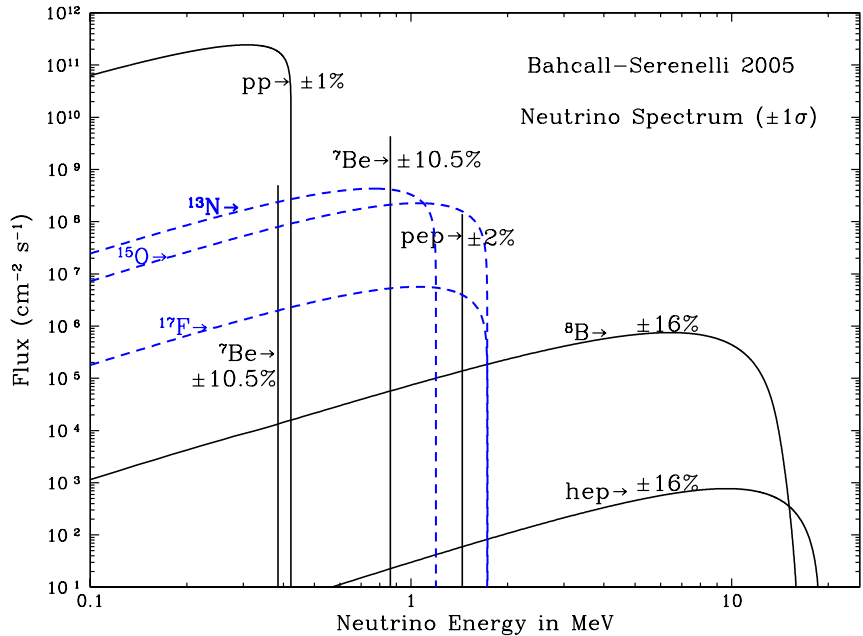}
\caption{Solar neutrino energy spectrum computed for a specific solar
model \cite{Bah05b}. Neutrino fluxes (in cm$^{-2}$ s$^{-1}$) correspond
to the predicted values on Earth. In the figure,
{\it pep} denotes the reaction $p+p+ e^- \rightarrow d + \nu$, while
{\it hep} corresponds to $^3$He(p,e$^+ \nu$)$^4$He.
The uncertainties associated with each neutrino
signal are indicated. Figure from
Bahcall, Serenelli \& Basu (2005) \cite{Bah05},
reproduced by permission of the American Astronomical Society.
\label{Fneut}}
\end{figure}

Models predict that a suite of neutrinos, with different fluxes and energies (Fig.~\ref{Fneut}), 
are produced in the solar interior \cite{Bah96,Bah05}. 
The dominant source is the p(p, e$^+$ $\nu$)d reaction, generating neutrinos 
with energies below
$\sim$0.4 MeV. These can be detected by both the GALLEX and SAGE gallium detectors. 
Recent theoretical estimates \cite{Ser09} of the solar 
 pp neutrino flux yield a value of $6 \times 10^{10}$ cm$^{-2}$ s$^{-1}$
with an associated uncertainty of 
$< 1\%$. A much smaller flux, $1.4 \times 10^{8}$ cm$^{-2}$ s$^{-1}$,
is expected for neutrinos produced in the $pep$ reaction with an energy
of 1.4 MeV, which are detectable by liquid scintillators (e.g., Borexino), or chlorine and water detectors.
Neutrinos produced by $^7$Be(e$^-$,$\nu$)$^7$Li have
discrete energies of 0.38 and 0.86 MeV, with a predicted flux of
$4.6 \times 10^{9}$ cm$^{-2}$ s$^{-1}$ ($\pm$11\%). These neutrinos should contribute substantially to the fluxes
measured in gallium and chlorine detectors. 
The flux expected from $^8$B($\beta^+ \nu$)$^8$Be neutrinos is very small, about 
$4.9 \times 10^{6}$ cm$^{-2}$ s$^{-1}$.
 However, because of their high energy (up to $\sim$ 15 MeV), the $^8$B
 neutrinos dominate the fluxes of many chlorine and water detectors, such as
 the pioneering Homestake mine experiment, or the more recent SNO and
 Super-Kamiokande experiments.  Unfortunately, the flux uncertainty associated with these neutrinos 
 ($\pm$16\%) is rather large. 
 Neutrinos produced in other reactions, such as 
 $^3$He(p,e$^+ \nu$)$^4$He ({\it hep}), 
 $^{13}$N($\beta^+ \nu$)$^{13}$C,
 $^{15}$O($\beta^+ \nu$)$^{15}$N,
 or $^{17}$F($\beta^+ \nu$)$^{17}$O, have estimated fluxes of
 $8.2 \times 10^{3}$, $2.1 \times 10^{8}$, $1.5 \times 10^{8}$, and
 $3.5 \times 10^{6}$ cm$^{-2}$ s$^{-1}$, respectively.
So far, the solar $^8$B neutrino flux has been directly measured by
 the SNO collaboration \cite{Ahm02,Aha05,Aha08} 
 and, more recently, by Borexino \cite{Bel10}. 
 Moreover, the $^7$Be neutrino flux has also been measured
 by Borexino, with a 10\% uncertainty \cite{Arp08}. Efforts to measure 
 the neutrino fluxes associated with the $\beta$-decays of $^{13}$N and
 $^{15}$O are currently ongoing at Borexino, Super-Kamiokande and the upcoming SNO+. 

One of the difficulties in solving the solar 
neutrino problem was related to the fact that the existing neutrino 
detectors were sensitive only to electron neutrinos. Indeed, the
detected number of electron neutrinos was only a third of the predicted number.
A clue 
came from Super-Kamiokande measurements in 1998, first suggesting a possible
transformation of muon neutrinos, created in the upper atmosphere by energetic 
cosmic rays, to tau neutrinos.
The first direct evidence of solar {\it neutrino oscillations} was found at SNO
\cite{Ahm01} in 2001. Solar neutrinos from $^8$B-decay were detected via
charged current reactions on deuterium (sensitive to electron neutrinos only)
and via elastic scattering on electrons (sensitive to all neutrino
flavors, i.e., electron, muon and tau neutrinos). The analysis of the data, after
a thorough comparison with the elastic scattering results obtained at 
Super-Kamiokande, revealed that only $\sim 35\%$ of the solar 
neutrinos arriving at Earth are actually electron neutrinos (the number depends 
on the exact neutrino energy). With this in mind, the experimental
neutrino flux agreed with the theoretical predictions.
In fact, the discovery of the
oscillatory nature of neutrinos firmly established the need for new
physics beyond the standard electroweak theory.

Several nuclear physics experiments, aimed at improving estimates of
solar neutrino fluxes have
been conducted in recent years, including measurements of $^7$Be(p,$\gamma$)$^8$B, 
$^3$He($\alpha$,$\gamma$)$^7$Be, and $^{14}$N(p,$\gamma$)$^{15}$O.
Indeed, the rate reduction by a factor of $\sim$2 inferred for the
$^{14}$N(p,$\gamma$)$^{15}$O reaction after new measurements performed
underground at LUNA \cite{For04,Imb05} and above ground at LENA
\cite{Run05}, had important implications for several astrophysical scenarios 
\cite{Cos09}. First, it reduces the expected flux of solar CNO neutrinos by a factor
of 2. Second, it affects the time at which stars leave the main sequence
after H-burning (so-called {\it turnoff} point). Indeed, the reduction of the 
$^{14}$N(p,$\gamma$)$^{15}$O rate translates into a smaller age for a given
turnoff luminosity. This results in an effective reduction of globular cluster ages  
by about 1 Gyr. Third,
it also provides better agreement between stellar models and observations of AGB stars
regarding the chemical abundance pattern for species heavier than carbon.

\subsection{The standard solar model: current status}
 The agreement between theoretical predictions
 and observations regarding the solar neutrino fluxes, achieved after 
  the discovery that neutrinos change flavor, gives credibility
  to the existing solar models. This discovery spawned a frantic activity 
  aimed at improving the input physics adopted in the solar models. 
  Solar abundances received particular attention. Early measurements
  of the photospheric solar oxygen content \cite{Alle01} paved the road
  for a major revision of the solar abundances that resulted in
  a dramatic reduction of the overall solar metallicity (in particular,
for CNO and Ne) with respect to previous estimates 
  \cite{Asp05}. Interestingly, the new solar models computed with the
  revised solar abundances showed severe discrepancies for the properties
  of the internal solar structure when compared with helioseismology 
  determinations. Ironically, shortly after the {\it solar neutrino problem} 
  was solved, a new and yet unsolved {\it solar abundance problem} appeared. 
  The reader is referred to Refs. \cite{Bah05b,Bas04,Del06,Bah06}
  for a thorough analysis of the impact of the new solar abundances
  on the properties of the solar interior.
 
  Standard solar models rely on the evolution of a 1 M$_\odot$  star
 from the pre-main sequence up to the current Solar System age of 4.57 Gyr.
 Models are constrained by the observed values for the present solar radius  
 and luminosity ($6.96 \times 10^{10}$ cm and $3.84 \times 10^{33}$ 
 erg s$^{-1}$, respectively). In addition, they are required to match the observed metal-to-hydrogen ratio in the solar photosphere. Emphasis is placed on 
 four important quantities that can be compared directly
 with helioseismology data:
 the surface helium mass fraction, $Y_S$, the depth of the convective
 envelope, $R_{conv}$, and the average relative differences
 of the sound speed and density profiles. 
 Whereas improved choices for the equation of state and 
for a number of nuclear reaction cross sections have shown 
 little impact on the overall properties of the models, the adoption
 of different abundance sets translated into significant differences \cite{Ser09}. For example, the abundances of Grevesse \& Sauval \cite{Gre98}, which were widely used in the past, yield 
values of  $Y_S$ = 0.2423 and $R_{conv}$ = 0.713 R$_\odot$, while  
 models computed with the improved abundances from Asplund et al. \cite{Asp05}
 give values of $Y_S$ = 0.2292 and $R_{conv}$ = 0.728 R$_\odot$.  
 For comparison, helioseismology yields $Y_S$ = $0.2485 \pm 0.0035$ 
 \cite{Bas04}, and $R_{conv}$ = $(0.713 \pm 0.001)$ R$_\odot$ 
 \cite{Bas97}. This clearly stresses that recent revisions of the solar abundances result in  
larger disagreements with helioseismology data.
Moreover, the most recent reanalysis of the solar abundances by Asplund et
al. \cite{Asp09}, which resulted in slightly higher CNO and Ne abundances compared to
their previous determinations \cite{Asp05}, does still 
 not restore the agreement between solar models ($Y_S$ = 0.2314 and $R_{conv}$ = 0.724 R$_\odot$) and helioseismology.  

Similar discrepancies are obtained between theoretical predictions 
from solar models with improved abundances and the corresponding
measurements of a suite of physical properties of the deepest
layers of the Sun (probed through helioseismology measurements with
high penetrative, low-degree modes), of neutrino fluxes, or even of 
the Solar System age. In all cases, models computed with higher solar
metallicities (as in Grevesse \& Sauval \cite{Gre98}) result in much
better agreement with helioseismology than those computed with 
lower metallicity values (as in Asplund et al. \cite{Asp05,Asp09}).
In particular, the former yield an age consistent with current estimates
for the Solar System ($\sim 4.6$ Gyr), whereas the latter suggest a
value between 4.8--4.9 Gyr \cite{ChD96}.

Titanic efforts have been invested in studying the suitable changes in the 
 physics adopted for solar models in order to achieve better 
 agreement with results from 
 helioseismology measurements.
 Particular attention deserves the 10,000 Monte Carlo simulations of 
 solar models with 21 input parameters that are randomly varied within expected
 uncertainties \cite{Bah06}. Many different solutions, but none satisfactory, have been proposed. The list includes
 changes in the solar composition (in particular, an enhancement 
 of the neon abundances \cite{Del06,Ant05,Bah05c}), enhanced microscopic
  diffusion \cite{Bas04,Mon04,Guz05},
  accretion of metal-poor material by the Sun \cite{Guz05,Cas07}, and changes 
  in the radiative opacities by 15\% \cite{Ser09,Ant05,Mon04,ChD09}.

\section{Red giants and AGB stars. The s-process\label{agb}}
\subsection{Evolution of massive stars and AGB stars}
A typical massive star of solar metallicity ($M\gtrsim 10M_{\odot}$) spends about 10  
My on the main sequence, fusing hydrogen to helium via the CNO cycles in its core.
 When the hydrogen in the center is exhausted, hydrogen burning continues in a shell.
 The core contracts and heats up until helium is ignited. This new source of nuclear 
energy heats the overlying hydrogen shell and the outer layers of the star expand 
greatly. The star shows up in the Hertzsprung-Russell diagram at the highest observed 
luminosities and becomes a supergiant. Examples are Rigel (blue supergiant) and 
Betelgeuse (red supergiant) in the constellation Orion. Core helium burning lasts 
for about 1 Myr. When helium is exhausted in the center, helium burning continues in a 
shell located beneath the hydrogen burning shell. Such massive stars are capable of 
initiating advanced burning stages in their cores, which are referred to as carbon 
burning, neon burning, oxygen burning, and silicon burning \cite{Ili07}. Each time
 a given fuel is exhausted in the core, the very same fuel continues to burn in a 
shell just outside the core. At the same time, the core contracts, thereby raising 
the temperature and pressure in order to stabilize the star. Eventually, the ashes
 from the previous core burning stage begin to fuse, providing a new source of 
nuclear energy and thus halting temporarily the contraction of the stellar core. 
The duration of each subsequent advanced burning stage decreases significantly. 
For example, while hydrogen burning may last several million years, core silicon
 burning may last only one day. The reasons are twofold. First, hydrogen burning 
releases far more energy per unit mass compared to subsequent burning stages and,
 hence, the hydrogen fuel is consumed much slower. Second, the manner by which 
the star radiates energy (or ``cools") changes dramatically. For hydrogen and 
helium burning, the nuclear energy generated in the core is carried by photons, 
which scatter many times on their way to the surface before being emitted from the star. 
Beyond helium burning, most of the stellar energy is radiated as neutrino-antineutrino pairs, 
produced via electron-positron pair annihilation ($e^+ + e^- \rightarrow \nu + \bar{\nu}$) 
or the photo-neutrino process ($\gamma + e^- \rightarrow e^- + \nu + \bar{\nu}$). 
Since the neutrino energy losses increase dramatically during the advanced burning 
stages and the neutrinos escape freely from the star, the evolution of the massive 
star accelerates rapidly. At the end of silicon burning the star consists of several 
layers of different composition, separated by thin nuclear burning shells, 
surrounding a core mainly composed of $^{56}$Fe (Sec.~\ref{sn}). 
We do not describe the advanced burning stages in detail here since a modern discussion can be found in 
standard texts (see, for example, Ref. \cite{Ili07}). However, note that the nucleosynthesis in terms of nuclear processes 
and final abundances is very similar to explosive nucleosynthesis in massive stars, which is discussed in Sec. 6.3. As will 
become clear below, the explosive burning stages operate at higher temperatures and also occur in reversed order compared 
to the advanced burning stages in hydrostatic equilibrium.
For more information 
on the evolution of massive stars, see Sec.~\ref{sn} and Refs. \cite{Lim00,Woo02}.

\begin{figure}
\includegraphics[scale=.70]{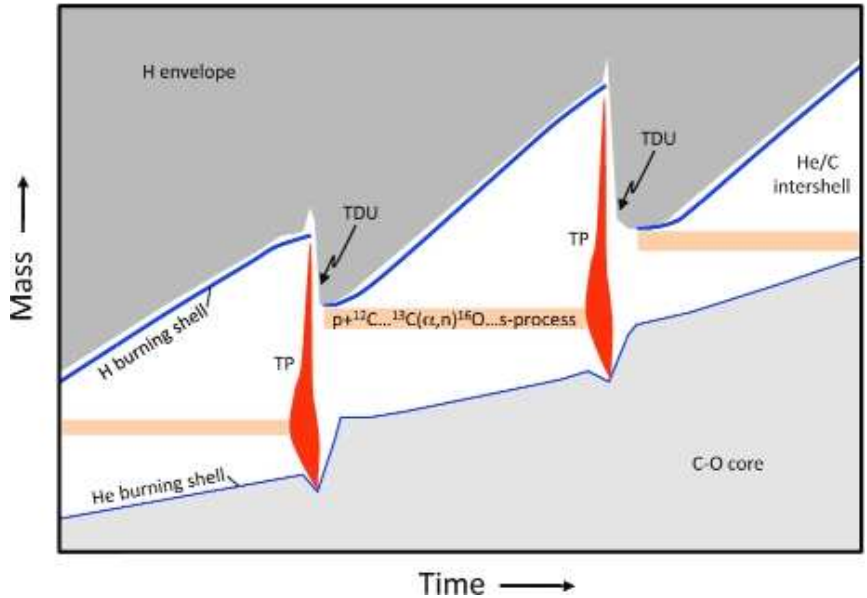}
\caption{Schematic representation (mass versus time) of a low- or intermediate-mass ($M\lesssim 10M_{\odot}$), thermally pulsing AGB star. The figure is not to scale. Shown are the convective H envelope (dark grey), the radiative $^4$He-$^{12}$C intershell (white), and the degenerate C-O core (light grey). The thick blue line indicates the H-burning shell, which is active between thermal pulses, and the thin blue line indicates the weakly active He-burning shell. 
For regular short periods of time the barely active He shell ignites in a thermal pulse 
(TP), giving rise to a convective region (red) that extends over the entire intershell, extinguishing in the process the H-burning shell. When the thermal pulse comes to an end, the convective shell disappears and the quiescent He-burning resumes. Important mixing episodes occur at the end of each thermal pulse: (i) the convective envelope reaches into the intershell so that synthesized matter is transported to the stellar surface (third dredge-up; TDU); (ii) protons diffuse from the base of the envelope into the intershell, where they are captured by $^{12}$C to produce (after $\beta$-decay) $^{13}$C. Neutrons are then released by the $^{13}$C($\alpha$,n)$^{16}$O reaction, producing in situ the main s-process component. During the subsequent thermal pulse, temperatures can be high enough to initiate the $^{22}$Ne($\alpha$,n)$^{25}$Mg neutron source. See text.\label{figagb}}
\end{figure}

Low-mass ($0.4M_{\odot} \lesssim M \lesssim 2M_{\odot}$) and intermediate-mass 
($2M_{\odot} \lesssim M \lesssim 10M_{\odot}$) stars of solar metallicity evolve 
very differently from their massive counterparts. After hydrogen is exhausted in
 the core, hydrogen burning continues via the CNO cycles in a shell and the star 
leaves the main sequence. The core contracts and the hydrogen burning shell becomes 
hotter. As a result, the surface expands dramatically and the star ascends the red giant branch. 
The envelope becomes convective, grows in size, and eventually dredges up matter 
to the surface that had experienced CNO processing (first dredge-up). 
When the temperature in the contracting core becomes sufficiently high, 
helium burning is initiated and the star now occupies the horizontal branch. 
Eventually the helium in the core is exhausted, the core contracts again, 
and the star ascends the asymptotic giant branch (AGB). At this time, stars with masses 
of $M\gtrsim 4M_{\odot}$ experience a second dredge-up event, which 
carries products of shell hydrogen and core helium burning to the stellar surface. 
Eventually, helium burns in a shell. The star now burns nuclear fuel in two shells:
 helium in a region surrounding a degenerate carbon-oxygen core, and hydrogen in a 
region surrounding the helium burning shell. The two shells are separated by a 
helium-rich intershell region. The details of the stellar models become rather 
complicated at this stage. The situation is represented schematically in Fig.~\ref{figagb}.
 It is sufficient to mention here that the hydrogen and helium burning shells, in fact,
 alternate as the main contributors to the overall luminosity. Most of the time,
 quiescent shell burning of hydrogen is the main nuclear energy source, giving rise 
to an increase of temperature and density in the barely active helium rich region 
above the carbon-oxygen core. The helium eventually ignites in a thermal pulse 
(instability). The sudden energy release produced by such a pulse drives convection 
within the helium-rich intershell and extinguishes the hydrogen burning shell. The
 star expands and cools until the thermal pulse quenches. At this point, the convective 
envelope reaches deep into the intershell region, carrying the products of helium
 burning (mainly carbon, but also elements heavier than iron; see below) to the stellar 
surface (third dredge-up). After a thermal pulse, the ashes of helium burning settle 
onto the core and the hydrogen burning shell reignites. The cycle described above may
 repeat many times, with a period on the order of about 10$^5$ y. This evolutionary 
stage is called the thermally pulsing asymptotic giant branch (TP-AGB). 
During the interpulse phase, in stars with masses of $M\gtrsim 4M_{\odot}$, 
the base of the convective envelope reaches 
down to the top of the hydrogen burning shell, with temperatures in excess of about 50 MK (hot bottom burning). 
Because the envelope is fully convective, it is completely cycled through this burning region and the products of hydrogen burning nucleosynthesis are enriched at the stellar surface. During their evolution, TP-AGB stars experience a significant mass loss via strong stellar winds, thereby enriching the interstellar medium with the products of their nucleosynthesis. The observational counterparts belong to the spectroscopic class M and are called MS, S, SC or C(N) stars, depending on their spectral properties. For example, C(N) stars are called carbon stars since they exhibit more carbon than oxygen, by number, in their atmospheres (since each thermal pulse and third dredge-up event increases the carbon abundance). Of great importance for nuclear astrophysics was the discovery of spectral lines from the element technetium in certain S stars \cite{Mer52}. All of the technetium isotopes are radioactive with half-lives of less than $\sim 4\times10^6$~yr. Such half-lives are very short compared to typical lifetimes of low- and intermediate-mass stars ($\gtrsim 10^7$~yr). Consequently, the discovery showed beyond doubt that the technetium must have been produced recently within the stars and that the products of nucleosynthesis could indeed reach the stellar surface via dredge-up processes. Asymptotic giant branch stars are believed to be the main sources of carbon and nitrogen in the Universe, and of many heavier nuclides, as will be seen below. For more information on AGB stars, see Refs. \cite{Hab04,Her05}.

\subsection{Helium burning\label{heburnsec}}
Helium burning in massive stars and AGB stars takes place in the temperature range of $T=0.1-0.4$ GK and starts with the fusion of two $\alpha$-particles. The composite nucleus $^{8}$Be lives for only $\sim 10^{-16}$~s and quickly decays back into two $\alpha$-particles. After a given time a tiny equilibrium abundance of $^{8}$Be builds up, sufficient to allow for capture of a third $\alpha$-particle to form $^{12}$C \cite{Sal52}. This process is referred to as triple-$\alpha$ reaction. It was pointed out by Fred Hoyle that the process would be too slow to account for the fusion of $^{12}$C, unless a resonance existed right above the $^8Be + \alpha$ threshold that is formed without inhibition by the centripetal barrier. The prediction and subsequent observation \cite{Dun53,Coo57} of the ``Hoyle state" in $^{12}$C represented a marvelous interplay of astrophysics and nuclear physics. The triple-$\alpha$ reaction is a (sequential) three-body interaction and thus has not been measured in the laboratory. From experimental knowledge of the nuclear masses and partial widths involved, the reaction rate can be estimated. Present uncertainties amount to about $\pm$15\%  at typical helium burning temperatures \cite{Ang99,Fyn05}, a relatively precise value for a process that has not be measured directly. Recently, a quantum three-body calculation of the triple-$\alpha$ reaction \cite{Oga09} 
resulted in a rate that differs at $T<0.25$ GK by large factors (for example, by a factor of $\sim 10^6$ at $T=0.1$ GK) from the established values. The proposed significant change of this important reaction rate, however, was found to be in 
 disagreement with fundamental observational evidence of stellar evolution \cite{Dot09} and energetics of type I X-ray bursts 
\cite{Pen10}.  

Clearly, a resolution of this controversy is of utmost importance.

Helium burning continues via the $^{12}$C($\alpha$,$\gamma$)$^{16}$O reaction. There is no resonance near or above the $\alpha$-particle threshold in $^{16}$O and thus this process must proceed via broad-resonance tails (including subthreshold resonances) and direct mechanisms. These amplitudes may interfere, causing problems in the extrapolation of the S-factor from directly measured energies ($\gtrsim 1$ MeV in the center of mass) to the astrophysically important energy range ($\sim$300 keV in the center of mass), which is at present not accessible experimentally. Knowledge of this reaction rate is of key importance. In massive stars, the ratio of the triple-$\alpha$ and $^{12}$C($\alpha$,$\gamma$)$^{16}$O reaction rates determines the ratio of $^{12}$C to $^{16}$O at the end of core helium burning. This ratio, in turn, strongly influences the subsequent hydrostatic 
burning stages, thus affecting the presupernova stellar structure, the explosive nucleosynthesis, and the nature of the remnant (neutron star or black hole) left behind after the core collapse (see below). At present the $^{12}$C($\alpha$,$\gamma$)$^{16}$O reaction rate at typical helium burning temperatures is uncertain by $\pm$40\% \cite{Kun02} and a more accurate rate (uncertainty of $<$10\%) is highly desirable. For more information on the status of this crucial reaction, see, for example, Ref. \cite{Buc06}.

A number of other important reactions occur during the helium burning stage. Since the stellar plasma most likely contains $^{14}$N (from CNO cycle operation during the preceding hydrogen 
burning stage), the reaction sequence $^{14}$N($\alpha$,$\gamma$)$^{18}$F($\beta^+\nu$)$^{18}$O($\alpha$,$\gamma$)$^{22}$Ne will be initiated. The subsequent $^{22}$Ne($\alpha$,n)$^{25}$Mg reaction takes place in massive stars and during a thermal pulse in AGB stars with $M_{AGB}\gtrsim 4M_{\odot}$ near temperatures of $T\sim$0.3--0.4 GK, giving rise to a flux of free neutrons. In AGB stars with $M_{AGB}\lesssim 4M_{\odot}$, after a thermal pulse and third dredge-up event, some protons from the convective envelope mix into the radiative intershell, which consists mainly of $^4$He and $^{12}$C. The nature of this mixing mechanism is not understood at present \cite{Str09} 
and its magnitude is usually described by a free parameter in models of 
low-mass TP-AGB stars. The protons fuse with the abundant $^{12}$C and form a thin $^{13}$C-rich pocket via the sequence $^{12}$C(p,$\gamma$)$^{13}$N($\beta^+\nu$)$^{13}$C near the top of the intershell. The subsequent $^{13}$C($\alpha$,n)$^{16}$O reaction occurs during the interpulse period under radiative conditions at temperatures near and in excess of $T\sim$0.09 GK, giving rise to a flux of free neutrons. Recent experimental work on the $^{13}$C($\alpha$,n)$^{16}$O reaction \cite{Hei09} reported a rate uncertainty of $\pm22$\% for a temperature near 0.1 GK. However, this reaction has been directly measured only for center of mass energies in excess of 280 keV \cite{Dro93}, while the astrophysically important energy range is centered near 190 keV. For the $^{22}$Ne($\alpha$,n)$^{25}$Mg reaction, the lowest lying measured resonance is located at 700 keV in the center of mass \cite{Jae01}, but typical temperatures of $T\sim 0.3$ GK translate to bombarding energies near 600 keV. Recent nuclear structure measurements of the $^{26}$Mg compound nucleus \cite{Koe02,Uga07,Lon09} resulted in a much improved reaction rate, with an estimated uncertainty of $\pm10$\% near $T=0.3$ GK \cite{Ili10}. For both of these neutron producing reactions, future direct measurements will be important.

\subsection{s-process\label{sprocsec}}
The neutrons released by the processes described above are of fundamental interest for the synthesis of about half of the elements heavier than iron. The heavier elements cannot be produced easily in stars via charged-particle reactions, since the rapidly decreasing transmission through the Coulomb barrier with increasing nuclear charge inhibits their formation. They are made instead by exposing lighter seed nuclei to a source of neutrons, such that neutron capture reactions can be initiated. The astrophysical environments discussed above give rise to relatively small neutron densities ($\sim 10^{6}-10^{15}$ neutrons per cm$^3$). Under such conditions, successive neutron captures by a chain of isotopes occur until a radioactive isotope is reached. In most cases, the $\beta$-decay probability of this species is much larger than the probability for another neutron capture to occur. Thus the radioactive nuclide will $\beta$-decay back to stability and another successive chain of neutron captures is initiated. This mechanism, referred to as the astrophysical s(low neutron capture)-process, gives rise to a nucleosynthesis path that runs close to the group of stable nuclides. The abundances synthesized by the s-process will depend on the magnitude of the neutron-capture cross sections and the total neutron flux. Nuclides with very small neutron-capture cross sections, for example, those with magic neutron numbers ($N=50$, 82 and 126) are expected to pile up in abundance, while those with large cross sections will quickly be destroyed and achieve only small abundances. This is the reason for the narrow peaks at the neutron magic numbers (corresponding to mass numbers of $A\sim$84, 138 and 208) in the solar system abundance distribution. At certain locations along the s-process path, depending on the conditions, the abundance flow encounters unstable nuclei with decay probabilities comparable in magnitude to the competing neutron capture probabilities. At these locations the s-process path splits into two branches. The analysis of  the abundance pattern of such branchings provides crucial constraints, not only for the physical conditions during the s-process (temperature, mass density and neutron density) \cite{Kae99}, but also for convective mixing (branchings at $^{128}$I) and the time scale (branchings at $^{85}$Kr or $^{95}$Zr) of a thermal pulse \cite{Kae09}. Neutron captures at the other extreme, that is, at very high neutron densities, will be discussed in Sec.~\ref{rprocsec}.

Detailed analyses of the solar system s-process abundance distribution indicated the existence of three distinct components, referred to as the main ($A\sim$90--205), weak ($A\lesssim90$) and strong ($A\sim$ 208) s-process component, which are characterized by the capture of about 10, 3 and 140 neutrons per seed nucleus, respectively \cite{Kae90}. Modern theories of stellar evolution have revealed the origin of these components. Low-mass TP-AGB stars ($M_{AGB} \lesssim 4M_{\odot}$) are believed to be the source of the main s-process component \cite{Arl99,Bus99}. Note that although in this case the $^{13}$C($\alpha$,n)$^{16}$O reaction represents the predominant neutron source, the $^{22}$Ne($\alpha$,n)$^{25}$Mg neutron source is marginally activated during a thermal pulse, thereby altering sensitively certain abundance ratios of nuclides belonging to s-process branchings. Low-mass, early-generation (metal-poor) TP-AGB stars are thought to be the source of the strong s-process component, with $^{13}$C($\alpha$,n)$^{16}$O providing most of the neutrons \cite{Gal98,Bis10}. Core helium burning and subsequent carbon shell burning in massive stars ($M\gtrsim 11M_{\odot}$) produces the weak s-process component \cite{Rai91,Pig10}, with $^{22}$Ne($\alpha$,n)$^{25}$Mg as the most important neutron source. 

The signatures of the s-process are imprinted not only in the solar system abundance distribution and in spectroscopic observations 
of stellar atmospheres, but also in meteoritic stardust (presolar grains). The latter observations are especially important for 
constraining models of AGB stars \cite{Lug03}, since the anomalous isotopic ratios in these grains can be measured with high 
precision (down to uncertainties of a few percent) \cite{Zin07}. In fact, low-mass AGB stars are predicted to be the 
most prolific sources of dust in the Galaxy \cite{Whi92} and the majority of discovered presolar grains 
indeed originate from such stars \cite{Lug10}. 
 Analysis of the isotopic composition of heavy elements in presolar grains from AGB stars can constrain,
 for example, the impact of the $^{22}$Ne neutron
source and the efficiency of $^{13}$C pocket formation \cite{Barz07}. Furthermore,          
predictions of models of dust production by AGB stars \cite{Zhu08} are in overall agreement with observations concerning low-mass stars. However, the models also predict a large production of oxygen-rich (from efficient hot bottom burning) dust in intermediate-mass AGB stars, which so far is not supported by observation. In fact, the complete lack of oxygen-rich presolar stardust grains from intermediate-mass AGB stars represents a major puzzle \cite{Lug07,Ili08}.

For accurate modeling of s-process scenarios, the neutron capture rates for a large number of nuclides in the 
$A\leq210$ mass region must be known, in addition to the rates of the neutron source reactions (see above) and the rates for 
neutron-consuming reactions (neutron poisons $^{12}$C, $^{14}$N, $^{16}$O, $^{22}$Ne and $^{25}$Mg). 
It must be emphasized that rather precise neutron capture rates are required for s-process models 
(uncertainties of less than a few percent) at energies ranging from $kT\sim$8~keV in low-mass AGB stars to $kT\sim$90~keV 
during shell carbon burning in massive stars. For many of the important neutron-capture reactions the required level
 of precision has been reached and the reliability of the current reaction rate data set is quite impressive \cite{Dil06}. 
Nevertheless, additional and more accurately measured cross sections are needed for many reactions, including those 
involving radioactive branching point nuclei. In addition, nuclear theory is indispensible for correcting the measured 
laboratory rates for effects of thermal target excitations in the hot stellar plasma. For more information on the 
s-process and related topics, see Ref. \cite{Kae11}. 

\section{Core-collapse supernovae (type II, Ib, Ic). The p-, $\nu$-, $\nu p$-, $\alpha$-, and r-processes\label{sn}}
\subsection{Classification}
The different types of supernovae are classified observationally according to their spectra. Type II supernova spectra contain hydrogen lines, while those of type I do not. Type I supernovae whose spectra show no absorption caused by the presence of silicon are referred to as  type Ib or Ic supernovae. (The distinction is based on a helium line feature in the spectrum). Type II, Ib and Ic supernovae tend to occur in the arms of spiral galaxies, but not in early-type galaxies. Since the spiral arms contain many massive and thus young stars and early-type galaxies do not contain such objects, the observations suggest that massive stars are the progenitors of type II, Ib and Ic supernovae.

Stellar model simulations and observations support the conclusion that these three supernova types result from the collapse of an iron core. Stars with initial masses less than $M\sim$20--30$M_\odot$ explode as a type II supernova and form a neutron star as a remnant. Stars with masses above this range (Wolf-Rayet stars), or less massive stars in binaries, that have lost their hydrogen envelopes are thought to be the progenitors of type Ib and Ic supernovae. It is not clear at present if the latter explosions leave a neutron star or a black hole behind as a remnant, mainly because of our incomplete knowledge of post-main sequence mass loss and the details of fallback of matter onto the central object during the explosion. As will become clear in the following, core-collapse supernovae are of outstanding importance for three broad topics: (i) they are predicted to be the most prolific sources of the elements in the Galaxy; (ii) they are the place where neutron stars are born; and (iii) they are a plausible source of shock waves that are believed to accelerate most Galactic cosmic rays.

\subsection{Evolution beyond core silicon burning}
\begin{figure}
\includegraphics[scale=.75]{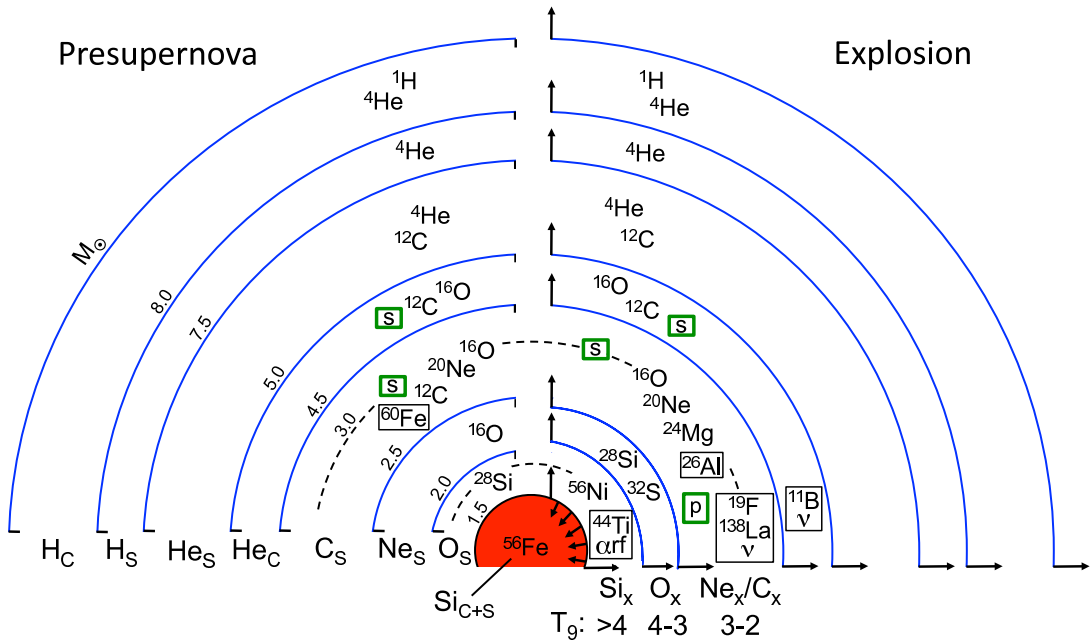}
\caption{Structure and evolution of a $25 M_{\odot}$ star of solar metallicity, as predicted by one-dimensional, spherically symmetric models \cite{Lim00,Lim10}, shortly before and after core-collapse (not to scale). Only the main constituents in each layer are shown. Minor constituents, among them important $\gamma$-ray emitters, are set in thin black boxes. Various nucleosynthesis processes are shown in green boxes: weak s-process (s); p-process (p); $\alpha$-rich freeze-out ($\alpha$rf); $\nu$-process ($\nu$). (Left) Snapshot of presupernova structure. Nuclear burning takes place in thin regions (burning shells) at the interface of different compositional layers, where each burning shell migrated outward to the position indicated by the blue lines. The compositions result from burning stages indicated at the bottom (subscripts {\it C} and {\it S} stand for core and shell burning, respectively). The diagonally arranged numbers indicate the interior mass (in solar masses) for each burning shell. (Right) Explosive nucleosynthesis resulting from passage of the shock wave through overlying layers, giving rise to explosive burning of silicon (Si$_x$), oxygen (O$_x$) and neon-carbon (Ne$_x$/C$_x$). Strictly speaking, this classification depends on the temperature range, not on the available fuel. Nevertheless, the names indicate approximately which compositional layers of the pre-supernova will usually be affected. Outside the outer dashed line, the composition is little altered by the shock. The inner dashed line indicates the approximate boundary of the part of the star that is ejected (mass cut). See text.\label{snfig}}
\end{figure}

After silicon has been exhausted in the core, a massive star ($M\gtrsim 10M_{\odot}$) consists of several layers of different composition (``onion-like" structure), separated by thin hydrostatic nuclear burning shells, surrounding a core mainly composed of $^{56}$Fe (Sec. \ref{agb}). Considering the declining number of stars of increasing mass, weighted by the fact that more massive stars eject a larger yield of heavy elements, one arrives at a ``typical" model star with a mass somewhere near $\sim$25$M_\odot$. For clarity sake, we will discuss in the following the evolution and explosion of such a star. It must be kept in mind that stars of different masses exhibit significant differences in structure, shell composition, location of nucleosynthesis processes, and so on. The situation is shown schematically in Fig.~\ref{snfig} (left side) for a 25$M_\odot$ star of initial solar composition, displaying the most abundant nuclides in each layer and, at the bottom, the nuclear burning stage that produced those particular ashes. (The subscripts {\it C} and {\it S} stand for core and shell burning, respectively.) For example, the most abundant nuclide in the core is $^{56}$Fe, which is the product of core and shell silicon burning. In the next layer, on top of the core, the species $^{28}$Si is the most abundant nuclide, which is the product of oxygen shell burning. The nuclear transformation of $^{28}$Si to $^{56}$Fe (silicon shell burning) continues at the intersection of the silicon layer and the iron core, the transformation of $^{16}$O to $^{28}$Si (oxygen shell burning) takes place at the intersection of the oxygen and silicon layers, and so on. Note the location of the weak s-process component (Sec.~\ref{agb}), both in the carbon-oxygen layer (resulting from helium core burning) and the oxygen-neon layer (resulting from carbon shell burning). The latter layer has been predicted \cite{Lim06} to be a main source of the important $\gamma$-ray emitter $^{60}$Fe.

At this point, the electron degenerate core has no further source of nuclear energy to its disposal. It grows in mass as a result of nuclear burning in the overlying shells. When the core mass reaches the Chandrasekhar limit ($M\sim$1.4$M_{\odot}$), the electron degeneracy pressure is unable to counteract gravity and, consequently, the core collapses freely at about a quarter of the speed of light. When the density reaches values near nuclear density 
($\rho\sim 10^{14}$ g cm$^{-3}$), nuclei and free nucleons begin to feel the short-range nuclear force, which is repulsive at very small distances. The collapsing inner core reaches high inward velocities, overshoots nuclear density, and rebounds as a consequence \cite{Bet79}. The rebounding part of the core encounters infalling matter, giving rise to an outward moving prompt shock wave. The hot and dense inner core has become a proto-neutron star with a mass of $M\sim$1.5$M_{\odot}$. While the shock moves outward through the outer core, it loses energy, both by photodisintegrating iron peak nuclei and by emission of neutrinos. About 1 s after core collapse, when the prompt shock reaches the outer edge of the core, it has lost its kinetic energy and stalls. How exactly the shock is revived and how it will ultimately propagate through the stellar layers beyond the iron core and disrupt the star in a core-collapse supernova is one of the most elusive problems in nuclear astrophysics \cite{Woo05}. We will address some current ideas later.

Since a self-consistent model of core-collapse supernovae is still lacking, current stellar models induce the shock wave artificially by depositing a significant amount of energy somewhere near the iron core. The magnitude of energy deposition is highly constrained, since the total predicted kinetic energy of the explosion should agree with observation. For example, observations of SN 1987A (a type II supernova with an estimated mass near 20$M_{\odot}$) and other type II supernova light curves result in explosion energies of $\sim (1-2)\times 10^{51}$~erg. The location of artificial energy deposition is also constrained by observation: it can neither be located inside the iron core or otherwise neutron-rich iron group nuclides are overproduced, nor can it be located beyond the oxygen burning shell or the resulting neutron star mass, after fall-back of matter, will be too large. In most simulations, the {\it mass cut}, that is, the boundary between ejected and fall-back matter, is located in the silicon layer (inner dashed line in top part of Fig.~\ref{snfig}). The shock wave moves through the star and heats matter to high temperatures for a time period of a few seconds. Subsequently, the hot and dense matter expands nearly adiabatically. As a result, the star experiences several episodes of explosive nucleosynthesis. The character of the explosive nuclear burning depends, among other things, on the location of the shock and the expansion time scale. During the explosion the nuclides that have been synthesized before and after the core collapse are ejected and are then mixed into the interstellar medium. In fact, core-collapse supernovae are predicted to be the source of the majority of nuclides ($A\gtrsim 12$) in the Galaxy. The situation is shown schematically in Fig.~\ref{snfig} (top right) and will be discussed in more detail below.

\subsection{Explosive nucleosynthesis\label{expnucsec}}
The first zone encountered by the shock, consisting mainly of $^{28}$Si (Fig.~\ref{snfig}), is heated to temperatures 
of $T \gtrsim 5$ GK, while the density exceeds $\rho \sim 10^8$ g cm$^{-3}$.
This matter undergoes complete explosive silicon burning, during which the rates for all forward and reverse strong or electromagnetic interactions achieve a global equilibrium. This condition is called {\it nuclear statistical equilibrium} (NSE) and has some interesting properties \cite{Cli65}. The abundance of any nuclide in NSE is independent of initial composition or thermonuclear reaction rates, but is instead given by its nuclear properties, such as binding energy, mass and spins. In fact, these properties are well-known for the nuclides participating in explosive silicon burning \cite{Aud03}. For known nuclear properties, the abundance of any nuclide in NSE is determined by three independent parameters: temperature, density, and neutron excess. The latter quantity is a measure for the number of excess neutrons per nucleon and can only change as a result of weak interactions. It is defined by $\eta \equiv \sum_i (N_i - Z_i)X_i / M_i$, where $N_i$, $Z_i$, $X_i$ and $M_i$ denote the number of neutrons, number of protons, mass fraction and relative atomic mass (in atomic mass units), respectively, of nuclide $i$ in the plasma\footnote{The neutron excess parameter is related to the electron mole fraction, $Y_e$, by $\eta = 1-2 Y_e$. For example, a plasma consisting only of nuclei with equal numbers of protons and neutrons has $\eta=0$ or $Y_e=0.5$.} 
In a 25$M_\odot$ pre-supernova star of initial solar composition the explosively burning silicon layer has a very small neutron excess of $\eta \sim$0.003. Under such conditions, NSE favors $^{56}$Ni as the main constituent \cite{Har85}, since it has the largest binding energy of any $N=Z$ species among the iron peak nuclides. Other iron group species are synthesized as well, such as $^{57}$Ni, $^{55}$Co and $^{54}$Fe. Note that exactly the same products would be obtained if the explosively burning nuclear fuel in this temperature regime would not consist mainly of $^{28}$Si, but instead of any other species with $\eta \sim$0.003. The fate of matter in NSE depends on the subsequent expansion as the gas quickly cools and, consequently, nuclear reactions will start to fall out of equilibrium at a certain freeze-out temperature. If the density and expansion time scale are sufficiently large, NSE predicts only a very small light particle (p, n, $\alpha$) abundances. As the temperature falls below the freeze-out temperature, there is not enough time to produce the light particles necessary to maintain NSE and the equilibrium will be terminated by a lack of light particles. The first reaction to drop out of equilibrium is the triple-$\alpha$ reaction. There are so few light particles that their subsequent capture during this {\it normal freeze-out} does not alter the NSE composition (mainly $^{56}$Ni and other iron peak nuclides). On the other hand, if the density or expansion time scale are sufficiently small, NSE predicts a relatively large light particle abundance, especially for $\alpha$-particles. When the temperature falls below the freeze-out temperature, the $\alpha$-particles cannot be converted fast enough to iron peak nuclei in order to maintain NSE. The equilibrium is terminated by an excess of $\alpha$-particles and their capture by nuclei during freeze-out will alter the NSE composition \cite{Woo72}. For a nuclear fuel with $\eta \sim$0, the main product of this {\it $\alpha$-rich freeze-out} is again $^{56}$Ni, but additional species are produced, for example, the important $\gamma$-ray emitter $^{44}$Ti. 

The next zone encountered by the shock, still composed mainly of $^{28}$Si, is heated to temperatures in the range of $T\sim$4--5 GK. Instead of global equilibrium among all species, equilibrium is established locally in two mass ranges: the region centered on silicon, and the iron peak. This local equilibrium in each of these regions is called {\it quasi-equilibrium} (QSE). The abundance of any nuclide in QSE, for example, with $^{28}$Si, is specified by four parameters (apart from nuclear binding energies, masses and spins): temperature, density, neutron excess, and the abundance of $^{28}$Si \cite{Bod68}. When the shock first encounters this zone, it starts to dissociate silicon and neighboring species, quickly forming a QSE cluster in equilibrium with the dominant species $^{28}$Si, which has the largest binding energy of all nuclides in this region. Soon thereafter another QSE cluster forms in the iron peak region, which consists of nuclides of even higher binding energies, mediated by reactions in the $A=40-44$ mass range linking the two clusters. If sufficient temperatures and time would be available, all of the silicon group species would be transferred to the iron peak and NSE would be established. However, the expansion causes the freeze-out to occur before NSE can be established. Since a significant amount of $^{28}$Si remains, the process is referred to as {\it incomplete silicon burning} \cite{Hix99}. The major products of this burning are iron peak species, $^{28}$Si, and intermediate-mass nuclides.

Subsequently, the shock reaches a zone composed mainly of $^{16}$O (Fig.~\ref{snfig}). Explosive oxygen burning occurs mainly at temperatures in the range of $T\sim$3--4 GK. The character of the burning is similar to incomplete silicon burning, in the sense that the fuel ($^{16}$O) is dissociated, giving rise to two QSE clusters in the mass regions of silicon and the iron peak. However, since the peak temperature is lower, less matter is converted to the iron peak and much more material remains locked in the silicon region. After freeze-out the most abundant nuclides in this zone are $^{28}$Si, $^{32}$S, $^{36}$Ar and $^{40}$Ca. Finally, the shock encounters a zone mainly composed of $^{16}$O, $^{20}$Ne and $^{12}$C, and heats it to temperatures of $T\sim$2--3 GK. As a result, $^{20}$Ne, and to a lesser extend $^{12}$C, burns explosively. In this case, the temperature and expansion time scale are too small for establishing QSE and the forward and reverse nuclear reactions operate far from equilibrium. Hence the abundance of a given species does not depend on a few parameters only, such as temperature, density, and neutron excess, but in addition is sensitively influenced by the initial composition and the magnitude of the thermonuclear reaction rates. After freeze-out the most abundant species in this zone are $^{16}$O, $^{20}$Ne, $^{24}$Mg and $^{28}$Si. Stellar models \cite{Lim06} also predict that explosive Ne/C burning is the main source for synthesizing the $\gamma$-ray emitter $^{26}$Al, which has been observed in the Galactic interstellar medium \cite{Mah82,Die95}. This observation was of paramount importance since it provided unambiguous {\it direct} evidence for the theory of nucleosynthesis in stars. The half-life of $^{26}$Al amounts to $7.17\times10^5$ yr \cite{Aud03}, which is small compared to the time scale of Galactic chemical evolution ($\sim 10^{10}$ y), and thus nucleosynthesis must be occurring currently throughout the Galaxy. The outer zones of the star are heated to peak temperatures of less than $T\sim$2~GK for very short periods of time and, consequently, do not experience significant nucleosynthesis. These layers, located beyond the outer dashed line in Fig.~\ref{snfig}, are ejected with their composition resulting from various hydrostatic burning stages prior to the explosion (left side of figure). About one hour after core collapse, the shock, traveling at an average speed of several thousand kilometers per second, reaches the stellar surface.

The sequence of events sketched above depends sensitively on all factors that influence the mass-radius relation (or density profile) of the pre-supernova star, since it determines the amount of matter exposed to the four explosive burning episodes. For example, the treatment of convection is important in this regard, because it impacts the size of the convective regions and the mixing efficiency in those zones. Also, as already mentioned in Sec.~\ref{agb}, the $^{12}$C($\alpha$,$\gamma$)$^{16}$O reaction rate determines the $^{12}$C/$^{16}$O abundance ratio at the end of core helium burning and hence impacts the amount of fuel available for the subsequent advanced hydrostatic shell burnings and the location of those shells. Furthermore, the magnitude of the time delay between core collapse and the revival of the shock impacts the amount of matter located in each compositional layer. Besides the mass-radius relation, the neutron excess (or $Y_e$) profile of the pre-supernova star is crucial for the outcome of those explosive burning episodes that take place under the conditions of NSE and QSE. The neutron excess profile, in turn, is influenced by the treatment of convection, the time delay between core collapse and shock revival, and the previous hydrostatic evolution of the star. 

\subsection{Observations}
Direct evidence for the nucleosynthesis predicted by current core-collapse supernova models can be obtained from 
observations of radioactivity and neutrinos. For example, the long-lived radioisotopes $^{26}$Al 
($T_{1/2}=7.17\times10^5$ yr \cite{Aud03}) and $^{60}$Fe ($T_{1/2}=2.62\times10^6$ yr \cite{Rug09}) are both predicted to
 be synthesized mainly by core-collapse supernovae. Their half-lives are very long compared to a typical frequency of about two 
Galactic core-collapse supernovae per century and, therefore, these nuclides accumulate after ejection from thousands of supernovae 
in the Galactic interstellar medium. The diffuse $\gamma$-ray emission from the decay of both species has been observed by detectors 
onboard satellites. The $^{60}$Fe/$^{26}$Al $\gamma$-ray line flux ratios measured by both the RHESSI instrument \cite{Smi04} and the
 SPI spectrometer onboard INTEGRAL \cite{Wan07} are near $\sim$0.15. Stellar model predictions \cite{Tim95,Lim06} seem to be 
consistent with observation, when the theoretical yields from single massive stars are folded with a standard distribution of
 stellar masses (initial mass function). However, it must be emphasized that the theoretical yields are sensitive, among other 
things, to current uncertainties in nuclear reaction rates \cite{Tur10,Ili11}. 

The radioisotope $^{44}$Ti has a half-life ($T_{1/2}=60$ yr \cite{Has01}) comparable to the Galactic supernova rate and thus probes 
individual supernovae. Gamma-rays from the decay of $^{44}$Ti have been detected (\cite{Ren06}, and references therein) in the 
330-year old supernova remnant Cassiopeia A, which may have been observed by the English astronomer Flamsteed in the year 1680, 
confirming the theoretical prediction that this nuclide is synthesized in the $\alpha$-rich freeze-out during explosive (complete) 
silicon burning in core-collapse supernovae. However, it is not clear at present why $^{44}$Ti has not been observed in any of the
 other supernova remnants. It has been speculated that perhaps Cassiopeia A resulted from an asymmetric explosion \cite{Nag98} or 
represents a peculiar event \cite{You06}. 

The best-studied supernova to date, SN 1987A, resulted from the core collapse of a progenitor star with a mass near 20$M_\odot$.
 A few months after the explosion, the evolution of the brightness (light curve) followed the radioactive decay of $^{56}$Co 
($T_{1/2}=77.2$ d) \cite{Ham88}, the daughter nuclide of radioactive $^{56}$Ni ($T_{1/2}=6.1$ d) that is the main product of 
explosive silicon burning at low neutron excess. Subsequently, $\gamma$-rays from $^{56}$Co decay were directly detected with the 
SMM satellite \cite{Mat88}. Together, both observations implied a total $^{56}$Co mass of $\sim$0.07$M_\odot$ in the ejecta \cite{Lei90}. 
Finally, a few hours before the light from SN 1987A reached Earth, 11 electron antineutrinos were recorded by the KamiokaNDE-II detector 
\cite{Hir87}, 8 electron antineutrinos by the Irvine-Michigan-Brookhaven (IMB) detector \cite{Bio87},  and 5 electron antineutrinos by the Baksan Neutrino 
Observatory \cite{Alexe88}. 
The number of detected neutrinos, their energies, and the measured burst duration are in agreement with theoretical predictions of the processes deep inside an exploding star (see below). 

\subsection{p-process\label{pproc}}
There are about 30 relatively neutron deficient nuclides with mass numbers of $A\ge 74$ (between $^{74}$Se and $^{196}$Hg) that cannot be
 synthesized by any of the known neutron capture processes. These species are called {\it p-nuclei}, and they all contain more protons relative to other stable isotopes of the same element. The mechanism responsible for the synthesis of the p-nuclei is called the {\it p-process}. It turns out that the p-nuclei are the rarest among the stable nuclides. In fact, no single element has a p-process isotope as a dominant component. In principle, there are two kinds of reactions that could allow for the production of neutron-deficient nuclei starting from heavy seed nuclei: proton captures, (p,$\gamma$), and photo-neutron emission, ($\gamma$,n). The first mechanism requires high densities, temperatures, and relatively long time scales, which are unlikely to exist in any hydrogen-rich zones of common stars \cite{Woo78}. Thus it is generally accepted that the main mechanism of p-nuclei production involves the photodisintegration of heavy seed nuclei in a hot photon environment (peak temperatures near $T\sim$2--3 GK), a hydrogen-exhausted stellar zone, and a short time scale. The abundance flow reaches from lead down to the iron peak, where further photodisintegrations become energetically unfavorable, all subject to the same hot photon environment. During this process, the seed nuclei are initially destroyed by several ($\gamma$,n) reactions along a given isotopic chain. This continues until a neutron deficient nucleus is reached that prefers a ($\gamma$,p) or ($\gamma$,$\alpha$) over a ($\gamma$,n) reaction, thus branching the abundance flow into an isotopic chain of a lighter element. Abundances tend to accumulate at these branching points, especially at nuclides with a closed neutron or proton shell structure (that is, at species with small photodisintegration rates). These {\it waiting points} become either directly p-nuclei (in the region of the lighter p-nuclei) or transmute via $\beta^+$-decays to p-nuclei after cooling, expansion and ejection of material (in the region of the heavier p-nuclei). It is important for any realistic site responsible for the synthesis of the p-nuclei that the combination of temperature and time scale guarantees the occurrence of some photodisintegrations, yet not so intense as to entirely reduce all nuclei to iron peak species. These arguments support the conclusion that the p-process occurs during stellar explosions with an associated rapid expansion and cooling of material.

Although the site(s) of the p-process have not been identified unambiguously yet, many investigations have assumed that it occurs in core-collapse supernovae, when the shock wave passes through the oxygen-neon zone (Fig.~\ref{snfig}) and heats the matter to temperatures between $\sim$2--3 GK \cite{Arn76}. During the explosion, different layers of this zone will undergo different thermodynamic histories and thus will achieve different peak temperatures. The weak s-process component operating during the preceding core helium and shell carbon burning stages (Sec.\ref{agb}) strongly enhances the p-process seed abundances in the $A\sim$60--90 mass region. It has been demonstrated that p-nuclei with masses of $A\lesssim92$, $A\sim$92--144 and $A\gtrsim 144$ are mainly produced in stellar zones with peak temperatures of $T\ge 3$ GK, $T\sim$2.7--3.0 GK and $T\le 2.5$ GK, respectively. In fact, each p-nuclide is synthesized in a relatively narrow temperature range only \cite{Ray90}. The simulations reproduce the solar system abundance of most p-nuclei within a factor of 2-3. However, there is a notable underproduction of the light p-nuclei $^{92}$Mo, $^{94}$Mo, $^{96}$Ru and $^{98}$Ru, as well as of the rare species $^{113}$In, $^{115}$In and $^{138}$La. Sites other than massive stars have also been considered and, interestingly, similar p-abundance distributions are obtained in each case. Some of the underproduced p-nuclides may be predominantly synthesized by a different mechanism (see below). For more information on the p-process, including a description of other proposed sites, see Ref. \cite{Arn03}.

A number of current nuclear physics uncertainties impact p-process simulations. For example, the $^{12}$C($\alpha$,$\gamma$)$^{16}$O rate influences the pre-supernova evolution of the massive star and hence the composition of the oxygen-neon zone prior to core collapse \cite{Ray95}. The $^{22}$Ne($\alpha$,n)$^{25}$Mg reaction (Sec~\ref{agb}) is important since it is responsible for the weak s-process component, which provides a fraction of the seed-nuclei for the p-process. With few exceptions, almost all of the many reactions ($>10,000$) in the region of the p-process, including the important ($\gamma$,n), ($\gamma$,p) and ($\gamma$,$\alpha$) photodisintegrations, have to be computed theoretically using the Hauser-Feshbach model of nuclear reactions. It has been shown that different prescriptions of the Hauser-Feshbach model ($\alpha$-nucleus and nucleon-nucleon optical potentials, and level densities) influence sensitively the final predicted p-nuclei abundances. Experimental (n,$\gamma$), (p,$\gamma$) and ($\alpha$,$\gamma$) rates on stable target nuclei in the mass range of $A>60$ play an important role in this regard. The experimental cross sections are used for adjusting Hauser-Feshbach model parameters, thus improving theoretical rate predictions for a multitude of unmeasured reactions. Furthermore, the reverse photodisintegration rates can be calculated from the measured forward rates using the reciprocity theorem. For a sensitivity study of p-process nucleosynthesis to nuclear reaction rates, see Ref. \cite{Rap06}.

\subsection{$\nu$-process}
The large energy release in the core collapse in form of neutrinos (see below) gives likely rise to the neutrino-induced synthesis of certain nuclides in the expanding mantle of the exploding star \cite{Dom78}. Specifically, neutrinos can interact with the nuclei via inelastic neutral-current neutrino scattering, ($\nu$,$\nu^\prime$). Since $\mu$ and $\tau$ neutrinos emerging from the proto-neutron star have larger predicted average energies compared to electron neutrinos (see below), and since the neutrino cross sections scale with the square of their energy, the interactions of the former neutrino species with nuclei predominate. On the other hand, electron neutrinos may interact with nuclei via charged-current interactions, ($\nu_e$,$e^-$) or ($\bar{\nu}_e$,$e^+$). All these neutrino interactions can populate excited nuclear levels that subsequently decay via emission of light particles (p, n, $\alpha$, etc.). The released light particles, in turn, undergo reactions with other nuclei in the high-temperature environment and thus contribute to the synthesis of certain nuclides. In any given layer of the star, neutrino induced nucleosynthesis may occur before the arrival of the shock, during explosive burning, or after shock passage when the material expands and cools. This mechanism is referred to as the $\nu$-process \cite{Woo90}. Simulations have shown \cite{Woo90,Yos04,Heg05} that it may contribute appreciably to the solar abundance of the rare species $^{11}$B (made in the carbon-oxygen layer; see Fig.~\ref{snfig}), $^{19}$F and $^{138}$La (both made in the oxygen-neon layer). Recall that the latter nuclide is underproduced in the p-process. Yield predictions from $\nu$-process calculations are sensitively influenced by current uncertainties in neutrino interaction cross sections, average neutrino energies of each flavor, total neutrino luminosity, as well as by the details of the adopted stellar evolution and explosion models.

\subsection{Core collapse revisited. Role of neutrinos}
As discussed above, core-collapse supernovae are powered by the release of gravitational energy in the collapse of a massive star core to a proto-neutron star. We will now discuss the core collapse in more detail. At the onset of the collapse (t=0), temperatures and densities in the center of the iron core amount to $T\sim$10 GK and $\rho \sim 10^{10}$ g cm$^{-3}$, respectively. Under these conditions, nuclear statistical equilibrium (NSE) is established. The silicon burning shell has continually increased the mass of the core, which is supported by electron degeneracy pressure. When the core exceeds the Chandrasekhar mass ($\sim$1.4$M_\odot$), it has no other thermonuclear energy source available to support the pressure and it becomes unstable to gravitational collapse. The collapse and the infall dynamics depend sensitively on two parameters\footnote{The electron mole fraction is equal to the electron-to-baryon ratio, $Y_e=N_e/\sum_i{N_i M_i}$, and hence to the proton-to-baryon ratio, $Y_p$; $N_e$, $N_i$ and $M_i$ denote the electron number density, nuclide number density and nuclide relative atomic mass, respectively, where the sum $i$ is over all nuclides. For a radiation dominated environment, the entropy per baryon, $s$, is related to the photon-to-baryon ratio, $\phi$, via $s\sim 10\phi \sim T^3/\rho$, where $T$ and $\rho$ denote the temperature and density, respectively. In simple terms, for a small value of the entropy per baryon, NSE favors a composition of iron peak nuclei, which are tightly bound. On the other hand, a large value of $s$ implies that many photons are available per baryon, which favors the photodisintegration of heavier nuclei into free nucleons \cite{Mey92}.}
:  (i) the electron mole fraction, $Y_e$, which is related to the neutron excess (see above); and (ii) the entropy per baryon, $s$. During the early stages, the core collapse is accelerated by two effects. First, as the (electron) density increases, electrons capture onto nuclei, ($e^-,\nu_e$), hence removing electrons that were contributing to the pressure. As a consequence of a decreasing $Y_e$, the neutron excess, $\eta=1-2Y_e$, increases in the core and, in addition, a burst of electron neutrinos is produced. Second, at the very high temperatures the thermal radiation becomes sufficiently energetic and intense that iron peak nuclei are photodisintegrated into lighter and less stable species, thus removing energy that could have provided pressure. Within a fraction of a second, the core with a size of several thousand kilometers collapses to a proto-neutron star of several tens of kilometer radius. The most important neutrino interactions during the collapse are (neutral current) elastic scattering on nuclei, $(\nu_e,\nu_e)$, electron-neutrino scattering, $e^-(\nu_e,\nu_e)e^-$, inverse $\beta$-decay, $(\nu_e,e^-)$, and inelastic scattering on nuclei, $(\nu_e,\nu_e^\prime)$ \cite{Bru91}. At $t\sim$0.1 s, when the density reaches a value near $\sim 10^{12}$ g cm$^{-3}$, the neutrino diffusion time becomes larger than the collapse time and, consequently, the neutrinos become trapped \cite{Bet90}. Inside this region, called {\it neutrino sphere}, the neutrinos are coupled to the matter and are in thermal equilibrium. Outside, the neutrinos escape almost freely, with an average energy that is determined at the radius of the neutrino sphere, $R_\nu$. (Note that the location of the neutrino sphere depends both on neutrino type and flavor, and on neutrino energy.) At $t\sim$0.11 s, the inner core ($M\sim$0.5$M_\odot$) reaches nuclear densities ($\sim 10^{14}$ g cm$^{-3}$), bounces, and drives a shock wave into the infalling matter. At the time of the bounce, the electron mole fraction in the inner core is relatively small ($Y_e \sim$0.3 for $M\lesssim 0.5M_\odot$), and increases gradually in the outer core ($Y_e \sim$0.45 near $M\sim$1.0$M_\odot$). At $t\sim$0.12 s, the prompt shock propagates outward, but loses severely energy ($\sim$9 MeV per nucleon) by dissociating iron peak nuclei into free nucleons. When the shock reaches the neutrino sphere, additional electron captures on free protons also remove energy from the shock, giving rise to a strong burst of electron neutrinos (prompt $\nu_e$ burst, corresponding to $\sim 10^{53}$ erg s$^{-1}$ for $\sim$10--20 ms). At $t\sim$0.2 s, the shock stalls at a radius of $\sim$100--200 km in the outer core. 

During core collapse, a gravitational binding energy of several 10$^{53}$~erg, representing a staggering $\sim$10\% of the iron core's rest mass, is released in form of neutrino radiation. Therefore, the stalled shock is thought to be revived by neutrinos and antineutrinos that emerge from the hot and dense proto-neutron star \cite{Col66,Bet85}. At the very high temperatures prevailing in the core, neutrinos and antineutrinos of all flavors (electron, muon and tau) are produced, mainly by electron-positron pair annihilation, $e^- + e^+ \rightarrow \nu_e + \bar{\nu}_e$, neutrino-antineutrino pair annihilation, $\nu_e + \bar{\nu}_e \rightarrow \nu_{\mu, \tau} + \bar{\nu}_{\mu, \tau}$, and nucleon bremsstrahlung, $N + N \rightarrow N + N + \nu_{\mu, \tau} + \bar{\nu}_{\mu, \tau}$ \cite{Bur03}. While the neutrinos diffuse out of the core, the luminosities and average energies for each of the different neutrino types evolve with time. First, muon and tau neutrinos have smaller opacities compared to electron neutrinos since their energies are too low for charged-current interactions, such as $\bar{\nu}_\mu + p \rightarrow n + \mu^+$ (because of the large masses of the $\mu$ and $\tau$ leptons). Thus, the muon and tau neutrinos decouple at smaller radii $R_\nu$ (at higher density and temperature) and hence emerge with higher average energies from the proto-neutron star compared to the electron neutrinos (see $\nu$-process). Second, there are fewer protons than neutrons in the outer layers of the proto-neutron star and thus the charged-current interaction $n + \nu_e \leftrightarrow p + e^-$ occurs more frequently compared to $p + \bar{\nu}_e \leftrightarrow n + e^+$. Hence, the electron neutrinos have a higher opacity than the electron antineutrinos and thus decouple at larger radii with smaller average energies.

\begin{figure}
\includegraphics[scale=.65]{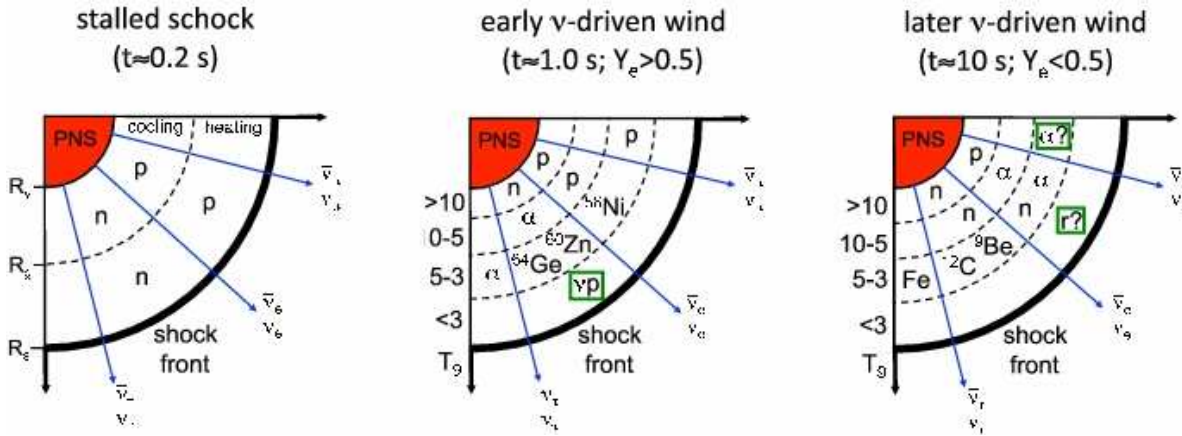}
\caption{Evolution of the region near the proto-neutron star (PNS) surface, as a result of neutrino-induced processes, expansion and cooling. The proto-neutron star surface can be defined by the radius of the energy-integrated electron neutrino sphere. The radii of the various zones differ from panel to panel and are not to scale. Approximate temperature ranges (in GK) are indicated. Only the main constituents are given in each zone. Symbols in green boxes indicate nucleosynthesis processes: $\nu p$-process ($\nu p$); $\alpha$-process ($\alpha$); r-process (r). (Left) Region between proto-neutron star surface and stalled shock, $t\sim$0.2 s after core collapse. The main constituents are free protons and neutrons. Inside the gain radius neutrino cooling prevails, while outside matter is effectively heated by neutrinos. Radii: $R_\nu$ (neutrino sphere); $R_g$ (gain radius);  $R_s$ (shock radius). (Middle) The early neutrino-driven wind, $t\sim$1.0, is proton-rich, $Y_e>0.5$, giving rise to the $\nu p$-process. (Right) The later neutrino-driven wind, $t\sim$10, may become neutron-rich, $Y_e<0.5$, giving rise to the $\alpha$- and r-processes. Note that the actual abundances of $^{9}$Be and $^{12}$C are very small since these species are destroyed almost as quickly as they are produced. The label ``Fe" denotes a distribution of seed nuclei in the $A\sim$50--100 mass range. See text.\label{corefig}}
\end{figure}

The situation at this stage is represented in Fig.~\ref{corefig} (left). The matter between the neutrino sphere and the stalling shock front consists mainly of free neutrons and protons. The {\it gain radius} divides this region into two parts: the first is located closer to the neutrino sphere and is characterized by the dominance of $p + e^- \rightarrow n + \nu_e$ and $n + e^+ \rightarrow p + \bar{\nu}_e$ over their reverse interactions, thus giving rise to effective neutrino cooling by neutrino emission; the second is located closer to the shock, where $n + \nu_e \rightarrow p + e^-$ and $p + \bar{\nu}_e \rightarrow n + e^+$ dominate over their reverse interactions, hence giving rise to effective heating via neutrino absorption. The continuous neutrino energy deposition in the latter region keeps the pressure high and could then rejuvenate the shock and cause the supernova explosion (delayed shock model \cite{May88,Wil93}). Only a fraction ($\sim$1\%) of the total gravitational binding energy, deposited by neutrinos as thermal energy of nucleons, leptons and photons in this region, would be required to initiate a powerful shock propagating through the stellar mantle and giving rise to an explosion. The success of this model depends crucially, among other things, on the product of neutrino luminosity and neutrino interaction cross sections (i.e., the square of the average neutrino energy). However, almost none of the most advanced, self-consistent core-collapse models have produced an explosion and the exact mechanisms by which neutrinos give rise to the explosion remain elusive. Instead, many models change artificially the neutrino properties, such as the charged-current interaction rates, in order to increase the neutrino energy deposition behind the shock. The problem is highly complex, involving energy-dependent neutrino transport in three dimensions, a convectively unstable region near a compact hot and dense object, possible diffusive instabilities, magneto-rotational effects, and so on. It appears that a self-consistent 3D simulation would require a computing power at least ten times as large as what is currently available.

We will consider below nucleosynthesis processes that effect the deepest layers to be ejected in a core-collapse supernova. After the rejuvenated shock has been launched, the strong electron neutrino and antineutrino fluxes drive a continuous flow of protons and neutrons from the region near the proto-neutron star surface. This outflow of matter is referred to as {\it neutrino-driven wind} \cite{Dun86}. Matter in this region expands and cools at high (possibly supersonic) velocity and eventually collides with the slower earlier supernova ejecta, resulting in a wind termination (or reverse) shock \cite{Bur95}. The nucleosynthesis in the early ejecta depends strongly on the entropy (photon-to-baryon ratio), expansion timescale, and electron mole fraction (electron-to-baryon ratio). The first two quantities depend on neutrino average energies and total luminosities, and the properties of the proto-neutron star (given by the equation of state). The latter quantity depends on the complicated interplay of the four (charged-current) electron neutrino and antineutrino interactions with free nucleons, $n + \nu_e \leftrightarrow p + e^-$ and $p + \bar{\nu}_e \leftrightarrow n + e^+$, in the wind region. If the neutrino-driven wind is proton-rich, then conditions are favorable for the synthesis of neutron-deficient nuclides, while a neutron-rich wind may give rise to the production of neutron-rich species. The outcome of the nucleosynthesis is highly sensitive to the still uncertain physical conditions of the neutrino-driven wind. 

\subsection{$\nu$p-process\label{nupprocess}}
Recent core-collapse simulations including energy-dependent neutrino transport have shown that the neutrino-driven wind is proton-rich ($Y_e>0.5$), either at very early times \cite{Bur06} or for an extended time period up to 20 s \cite{Fis09}. The is shown schematically in Fig.~\ref{corefig} (middle). The wind is ejected at temperatures above 10 GK, consisting of free neutrons and protons in nuclear statistical equilibrium (NSE). Four distinct phases can be identified during the nucleosynthesis: (i) expansion and cooling through the temperatures range of $T\sim$10--5 GK causes all neutrons to combine with protons, leaving a composition consisting of $\alpha$-particles and an excess of protons; (ii) further cooling through temperatures of $T\sim$5--3 GK allows the $\alpha$-particles to combine to heavier nuclei, giving mainly rise to the production of the nuclides $^{56}$Ni, $^{60}$Zn and $^{64}$Ge, which consist of an equal number of neutrons and protons ($N=Z$). Since the slowest link in this sequence is the triple-$\alpha$ reaction, the number of heavy seed nuclei synthesized is equal to the number of $^{12}$C nuclei produced. Abundance flows beyond $^{64}$Ge are inhibited by strong reverse flows and by its long half-life ($T_{1/2}=64$ s in the laboratory \cite{Aud03}); (iii) in the temperature range of $T=3-1.5$ GK the (charged-current) interaction $p + \bar{\nu}_e \rightarrow n + e^+$ on the abundant protons produces free neutrons (typically 10$^{14}$ per cm$^3$ for several seconds), which then participate in the nucleosynthesis. This aspect is very important, because fast (n,p) reactions on waiting points, such as $^{56}$Ni, $^{60}$Zn and $^{64}$Ge, and subsequent (p,$\gamma$) reactions allow for a continuation of the abundance flow toward heavier nuclides on the neutron-deficient side of the valley of stability. Note that a (n,p) reaction connects the same pair of nuclei as a corresponding $\beta^+$-decay; (iv) when the temperature falls below $T\sim$1.5 GK, the (p,$\gamma$) reactions freeze out and (n,p) reactions and $\beta^+$-decays convert the heavy nuclei to stable, neutron-deficient, daughters. This nucleosynthesis mechanism is called the $\nu p$-process \cite{Fro06,Wan06,Pru06}. 

The termination of the $\nu p$-process is caused by the cooling of matter to temperatures below $T\sim$1 GK, and not by the exhaustion of free protons. A few studies (for example, Ref.~\cite{Wan10}) have shown that the $\nu p$-process may account for the solar abundance of the light p-nuclides up to $^{108}$Cd, including $^{92}$Mo, $^{94}$Mo, $^{96}$Ru and $^{98}$Ru, which are significantly underproduced in the p-process or other proposed scenarios (Sec.~\ref{pproc}). However, other studies \cite{Pru06,Fis09b} could not reproduce the solar abundance of the light p-nuclides and thus the issue remains controversial at present. It is also important to stress that, although many model simulations predict an early proton-rich neutrino-driven wind, the $\nu p$-process is highly sensitive to the details of the explosion mechanism, the mass, and possibly the rotation rate of the proto-neutron star. Thus it is expected that the nucleosynthesis may vary considerably from event to event. Currently, the key uncertainties originate from the neutrino luminosities and average energies, which determine the electron mole fraction, $Y_e$, in the neutrino-driven wind. A larger value of $Y_e$ will generally give rise to the synthesis of heavier p-nuclei. For example, models with $Y_e\sim$0.6 (i.e., at the onset of the $\nu p$-process, when $T\sim$3 GK) produce p-nuclei up to $A=152$. For even larger, perhaps excessive, values of $Y_e$, the density of free neutrons would increase as a result of the charged-current interaction between electron antineutrinos and the large proton abundance. Consequently, the (n,p) reactions would not only carry the abundance flow to the 
group of stable nuclides, but (n,$\gamma$) reactions will, in fact, carry the flow to the neutron-rich side. The protons do not participate directly in the nucleosynthesis, but mainly act as a source of free neutrons \cite{Pru06}. 

The $\nu p$-process is also sensitive to nuclear physics uncertainties. For example, the slow triple-$\alpha$-reaction (Sec.~\ref{heburnsec}) represents a bottleneck for the synthesis of the heavy seed nuclei (before the start of the $\nu p$-process, near $T=3$ GK). Varying this rate will impact the production of the p-nuclei in the $A=100-110$ mass region. Changing the rates of the (n,p) reactions on the waiting point nuclei that control the abundance flow to heavier nuclides, especially for $^{56}$Ni(n,p)$^{56}$Co and $^{60}$Zn(n,p)$^{60}$Cu, strongly influences the production of the p-nuclei in the $A>100$ mass region. In the most recent simulations \cite{Wan10} these (n,p) rates are based on theory (Hauser-Feshbach model), since experimental data are lacking at present. A study of the (modest) impact of nuclear mass uncertainties on the path of the $\nu p$-process can be found in Ref.~\cite{Web08}.

\subsection{r-process\label{rprocsec}}
About half of the nuclides beyond the iron peak, including gold, silver, platinum and uranium, are believed to be synthesized via neutron-induced reactions, under extreme conditions that are very different from those giving rise to the s-process (Sec.~\ref{sprocsec}). The abundances observed in the solar system and in (old) stars of very low metallicity hint at a mechanism that started to occur early in the history of the Galaxy, providing very high neutron densities at large temperatures for a very short period of time. This mechanism is called the astrophysical r(apid neutron capture)-process and its phenomenology has been understood for a long time (see, for example, Ref.~\cite{See65}). One of the most attractive sites, which has been studied intensively over the past two decades, in fact, is the neutrino-driven wind. We will first describe the phenomenological model and then return to the connection between the r-process and the neutrino-driven wind.

If one subtracts the main s-process component, estimated using modern models of AGB stars (Sec.~\ref{agb}), from the measured solar system abundances in the $A\gtrsim 100$ mass region, then the residual distribution exhibits two prominent peaks, at mass numbers near $A=130$ and 195, about ten mass units removed from the s-process peaks near $A=138$ and 208. The existence of the r-process peaks and of the long-lived radioisotopes $^{232}$Th ($T_{1/2}=1.4\times 10^{10}$ y), $^{235}$U ($T_{1/2}=7.0\times 10^8$ y), and $^{238}$U ($T_{1/2}=4.5\times 10^9$ y), which are located beyond the endpoint of the s-process, provide strong evidence for the existence of the r-process. The abundance peaks can be explained by the neutron magic numbers at N=82 and 126. Unlike the s-process, one may imagine a very strong neutron flux, such that neutron capture is more likely than $\beta^-$-decay. Thus the abundance flow is driven away from the stability valley toward the neutron-rich side. Here, the abundance accumulates at nuclides with neutron magic numbers \cite{Ili07}. After termination of the neutron flux in the r-process, these neutron-magic nuclei undergo series of $\beta^-$-decays along isobaric chains ($A=const$) until the most neutron-rich stable (or very long-lived) isobar is reached. Hence the r-process gives rise to abundance maxima in mass regions located {\it below} the corresponding s-process abundance peaks. 

Suppose that seed nuclei (for example, iron) are exposed to high neutron densities ($N_n\gtrsim10^{20}$ cm$^{-3}$) and temperatures ($T\gtrsim 1$ GK). Under such conditions, both (n,$\gamma$) and ($\gamma$,n) reactions are much faster than competing $\beta^-$-decays. Consequently, thermal equilibrium is quickly established along each chain of isotopes, implying that the abundance of a given isotope depends mainly on (n,$\gamma$) reaction Q-values, temperature and neutron density, but is independent of neutron capture rates. Furthermore, the systematic behavior of the Q-values when moving from the stability valley toward the neutron drip line gives rise to an abundance maximum at one or at most two species in each isotopic chain. These nuclides represent waiting points for the abundance flows and, therefore, the (n,$\gamma$)$\leftrightarrow$($\gamma$,n) equilibrium condition is referred to as the {\it waiting point approximation}. The abundance flow must then continue with a slow $\beta^-$-decay, connecting one isotopic chain to the next, where again an independent equilibrium within the new isotopic chain is established. This repetitive sequence of events gives rise to the r-process path. The situation is analogous to a radioactive decay chain: in both cases the abundance flow will attempt to achieve a constant $\beta^-$-decay flow from one isotopic chain to the next. This condition is referred to as {\it steady flow approximation}. In this model, for given values of neutron density, temperature and time duration, the r-process path and r-process abundances can be calculated precisely, if the nuclear properties are known: neutron capture Q-values (or neutron separation energies), $\beta$-decay half-lives, branching ratios for $\beta$-delayed neutron emission, normalized partition functions, and so on. 

The phenomenological model discussed above is referred to as the {\it classical r-process model}. It is rather simple because it assumes: (i) a constant temperature, $T$, and neutron density, $N_n$, (ii) an instantaneous termination of the neutron flux after time $\tau$, and (iii) the waiting point and steady flow approximations. Nuclear masses play a central role for the r-process since they determine directly or indirectly most of the nuclear properties listed above. The (n,$\gamma$) Q-values (which are given by mass differences) enter exponentially in the determination of the equilibrium abundances of each isotopic chain and thus must be known rather accurately. It is obvious that the nuclear properties needed are for isotopes that are located far away from the valley of stability. With few exceptions, the required information is experimentally unknown since most of the nuclei on the r-process path cannot be produced yet in the laboratory. There is little alternative at present but to estimate the required nuclear properties by using nuclear models. 
In practice, attempts are made to derive semi-empirical formulas from the known properties of nuclei close to stability that can be extrapolated into the region covered by the r-process path. Such extrapolation procedures are subject to significant uncertainties even for the most sophisticated models (see, for example, Ref. \cite{Moe03}). Obviously, any deficiencies in current nuclear models will have a direct impact on r-process predictions. The associated nuclear physics uncertainties affect most discussions of the r-process. 

Attempts to describe the entire observed distribution of solar system r-abundances by using the classical r-process model with a single set of $T$-$N_n$-$\tau$ conditions were unsuccessful \cite{Kra93}. It was found that at least three different sets of $T$-$N_n$-$\tau$ conditions, each corresponding to a specific r-process path, are required in different mass regions in order to reproduce the observed solar-system abundance distribution. Temperatures, neutron densities and time scales for these components were in the ranges of $T\sim$1.2--1.35~GK, $N_n\sim 3\times 10^{20} - 3\times 10^{22}$~cm$^{-3}$ and $\tau \sim$1.5--2.5~s, respectively. Furthermore, it was demonstrated that the steady flow approximation applies {\it locally} in each of these mass ranges, but not globally over the entire mass region. The overall implication is that the solar system r-abundance distribution results from a superposition of components representing different r-process conditions. This may be caused, for example, by varying conditions in different zones of a given site. Clearly, the phenomenological model described above makes no assumption regarding the site of the r-process, but provides a general idea regarding some of its properties: very high neutron densities and short time scales. On the other hand, the temperature should not be too high or otherwise the heavy nuclei will be either destroyed by photodisintegration or the waiting point abundances shift too close to the stability valley, where the $\beta^-$-decay half-lives are too slow to allow for efficient r-processing. The classical r-process has also been applied for reproducing or predicting abundance ratios of neighboring nuclides, for example, for nuclear chronometers \cite{Kra04} or isotopic anomalies in primitive meteorites \cite{Kra01}. Such abundance ratios are most likely influenced by nuclear properties rather than by the details of the astrophysical r-process site. For an extensive discussion of the limitations of the classical r-process model, see Ref. \cite{Kra07}.

It is worthwhile to visualize the results of a {\it dynamic} r-process calculation as opposed to the static model described above. Clearly, in reality the temperature and neutron density will change with time. Early during the r-process, the temperature and neutron density will be sufficiently large to ensure that (n,$\gamma$)$\leftrightarrow$($\gamma$,n) equilibria hold in all isotopic chains. Suppose now that the temperature and neutron density decrease with time. The abundance flow will continuously adjust to the new conditions according to the waiting point approximation. Thus the r-process path, defined by $T$ and $N_n$, must continuously move, starting from a location closer to the neutron drip line to one that is located closer to stability. For each location of the path, the $\beta^-$-decay half-lives are different. When the neutron flux ceases, the r-process nuclei decay towards stability. It is apparent that only the r-process path just before freeze-out, and in particular the sections near neutron magic nuclei, matters for the observed final distribution of r-abundances. In other words, at freeze-out the r-process has mostly forgotten its earlier history.
Furthermore, if the condition of an instantaneous termination of the neutron flux is relaxed, additional nuclear processes that may occur during the freeze-out will likely become important. For example, neutron capture reactions and $\beta$-delayed neutron decays (apart from $\beta^-$-decays) may alter the final composition.

\begin{figure}
\includegraphics[scale=.60]{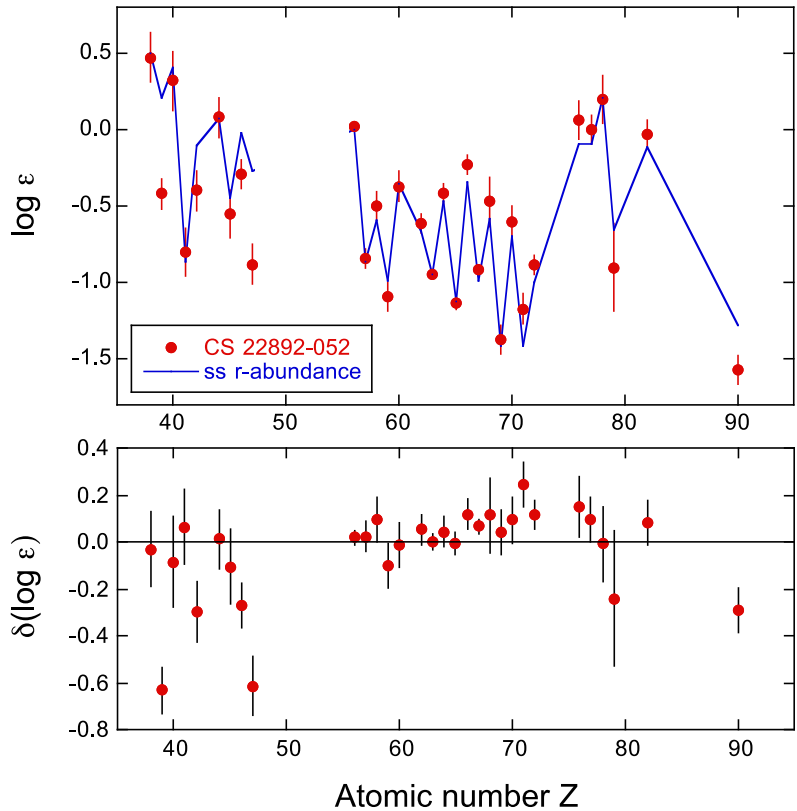}
\caption{(Top) Elemental abundances of the low-metallicity halo giant star CS 22892-052 (solid circles) compared to the solar system r-process abundances (solid line). The abundances are given in standard spectroscopic notation, where $\log \epsilon(A)\equiv \log_{10}(N_A/N_H) + 12.0$, and $N_i$ denotes number abundance. The stellar data are from Ref. \cite{Sne08}. The solar system r-abundances (total solar system abundance minus s-abundances) are from Ref. \cite{Sim04} and have been normalized to the stellar Eu abundance. Note that the graph does not include the uncertainties introduced by the deconvolution of s- and r-abundances. (Bottom) Difference between total stellar and scaled r-process solar system abundances. The large and small scatter in the regions $Z<50$ and $Z>50$, respectively, is apparent. See text.
\label{figrprocess}}
\end{figure}

In addition to studying the solar abundance distribution, important clues regarding the r-process are provided by stellar spectroscopy of metal-poor 
(and thus very old) Galactic halo stars \cite{Sne08,Roe10}. For example, observed elemental abundances for the remarkable halo giant star 
CS 22892-052 are shown in Fig.~\ref{figrprocess}. It was found in this case, and for several other low-metallicity halo giant stars, that 
for elements of barium ($Z=56$, $A\sim$135) and beyond the observed heavy element abundances (data points) agree remarkably well with 
relative solar system r-abundances (solid line). This implies that most of the heavy elements observed in those stars were synthesized 
by the r-process early during the evolution of the Galaxy, with no apparent contribution from the s-process. Since the r-process elements
 could not have been synthesized in the halo stars themselves, they must have been produced by progenitors that evolved very rapidly and that 
ejected the matter into the interstellar medium before the formation of the currently observed halo stars. 
In fact, these stars formed so early in the history of the Galaxy that they may have received contributions from only one, or at most a few, 
r-process events. The most likely r-process sites seem to be associated with massive stars since low-mass or intermediate-mass stars evolve 
on much longer time scales. The agreement in abundances above $A\sim$135 shows that the r-process mechanism is robust, in the sense that
 a similar abundance distribution pattern is produced in each r-process event. Interestingly, the good agreement between solar r-process 
and stellar abundances does not extend to the lighter elements between strontium (Z=38) and barium (Z=56). There have been many suggestions to explain their synthesis,
including a ``weak r-process" \cite{Kra07}, a ``Light Element Primary Process" (LEPP) \cite{Tra04,Mon07}, and the $\alpha$-process  \cite{Qia07}. 
Clearly, more observations and studies will be needed to understand the synthesis of these elements.

We will now describe the nucleosynthesis that may occur in the late neutrino-driven wind, {\it assuming} that conditions are favorable for the r-process. Four different phases can be identified, as shown schematically in Fig.~\ref{corefig} (right): (i) above a temperature of $T\sim$10 GK, the wind consists of free neutrons and protons in nuclear statistical equilibrium (NSE); (ii) as the wind expands and cools below $T\sim$10~GK, the nucleons start to combine to $\alpha$-particles. Near $T\sim$7~GK, $\alpha$-particles become the dominant constituent, leaving behind an excess of neutrons. Furthermore, some of the $\alpha$-particles begin to assemble into nuclei via the strongly density-dependent process $\alpha + \alpha + n \rightarrow$$^{9}Be$, followed by $^9Be(\alpha,n)^{12}C$; (iii) near $T\sim$5 GK, nuclear statistical equilibrium starts to break down because the expansion time scale becomes shorter than the time required to maintain nuclear statistical equilibrium under conditions of high temperature, low density and large $\alpha$-particle abundance. The slow process $\alpha + \alpha + n \rightarrow$$^{9}Be$ is the first reaction to fall out of NSE. The situation is similar to the $\alpha$-rich freeze-out (Sec.~\ref{expnucsec}). There is again an excess of $\alpha$-particles and these cannot be consumed fast enough by the slow helium-induced reactions in the time available. The important difference, however, is that there is now a neutron excess. Both the abundant $\alpha$-particles and the neutrons will participate in the buildup of heavier nuclei, starting with the sequence $^{12}C(n,\gamma)^{13}C(\alpha,n)^{16}O(\alpha,\gamma)^{20}Ne$, and so on. Without the presence of neutrons, the $\alpha$-rich freeze-out produces mainly the $N = Z$ isotope $^{56}$Ni, while the flows beyond the iron peak would be negligible (Sec.~\ref{expnucsec}). The capture of neutrons has the important effect of shifting the abundance flow towards the stability valley, where the nuclei are more neutron-rich, thus extending the flow beyond the iron peak up to $A\sim$110. The most important processes in the buildup of these heavy nuclei are ($\alpha$,n) and (n,$\gamma$) reactions. This neutron-rich, $\alpha$-rich freeze-out is referred to as the {\it $\alpha$-process}; (iv) at a temperature near $T\sim$3~GK, the $\alpha$-particle induced reactions become too slow to change the composition of the matter and the $\alpha$-process ceases. At this time, the composition resides  in $\alpha$-particles (mostly), neutrons and seed nuclei in the $A\sim$50--110 mass range. If the neutron-to-seed ratio is sufficiently large, an r-process can be launched while the matter further expands and cools. 

The r-process model described above has the advantage that the properties of the neutrino wind are determined by the proto-neutron star, not by the pre-supernova evolution. On the one hand, this implies that r-processing in this site should produce similar abundances for events involving proto-neutron stars of the same mass. On the other hand, the r-process will be sensitive to the proto-neutron star evolution and the neutrino interactions, and thus represents an important diagnostic of the event. It is crucial for a successful r-process that the preceding $\alpha$-process is not too efficient. Otherwise, too many heavy seed nuclei are produced and too many neutrons are consumed, resulting in insufficient neutron-to-seed ratios for synthesizing nuclides up to $A\sim$200 during the subsequent r-process. This requirement translates into relatively high entropies (or low densities) in the neutrino-driven wind so that the $\alpha + \alpha + n \rightarrow$$^9Be$ reaction is less efficient in converting $\alpha$-particles to heavier nuclei. A fast expansion time scale limits the duration over which the freeze-out operates and is also helpful for reaching a high neutron-to-seed ratio. To summarize, an appropriate combination of low electron mole fraction, high entropy (i.e., high temperatures and low densities), and short expansion time scales is a necessary condition for a successful r-process in the neutrino-driven wind model. 

Some studies have attempted to reproduce the solar system r-process abundance distribution by assuming a superposition of many individual contributions in the neutrino-driven wind, each representing an appropriate combination of electron mole fraction, entropy and expansion time (see, for example, Ref.~\cite{Far10}). It was found that the component with the heavier r-abundances ($A\gtrsim 120$), including Th and U, could be reproduced for electron mole fractions below $Y_e\sim$0.48 for modest entropies ($s\lesssim$300). Furthermore, it could be confirmed that the waiting point and steady flow approximations of the classical r-process apply locally (but not globally) over certain mass regions in the dynamic simulations. Recent studies \cite{Far09} (see also Ref.~\cite{Hof96}) have found that many (though not all) nuclides between zinc ($Z=30$, $A\sim$65) and ruthenium ($Z=44$, $A\sim$100) can may be synthesized in the neutrino-driven wind, under suitable {\it low-entropy} conditions, via the $\alpha$-process when the abundance of free neutrons at charged-particle freeze-out is negligible (i.e., no subsequent r-process occurs). The results are interesting because this scenario {\it co-produces} via a charged-particle process a fraction of the $A\sim$65--100 nuclides, which are commonly thought to originate from different processes, that is, the s-, r- and p-processes. Clearly, more abundances from stellar spectroscopy and more extensive nucleosynthesis simulations are called for.

Calculations of r-process nucleosynthesis require an extensive input of nuclear physics information, 
such as nuclear masses, $\beta^-$-decay half-lives, branching ratios for $\beta$-delayed neutron decay, 
partition functions, and so on. If the freeze-out from equilibrium is followed explicitly 
(that is, if the waiting point and steady flow approximations are not applied and the reaction network is solved numerically) then reaction rates for neutron captures and photodisintegrations are required as well. All of this information is needed for neutron-rich isotopes that are located far away from stability. If the $\alpha$-process needs to be followed explicitly, then another large data set consisting of rates for charged-particle reactions and neutron-induced processes, such as (n,$\alpha$) and (n,p), is required. 
Needless to say that almost all of this information must be obtained from global, semi-empirical, models for nuclear masses 
and $\beta^-$-decays, and from Hauser-Feshbach calculations. 
 Information on directly measured properties, mainly life times and $\beta$-delayed neutron emission probabilities, of nuclei located near or on the r-process path has been obtained in some 
cases (for recent work, see Refs. \cite{Kra05,Kra06,Baru08,Pere09,Hosm10,Nish11}).
A notable measurement of the $Q_\beta$ ($\beta$-decay end point) value of the waiting point nuclide $^{130}$Cd is described in Ref. \cite{Dill03}.
There is reasonable hope that many crucial nuclear properties of very neutron-rich, unstable nuclides can be measured in the near future at radioactive-ion beam facilities. 

Although the neutrino-driven wind represents the most popular r-process site, recent long-term simulations of 
core-collapse supernovae predict unfavorable conditions (electron mole fractions, entropies, expansion time scales) 
for an r-process \cite{Fis09,Hue10}. In particular, these studies obtained proton-rich ($Y_e>0.5$) neutrino-driven 
winds for an extended time period up to 20 s. Clearly, this represents a severe difficulty for r-process scenarios 
in core-collapse supernovae. Thus the quest for the site(s) of the r-process remains an unsolved puzzle in nuclear 
astrophysics. An alternative r-process scenario will be discussed in Sec.~\ref{mergsec}. For additional information 
on topics related to the r-process, see Ref.~\cite{Far10,Arn07}.

\subsection{The gamma-ray burst--supernova connection\label{GRBSN}}
Gamma-ray bursts (GRBs) are brief,
 intense flashes of electromagnetic radiation with typical photon energies 
 ($\sim$100 keV) in the $\gamma$-ray domain. Indeed, they constitute
 the most luminous electromagnetic events in the Universe.
 This initial $\gamma$ flash is usually followed by a longer-lived {\it
 afterglow} emitted at longer wavelengths (X-rays, ultraviolet, optical, 
 infrared and radio).  Gamma-ray bursts are isotropically distributed on the sky and
 hit the Earth from unpredictable locations several times per day
 \cite{Geh09}. To date, no repeated burst has been
detected from the same source. 

Gamma-ray bursts were serendipitously discovered in 1967 by the military U.S. Vela 
satellites, originally designed to detect secret nuclear weapon tests 
(suspectedly carried by the former USSR, after signing the Nuclear Test Ban
Treaty in 1963). On July 2, 1967, the Vela 3 and 4 satellites detected 
a flash of $\gamma$ radiation unlike any known nuclear weapon imprint. 
 As other Vela satellites were launched, additional 
 $\gamma$-ray flashes were detected. The analysis of the different arrival 
 times of the bursts, performed by researchers from Los Alamos National 
 Laboratory led by Klebesadel, ruled out either a terrestrial or a
 Solar System origin. The work was later declassified and revealed to the
 scientific community \cite{Kle73}. 

In contrast with many other stellar variables and transients, 
which exhibit a characteristic time evolution, with a rapid rise to peak
luminosity followed by an exponential-like decay, $\gamma$-ray burst light curves show
very complex and diverse patterns. Indeed, no two $\gamma$-ray burst light curves
look alike. Some $\gamma$-ray bursts are preceded 
by a weak burst ({\it precursor}) that anticipates the true bursting 
episode by seconds or minutes. The light curve spans for an overall
duration ranging from milliseconds to tens of minutes, being single peaked
or showing several individual 
minipulses. Moreover, the peaks themselves are sometimes symmetric or 
characterized by a fast rise and a very slow fading. 
In spite of the large differences found in 
the $\gamma$-ray burst light curves (also in the spectra), two broad 
categories have been established on 
the basis of the burst duration and its power spectrum:
{\it short} (hard) $\gamma$-ray bursts, with a typical duration of 0.3 s, and 
{\it long} (soft) bursts with a duration of about 20--30 s.
Moreover, the former are characterized by a 50\% larger average peak energy (360 keV)
than the long $\gamma$-ray burst class (220 keV) \cite{Hak00}.

More than hundred theoretical models aimed at explaining the nature of
$\gamma$-ray bursts have been  proposed since their discovery (including collisions 
between comets and neutron stars \cite{Nem94}). Extensive reviews on the different 
mechanisms proposed for the engines of $\gamma$-ray bursts can be found in Refs. 
\cite{Mes02,Pir05,WB06}. 
Current models 
 require the acceleration of tiny amounts of matter to
ultrarelativistic speeds \cite{Mes02}, as well as a beamed 
emission focalized into a very small fraction of the sky. 
Once this beaming effect is taken into account, 
the total energy emitted in $\gamma$-rays (at least, in the long-lasting 
$\gamma$-ray bursts) is inferred to be $\sim 10^{51}$ erg, 
roughly equal to the overall kinetic energy in core-collapse supernovae.
Despite this similarity, no evident link between the two phenomena was
established for decades. An exception can be found in Ref. \cite{Col68}, 
perhaps the only work that anticipated the existence of $\gamma$-ray bursts before 
their discovery, in the context of relativistic shocks from supernovae.
Actually, the most widely accepted scenario for $\gamma$-ray bursts in the 1990s involved
the merging of degenerate binary systems, such as double neutron
stars \cite{Bli84,Pac86} or binaries containing  
a neutron star and black hole \cite{Nar91,Pac91}.  At that time, nothing
was pointing towards a possible connection between $\gamma$-ray bursts and supernovae. 
An important achievement in this direction was the 
 detection of the first X-ray and optical afterglows in 1997, as well as
 the direct measurement of their redshifts using optical spectroscopy. 
 This put the merger hypothesis aside, while clearly favoring an origin
associated with the death of young, massive stars. 
 The discovery of GRB 980425 in conjunction with one of the most unusual 
 supernovae ever observed, SN 1998bw \cite{Gal98b,Iwa98}, 
 represented a breakthrough in the history of $\gamma$-ray bursts. 
 This has been followed by the discovery of at least two other clearly connected GRB-SN pairs:
 GRB 030329 -- SN 2003dh \cite{Sta03} and 
 GRB 031203 -- SN 2003lw \cite{Mal04}. In such systems,
 the detected supernova  seem to correspond to the type Ic class, 
  that is, supernovae involving massive progenitors, showing no hydrogen in the 
spectra and also lacking strong lines of He I or Si II. 
 Even though some models suggest that all long $\gamma$-ray bursts might be accompanied by 
 supernova explosions, it is clear that not all supernovae (not even the 
 type Ic class) give rise to $\gamma$-ray bursts. The reason why massive stars {\it choose}
 one path (a SN-GRB pair) or another (a SN explosion only) is not
 yet clear. It has been claimed, however, that 
$\gamma$-ray bursts may result only from the most rapidly rotating and massive 
stars. It is worth noting 
that in the cases of SNe 2003dh and 1998bw, the explosion yielded 
$10^{52}$ erg of kinetic energy into a solid angle of $\sim 1$ radian.
This implies about $\sim 10$ times more energy than a regular supernova.
Such energetic explosions have been unofficially coined as {\it hypernovae}
\cite{Nom11}.

About 30\% of the sample of $\gamma$-ray bursts detected by the BATSE
instrument on-board NASA's Compton Gamma-Ray Observatory (CGRO)
correspond to the short GRB class.
However, there are reasons to believe that because of observational bias,
this may be the most frequent variety of $\gamma$-ray bursts in the cosmos.
Indeed, short $\gamma$-ray bursts tend to release smaller amounts of energy than
long bursts, and this may limit their detectability beyond a certain 
distance. Moreover, short $\gamma$-ray bursts for which well-localized counterparts
have been determined, are associated with elliptical galaxies. 
This clearly points towards old progenitors, for example, the merging of
binary neutron stars, as the likely sources powering short $\gamma$-ray bursts.
It has also been claimed that a small subset of this GRB class is 
probably associated with giant flares from soft-$\gamma$ repeaters,
that is, sources of short ($\sim 100$ ms), repeating bursts of soft 
$\gamma$-radiation ($<100$ keV), located in nearby galaxies.

Some of the models aimed to explain long $\gamma$-ray bursts involve the rotation
(near breakup) of a highly magnetized neutron star (the {\it magnetar} 
model \cite{Whe00,Lyu01,Dre02}), 
or the collapse of the innermost layers of a massive
star forming a rapidly rotating black hole surrounded by a disk  
(the {\it collapsar} model \cite{Woo93,MFW99}). It has been argued, however,
that all magnetar models computed so far systematically disregarded the 
role played by mass accretion (at a rate $\sim 0.1$ M$_\odot$ yr$^{-1}$)
onto the proto-neutron star before its 
 rotation is fully established. It has also been suggested that these
models may require the initial
operation of a successful neutrino-powered explosion before they can function.
Provided that a disk and a black hole form, the greatest uncertainty associated with the latter 
models is the mechanism for turning disk binding energy (or black-hole rotation 
energy) into confined relativistic outflows. Suggested 
mechanisms include neutrinos \cite{Woo93},  
magnetic disk instabilities \cite{BP82}, or a magnetohydrodynamic 
extraction of the black hole rotational energy \cite{BZ77}.

The nucleosynthesis contribution of $\gamma$-ray bursts has been explored by different
groups. In the framework of the collapsar model, two possible explanations
have been given for the synthesis of Ni that likely powers the light curves.
In the first one, as in a regular supernova, it may be linked to
a strong shock traversing the stellar mantle and exploding
the star. Nomoto and collaborators \cite{Mae03,Nom08} 
discussed the nucleosynthesis consequences of this scenario, showing 
that a very energetic shock, driven by bipolar jets in hypernova explosions, may 
yield a peculiar abundance pattern resembling that inferred from some
 extremely metal-poor stars (in particular, C-rich metal-poor stars,
 and hyper metal-poor stars). Notice, however, that a recent reanalysis
performed in 3-D \cite{Roc08} 
revealed a lower $^{56}$Ni production than in previous work. 
The second possibility involves Ni production 
in the wind blown off from the accretion disk surrounding the black hole
\cite{MFW99,McF03,Pru04}.  It has been found that matter
ejected from the disk can exhibit a rather rich range of
nucleosynthesis, depending on the 
specific mass accretion rate of the disk, the velocity and entropy of the 
outflow, and the radius at which matter is released \cite{Sur06}. 
Hence, the corresponding outflow can be either neutron- or proton-rich.
This has been analyzed in detail in a number of papers
\cite{Sur06,PSM04,Sur05},
showing that the outflows from rapidly accreting disks
($\dot M \sim 10$ M$_\odot$ s$^{-1}$) may produce r-process elements (Sec.~\ref{rprocsec}).
Moderate mass-accretion rates can yield $^{56}$Ni and a suite of light
elements, such as Li or B. For disks with low mass-accretion rates
($\dot M \sim 1$ M$_\odot$ s$^{-1}$), the main products are likely
$^{56}$Ni, $^{49}$Ti, $^{45}$Sc, $^{64}$Zn and $^{92}$Mo. 
The levels at which those elements are produced make 
$\gamma$-ray bursts likely contributors to the Galactic inventory.  
A recent reanalysis \cite{Kiz10} has shown that disk
models with mass accretion rates near 1 M$_\odot$ s$^{-1}$ may result, for a certain combination
of parameters, in
$\nu$p-process nucleosynthesis products (Sec.~\ref{nupprocess}). Nevertheless, the study revealed that the time during which 
matter is subjected to neutrino interactions turns out to be critical, 
and may even halt $\nu$p-process nucleosynthesis. In order to shed 
light into this issue, state-of-the-art hydrodynamic simulations will be 
needed.

\section{Classical and recurrent novae}
Novae have captivated the interest of astronomers since more than 
two Millennia. Indeed, their own name, from the latin {\it stella nova} 
(or ``new star"), reflects the reasons that lay behind such interest:
the sudden appearance of a luminous object in the sky (at a 
spot where nothing was clearly visible before), that fades away, back 
to darkness, in a matter of days to months. 

Until the astronomical scale distance was soundly established, 
 (classical) novae and (type I/II) supernovae were 
both associated with the same celestial phenomenon, 
generically designated as ``nova". 
Shapley and Curtis were among the first astronomers to question 
the real distance to spiral ``nebulae", where novae had been
serendipitously discovered, suggesting their extragalactic nature.
A major step forward was achieved with the discovery and analysis of 
Nova S Andromedae (1885) by Hartwig, as well as 
by the new scale distance of Hubble that placed Andromeda outside the Milky 
Way. This pushed Nova S Andromedae far away 
and revealed that the underlying stellar object was 
characterized by a huge intrinsic luminosity. 
Thus the road was paved for the hypothesis  
that two distinct classes of {\it stellae novae} exist, which are in modern parlance referred to as
{\it novae} and {\it supernovae}. 

The understanding of the physics that gives rise to the sudden luminosity
burst characterizing novae 
has motivated vivid discussions. Probably the first 
explanation that underscored the real mechanism powering nova explosions 
appears in Newton's {\it Principia Mathematica}: 
{\it ``So fixed stars, that have been gradually wasted by the light and vapors 
emitted from them for a long time, may be recruited by comets that fall upon 
them; and from this fresh supply of new fuel those old stars, acquiring new 
splendor, may pass for new stars."}
Indeed, the concept of revitalization of old stars by fresh supply of new 
fuel (although not by comets) is at the base of the thermonuclear runaway 
model, in which mass accretion plays a central role. But the basic
picture that define nova outbursts emerged from  a number
of observational breakthroughs in  spectroscopy
(the discovery of neon lines in Nova Persei 1901 \cite{Sid1901,Sid1901b}, 
revealing the existence of several nova types)
and photometry (the interpretation of the dips in the optical light curves as
caused by dust forming episodes \cite{Stra1939}, 
or the discovery of the binary nature of nova systems \cite{Wal54,Kra64}).
All in all, even though the observed features 
were soon interpreted as produced by the ejection of a shell from a star 
\cite{Pic1895}, the explanation of the mechanism that causes the burst
had to wait 
until mid 20th century.  Indeed, its thermonuclear origin was first theorized by 
Schatzmann \cite{Sch49,Sch51}, and Cameron \cite{Cam59} (see also 
Gurevitch \& Lebedinsky \cite{GL57}, and references therein), while the 
first hydrodynamic 
simulation of a nova outburst was performed by Sparks \cite{Spa69}.

Novae have been observed in all wavelengths, except
in $\gamma$-rays\footnote{An unexpected emission at $>$100 MeV
detected in the symbiotic binary V407 Cygni, attributed to shock 
acceleration in the ejected nova shell after interaction with the wind of
the red giant, was reported by the Fermi-LAT Collaboration \cite{Abd10}.}.
They constitute a very common phenomenon, representing the 
second most frequent type of thermonuclear explosion in the 
Galaxy after type I X-ray bursts. Although only a handful of novae 
(3 to 5) are 
discovered every year mainly by amateur astronomers, a Galactic  
rate of about $30 \pm 10$ yr$^{-1}$ \cite{Sha02} has been predicted through 
extrapolation of the number of events observed in other galaxies. The reason 
for the scarcity of detections in our Galaxy is interstellar extinction by 
cosmic dust. 

Classical novae take place in close stellar binary systems, with orbital
periods in the range 1 -- 12 hr. They  
consist of a white dwarf star and a low-mass main sequence companion 
(a K-M dwarf, although evidence of more evolved companions also exists). 
Contrary to type Ia supernovae (Sec.~\ref{sn1a}), in which the white dwarf is fully disrupted 
by the violence of the explosion, a nova outburst is basically confined to the outer layers.
The star is not destroyed by the explosion and hence all classical novae are expected to recur, 
with periodicities of the order of $10^4-10^5$ yr. Notice, however, that in 
the explosions that occur in very massive white dwarfs, the 
so-called {\it recurrent novae}, the expected periodicities range typically 
between 10--100 yr. Both novae and supernovae are characterized by a 
remarkable energy output, with peak luminosities reaching $10^5$ and $10^{10}$ 
L$_\odot$, respectively. These explosive phenomena also differ greatly in
 the amount of mass ejected (the whole star in a thermonuclear supernova compared to 
 only $10^{-4}-10^{-5}$ M$_\odot$ in a nova) and in the mean ejection 
velocity ($\sim 10^4$ km s$^{-1}$ in a supernova compared to 
$\sim 10^3$ km s$^{-1}$ in a classical nova).  
 Interested readers are referred to some of the recent monographs
 and reviews on this subject 
 \cite{Bod89,Bod08,Her02,JHI06,JH08,Ise11}. 

Under nova conditions, the early evolution of the burst is dominated by 
the operation of both the proton-proton chains and the cold CNO cycle 
(mainly through $^{12}$C(p,$\gamma$)$^{13}$N($\beta^+ \nu$)$^{13}$C(p,$\gamma$)$^{14}$N). As the temperature increases, 
the characteristic timescale for proton captures onto $^{13}$N becomes 
shorter than its $\beta$-decay time, favoring a 
number of reactions 
of the hot CNO-cycle, such as $^{13}$N(p,$\gamma$)$^{14}$O, together 
with $^{14}$N(p,$\gamma$)$^{15}$O, 
or $^{16}$O(p,$\gamma$)$^{17}$F. Since such nuclear processes take 
place in an envelope that 
operates in degenerate conditions, with a pressure depending only on 
the density rather than on temperature, the star cannot react to the 
temperature increase with an expansion. This paves the road for a 
thermonuclear runaway (TNR), in which convection plays a critical role.
Actually, 
convective transport sets in when temperatures exceed $\sim$20 MK and
carries a substantial fraction of the  short-lived, $\beta^+$-unstable 
nuclei $^{14,15}$O, $^{17}$F, $^{13}$N, synthesized in the CNO-cycle, to 
the outer and cooler layers 
of the envelope, where they escape destruction by proton captures. In fact,
it is the sudden release of energy from 
these short-lived species, which become the most abundant elements in the
envelope after H and He, that powers the expansion and ejection stages in 
a nova outburst \cite{Sta72}. As a result, the ejecta will likely be 
enriched in their 
daughter nuclei $^{15}$N, $^{17}$O, and $^{13}$C, and hence these species are expected to be the
fingerprints of nova explosions in the chemical abundance pattern
\cite{Sta98,Sta09,JH98,Kov97,Yar05}.
This basic picture emphasizes 
the key role played by the initial $^{12}$C content as the main trigger of 
the thermonuclear runaway. Indeed, the reaction $^{12}$C(p,$\gamma$)$^{13}$N determines the amount of mass accreted 
in the envelope and the pressure at the envelope's base, which 
in turn determines the strength of the outburst (i.e., peak temperature, 
mass and velocity of the ejected shells). Note that the rates for this reaction seem to be well established at all 
temperatures \cite{Adel11,Azum10}. 

	From the nuclear physics viewpoint, novae are unique stellar 
explosions. Their limited nuclear activity, which involves about a hundred 
relevant species in the $A<40$ mass range, linked by a few hundred nuclear reactions, 
as well as the limited range of temperatures achieved in such explosions 
(10--400 MK), allows us to rely primarily on experimental nuclear physics input
\cite{JHI06}. In contrast to related astrophysical explosive sites that operate at much higher temperatures (e.g., supernovae, X-ray bursts), during a 
classical nova outburst the main nuclear path runs close to the valley of 
stability and is driven by (p,$\gamma$) and (p,$\alpha$) reactions, as well as by $\beta^+$-decays. Simulations also show that contributions 
from any neutron- or 
$\alpha$-capture reactions, including  $^{15}$O($\alpha$, $\gamma$)$^{19}$Ne, become 
negligible. Different studies have analyzed the role played by nuclear 
reaction rate uncertainties affecting the overall nova nucleosynthesis pattern
\cite{Ili02}. This sparked an extraordinary activity in many nuclear 
physics laboratories worldwide. Notable measurements were performed for the reactions $^{17}$O(p,$\gamma$)$^{18}$F, using stable beam facilities \cite{Fox05,New07,Cha07,New10}, 
and for the reaction $^{21}$Na(p,$\gamma$)$^{22}$Mg, using the ISAC radioactive ion beam facility at TRIUMF, Vancouver \cite{Dau04}. 
The former process strongly impacts production of the radioisotope $^{18}$F and the galactic synthesis of $^{17}$O, while the latter influences the production of $\gamma$-rays from $^{22}$Na radioactive decay.
Actually, the list of 
reactions whose uncertainty still has a strong impact on nova yields 
has been dramatically reduced. The main interest is now focused 
on measuring the challenging reactions $^{18}$F(p,$\alpha$)$^{15}$O, $^{25}$Al(p,$\gamma$)$^{26}$Si, 
and $^{30}$P(p,$\gamma$)$^{31}$S. 

\begin{figure}
\includegraphics[scale=.25]{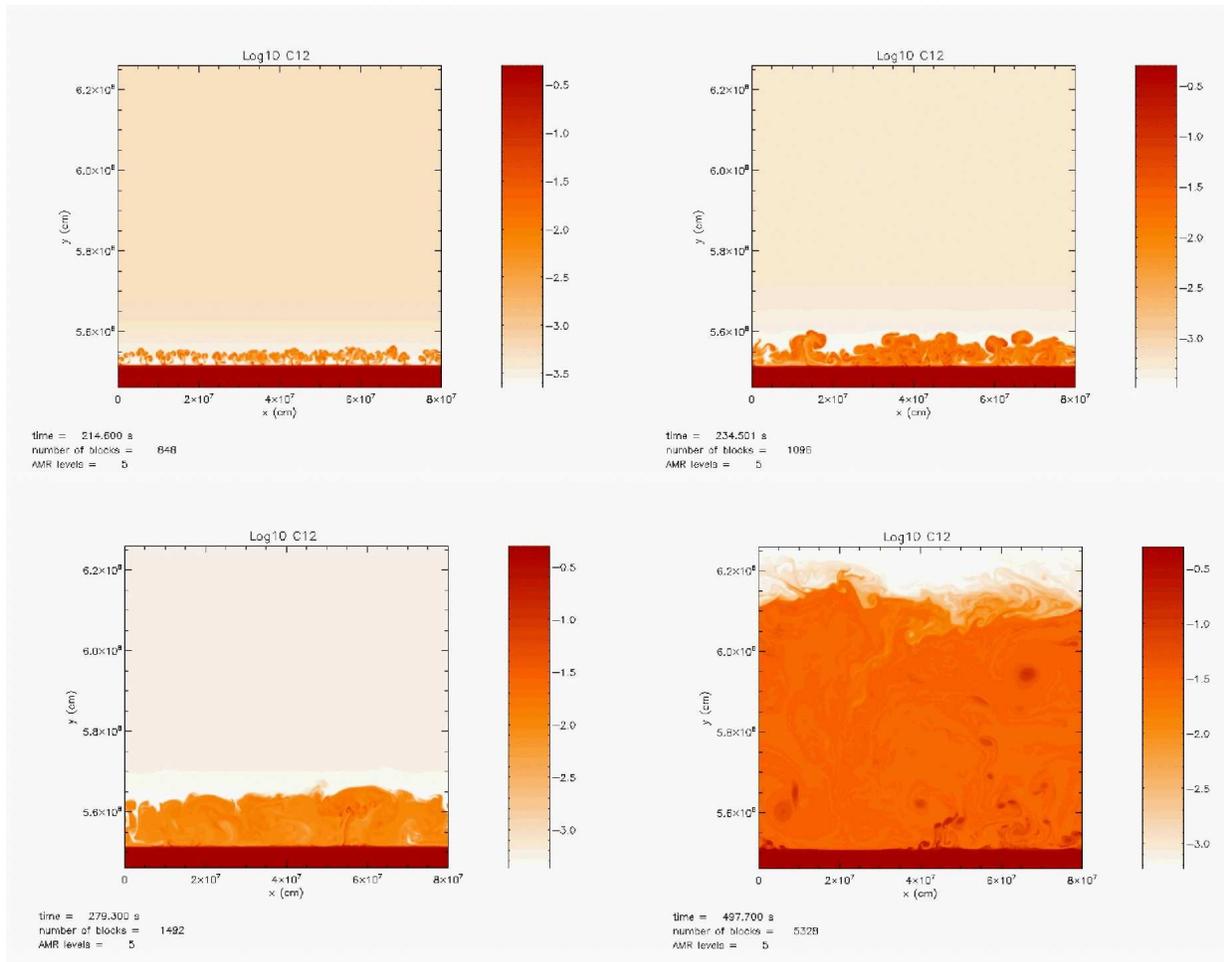}
\caption{Snapshots of the development of Kelvin-Helmholtz instabilities at
the core-envelope interface in 2-D simulations of nova outbursts, 
at t = 215 s (upper left panel), 235 s (upper right), 279 s (lower left), 
and 498 s (lower right), shown in terms of $^{12}$C mass fraction along
the computational domain. Notice how the
 Kelvin-Helmholtz 
instabilities dredge-up and transport unburnt CO-rich material from the outmost
layers of the white dwarf core and inject it into the envelope.
The mean CNO abundance in the envelope 
at the end of the simulations reaches $\sim$0.30, by mass \cite{Cas10}. 
\label{2DNOVA}}
\end{figure}

Models of nova nucleosynthesis point toward Ca as the likely nucleosynthesis 
endpoint, in agreement with observations. Despite 
the current problems associated with the modeling of nova outbursts (mainly, 
the discrepancy in the amount of mass 
ejected between models and observations \cite{Sta98,JS08}), no large
differences are found between the overall abundance 
pattern inferred from observations and that derived from numerical 
calculations.  It is worth noting however that abundances are not directly
determined  but inferred, and their associated uncertainty is often quite large.
Moreover, 
observations reveal that 
 the ejecta accompanying classical novae are significantly enriched in 
metals: on average, CO-rich novae show a mean mass fraction of about 0.25,
while ONe-rich novae are characterized by values near 0.5. Because of the moderate 
peak temperatures achieved in these explosions, it is unlikely that such 
metallicity enhancements could result from thermonuclear processing of 
solar-like material. Instead, mixing at the core-envelope interface has 
been proposed as the likely explanation. Several mixing mechanisms have 
been proposed, including multidimensional processes. 
It has to be emphasized, however, that all multidimensional nova simulations
performed so far rely on extremely limited nuclear reaction networks, 
containing only 13 isotopes to properly account for the energetics of the 
explosion. These studies are also restricted to some specific stages of the thermonuclear runaway,
following the development of convection and the onset of the explosion 
for several hundred seconds only. 
Bearing in mind that the overall process, from accretion to ejection, 
lasts for about
$\sim 10^5$ yr, it is clear that such multidimensional models cannot provide presently
nucleosynthesis constraints.
The only two-dimensional studies of mixing in novae available until recent
 years \cite{Gla95,Gla97,Gla05,Gla07,Ker98} reached totally different
 conclusions regarding the feasibility of mixing by 
hydrodynamic instabilities. 
 To solve the existing controversy, other multidimensional simulations
have been performed recently with the hydrodynamic code FLASH.
The new 2-D studies \cite{Cas10,Cas11} showed   
 large convective eddies naturally developing close to
the core-envelope interface through {\it Kelvin-Helmholtz instabilities}
(instabilities resulting from a velocity shear between two fluids), with a 
size comparable to the height of the 
envelope (Fig. \ref{2DNOVA}), 
which dredge-up CO-rich material from the outermost layers of the 
underlying white dwarf into the accreted envelope. The mean metallicity 
achieved in the envelope, Z$\sim$0.20--0.30, is in agreement with observations 
of CO nova ejecta. Similar results have been found in a 3-D framework by
the same group (in contrast to the results reported in Ref. \cite{Ker99}),
despite of the different way in which convection develops and extends throughout the envelope in the 2-D and 3-D simulations.

Infrared and ultraviolet observations have significantly contributed to the
understanding of the nova phenomenon, frequently revealing dust forming 
episodes in the ejecta \cite{Geh98}.  
Since the pioneering studies of dust formation in novae by Clayton \& Hoyle 
\cite{CH76} (a concept already suggested by Cameron in 1973 
\cite{Cam73}), all attempts
devoted to the identification of possible nova grains relied mainly on 
the search for low $^{20}$Ne/$^{22}$Ne ratios. Since noble gases, such as 
Ne, do not condense into grains, $^{22}$Ne was attributed to {\it in situ}
$^{22}$Na decay, a clear imprint of a classical nova explosion. Indeed, 
Clayton and Hoyle pointed out several isotopic signatures (large 
overproduction of $^{13,14}$C, $^{18}$O, $^{22}$Na, $^{26}$Al or $^{30}$Si), 
that may help to identify such putative nova grains. About fourty 
years later most, but not all, of these signatures still hold in view of our current 
understanding of such explosions (except for $^{14}$C, which is bypassed by the main 
nuclear path in novae, and $^{18}$O, which is slightly overproduced in novae 
although grains nucleated in such environments are expected to be
more anomalous in $^{17}$O.).
A major step forward in the identification of presolar nova grains 
was achieved by Amari et al. \cite{Ama01,Ama02}, who reported on 
several SiC and graphite grains, isolated from the Murchison and Acfer 094 
meteorites, with abundance patterns qualitatively similar to those obtained 
from nova models: low $^{12}$C/$^{13}$C and $^{14}$N/$^{15}$N ratios, high 
$^{30}$Si/$^{28}$Si, and close-to-solar $^{29}$Si/$^{28}$Si (plus high 
$^{26}$Al/$^{27}$Al and $^{22}$Ne/$^{20}$Ne ratios for some of the grains). 
However, in order to quantitatively match the grain data, one had to assume  
mixing between material synthesized in the outburst with
more than ten times as much unprocessed, isotopically close-to-solar 
material before grain formation. One possible source of dilution might be
mixing driven by the collision of the ejecta either with the accretion 
disk or the outer 
layers of the stellar companion \cite{Cam11}. 
Concerns about the likely nova paternity of the Amari et al. grains have been raised \cite{NH05},
after three additional $\mu$m-sized SiC grains 
were isolated from the Murchison meteorite with similar trends, in 
particular, low $^{12}$C/$^{13}$C and $^{14}$N/$^{15}$N ratios, but also
non-solar Ti ratios. A supernova origin cannot therefore be excluded for 
at least some of the grains, although
additional work on this subject seems mandatory to disentangle the controversy. 
Recently, other nova candidate (oxide) grains have been identified \cite{Gyn11}.

Among the species synthesized in classical nova outbursts, several 
radioactive nuclei have deserved particular attention as potential 
$\gamma$-ray emitters \cite{CH74,Cla81,LC87,Gom98}. The species
 $^{13}$N ($T_{1/2}=598$~s) and $^{18}$F ($T_{1/2}=1.83$ hr) 
 may provide a prompt 
$\gamma$-ray signature at 511 keV (through electron-positron 
annihilation), 
plus a continuum at lower energies due to Comptonized photons (with a 
cut-off at 20-30 keV because of photoelectric absorption) \cite{Her99}.
Unfortunately, this emission is short-lived, lasting only a few hours
and takes place before the nova optical
discovery. Hence, this signature must be detected {\it a posteriori}, through 
data mining by analyzing the data collected
by any $\gamma$-ray satellite that was looking at the 
right spot on the sky at the right time.
Even though the 511 keV line (and the lower energy continuum) has 
the largest predicted flux of all $\gamma$-ray
signals, better chances of observing $\gamma$-rays from novae may be
achieved
with two other (long lasting) signatures: a 478-keV line from $^7$Be decay 
($T_{1/2}=53.3$ days),
 and a 1275-keV line from $^{22}$Na decay ($T_{1/2}=2.6$ yr).
  The species
 $^{26}$Al is another important radioactive isotope that may
be synthesized during nova outbursts, although only its cumulative 
emission 
can be observed because of its slow decay ($T_{1/2}=7.16 \times 10^5$ yr). 
 More details on the predicted $\gamma$-ray emission from novae can be 
found in Ref. \cite{Her08}.

\section{Thermonuclear (type Ia) supernovae\label{sn1a}}
Type Ia supernovae (SN Ia) have become valuable cosmological tools. 
Through calibrated light curve analysis, they have been used as probes to 
outline the geometrical structure of the Universe, unraveling 
its unexpected acceleration stage \cite{Rie98,Per99}.
Moreover, their leading role in the chemical
and dynamical evolution of the Galaxy has motivated titanic efforts in the modeling of such
stellar beacons.

\subsection{Observational constraints}
Type Ia supernovae are observed in all types of galaxies, 
in sharp contrast with other supernovae types (SN Ib, SN Ic or SN II), which are
only found in spiral and irregular galaxies. This suggests that  
SN Ia are likely associated with old stellar populations and that the progenitors of SN Ib/c and SN II are much younger. 

Photometrically, type Ia supernovae exhibit a sudden rise in  
luminosity, up to a maximum visual absolute magnitude\footnote{The apparent magnitude of a star, $m$,
indicates its brightness as seen from Earth. It is based on  
a logarithmic measure of luminosity, such that a decrease 
in magnitude, $\Delta m$, translates into an increase in luminosity (from L 
to L$^\prime$) as L$^\prime$/L = 10$^{\Delta m/2.5}$. The absolute magnitude, $M$, is defined as
the apparent magnitude that a star would have if it were located at a distance of 10 parsecs (32.62 light-years) from Earth.
The connection between apparent and absolute magnitudes of a star can be expressed as
m - M = 5 log (d/10), where $d$ is the distance to the star in parsecs. 
The absolute magnitude and the luminosity of a star are related through M = M$_{sun}$ - 2.5 log(L/L$_\odot$), where
M$_{sun}$ = 4.76 and L$_\odot = 3.83 \times 10^{33}$ erg s$^{-1}$ are the absolute magnitude and the luminosity
of the Sun.} 
of $M_v \sim -19.3^{mag}$ (L$_{peak} \sim 10^{10}$ L$_\odot$) 
in about 20 days. This phase is
followed by a steep decline in brightness by about 3 magnitudes 
in $\sim$ 30 days, and later by a second, smoother decline over a period of
$\sim$70 days \cite{Kir90,HW90}.

The energy required to power a supernova explosion can be inferred from
determinations of the kinetic energy of the expanding ejecta (at typical
velocities ranging between 5,000--10,000 km s$^{-1}$), E$_{kin} \sim
10^{51}$ erg, as well as from the energy integrated over the
light curve,  E$_{rad} \sim 10^{49}$ erg. 
This energy budget obviously requires a powerful source. Early models attributed
this to thermonuclear processing of CO-rich material into Fe-peak elements
inside a white dwarf \cite{HF60}. Indeed,  
the incineration of $\sim 1$ M$_\odot$ of a C-O mixture, 
which releases $\sim 10^{18}$ erg g$^{-1}$, accounts for the required
$10^{51}$ erg. 

During the late stages the light curve is powered by the radioactive decay chain
$^{56}$Ni($\beta^+ \nu$)$^{56}$Co($\beta^+ \nu$)$^{56}$Fe 
 \cite{Tru67,Col69}, with the two different
 slopes attributed to the different half-lives of the corresponding
nuclear decays, $T_{1/2}(^{56}Ni) = 6.1$ days, 
and $T_{1/2}(^{56}Co) = 77.2$ days.
 Estimates of the $^{56}$Ni mass produced 
 in type Ia supernovae can be 
 determined empirically from modeling the late-time (nebular) spectrum
 and through application of {\it Arnett's rule}
\cite{Arn82}: the peak luminosity 
 is equal to the energy inputs from radioactive
decays within the expanding ejecta. Assuming a rise time to peak luminosity of 19 days,
Arnett's rule gives for the peak luminosity
L$_{peak} = 
2 \times 10^{43}$ $\left(M_{Ni}/M_\odot\right)$ erg s$^{-1}$ \cite{SL05}. 
Estimates for the $^{56}$Ni masses from 17 type Ia supernovae, 
inferred from these methods, are in the range of
0.1--1 M$_\odot$ \cite{Str06}. 
It is worth noting that a large fraction of the energy released in the
decays from $^{56}$Ni and $^{56}$Co is in the form of $\gamma$--rays,
with energies of 158, 812, 750 and 480 keV for 
$^{56}$Ni($\beta^+ \nu$)$^{56}$Co,
and 847 and 1238 keV for $^{56}$Co($\beta^+ \nu$)$^{56}$Fe.
Hence, supernovae have become priority targets for many $\gamma$-ray missions,
like INTEGRAL, or in mission proposals, such as DUAL or GRIPS.

Type Ia supernovae are spectroscopically identified by the absence of
hydrogen emission Balmer lines and the presence of a prominent blue-shifted silicon II (P Cygni) 
emission feature near maximum light \cite{HW90}.  Whereas
the first observational constraint poses limits on the maximum amount of
hydrogen that can be present in the expanding atmosphere of the star
(i.e., $M_{\rm H} \leq 0.03 - 0.1 M_\odot$), the second feature reveals that 
nuclear processing must be taking place during the event. 
Spectroscopic snapshots at different epochs provide valuable clues on the
physical mechanism powering such events: the spectra at maximum 
light are dominated by the presence of neutral or singly ionized lines of
Si, Ca, Mg, S, and O, moving at high speeds (in the range of $v \sim$ 8,000--30,000 
km s$^{-1}$) \cite{Fil92}.  Around two weeks after peak luminosity prominent, 
permitted Fe II lines are clearly seen. Finally, a month after maximum,
the spectrum is dominated by the presence of forbidden emission lines from
Fe II, Fe III and Co III.
The overall picture clearly stresses key nuclear physics issues:
the presence of intermediate-mass
elements in the early spectra, when only emission from the outermost ejected 
layers is seen, reveals that the thermonuclear explosion did not incinerate 
the whole star (incomplete burning). This, in turn, provides clues on the
flame propagation regime (a yet unsolved issue in supernova theory), since
the presence of intermediate-mass elements rules out a pure detonation
(Sec. 8.3).
In sharp contrast, when inner regions of the star become later on 
accessible, the prominence
of the Fe lines clearly points towards complete nuclear processing
of matter to Fe-peak elements. Finally, the presence of Co lines
at late stages strongly supports the hypothesis of a light curve tail
powered by $^{56}$Co-decay. 

\subsection{Progenitors}
The most widely accepted scenario for type Ia supernova explosions 
involves a carbon deflagration or detonation of a CO white dwarf that reaches
its maximum mass (the so-called {\it Chandrasekhar limit}, $M_{Ch}\sim  
1.4$ M$_\odot$) by accretion from a companion star.
Helium white dwarfs have been ruled out
as potential progenitors, since these objects would undergo extremely violent and
too energetic detonations \cite{Nom77,Woo86},
and no intermediate-mass elements would be produced.  
In turn, ONe(Mg) white dwarfs are expected 
to collapse, because of the efficiency of electron captures on $^{24}$Mg 
\cite{Nom87,NK91,Can92,Gut06}.

Two different progenitor systems have been proposed so far:
a single degenerate one, dealing with a main sequence or
giant companion that transfers matter onto a white dwarf star \cite{WI73};
and a double degenerate system composed by two white dwarfs \cite{IT84,Web84}.  In the
latter scenario, the explosion takes place when the two white dwarfs merge as a 
result of angular momentum loss driven by the emission of 
gravitational waves.
During the last decades, both scenarios were regarded as
competing, since SN Ia were considered a quite homogeneous
class of objects, both at maximum light \cite{Bar73}
and at late stages \cite{Bra92}. Hence, a unique progenitor was assumed. 
 Arguments in favor of double-degenerates
include the fact that those systems
 account in a natural way for the lack of hydrogen in the spectra. 
A major problem, however, involves the statistics derived from the search of 
systems with 
double-degenerates: only KPD 0422+5421 \cite{Koe98} 
has a mass that could drive a Chandrasekhar-mass explosion after a merging episode lasting less than the age of the Universe. Although theoretical estimates 
based on population synthesis codes provide an appropriate frequency of 
massive mergers to account for the
observed type Ia supernova rate \cite{Liv00}, numerical
simulations do not necessarily support an explosive fate for such
systems (see Refs. \cite{NoIb85,WW86,ML90}, and Sect. 10).
 These problems have revitalized 
the interest in single degenerate scenarios, with potential candidates like cataclysmic variables 
(formed by a white dwarf and a low-mass main sequence star) or symbiotic 
variables (a white dwarf plus a red giant star).
In the first case, the accreted matter is H-rich. 
This poses problems since, even though some hydrogen should have been
stripped from the secondary star during the explosion, this element
is not seen in the spectra of type Ia supernovae \cite{MBF00,MCH07}.
But its major drawback is related to its path toward the Chandrasekhar mass:
the theory of stellar evolution predicts that the maximum mass for a 
CO-white dwarf is about $\sim 1.1$ M$_\odot$ (see Ref. \cite{Alt10} and
references therein). 
If a CO-white dwarf is required for a successful explosion, 
such objects would need to accumulate roughly 0.3 M$_\odot$.
This is difficult to achieve and, in particular, mass-accretion rates
leading to nova-like explosions, implying loss rather than accumulation of mass,
must be avoided (Sect. 7).
In symbiotic variables, the accreted matter is instead 
He-rich. Several works \cite{MuRe92,Mun94}
 have estimated a very high frequency
of such systems in the Galaxy. Actually, to account for the
observed Galactic SN Ia rate it would be sufficient if $1-4 \%$ of all
symbiotic variables 
 end their evolution with the accreting white dwarf reaching the
Chandrasekhar limit. However, it has been pointed out \cite{Ken93}
that white
dwarfs in symbiotic binary systems scarcely reach the Chandrasekhar mass.
Calculations yield about $0.1$ M$_\odot$ accreted during an
assumed lifetime of $\sim5 \times 10^6$ yr,
suggesting that some white dwarfs may even explode
 before reaching the Chandrasekhar limit.

\begin{figure}
\includegraphics[scale=.80]{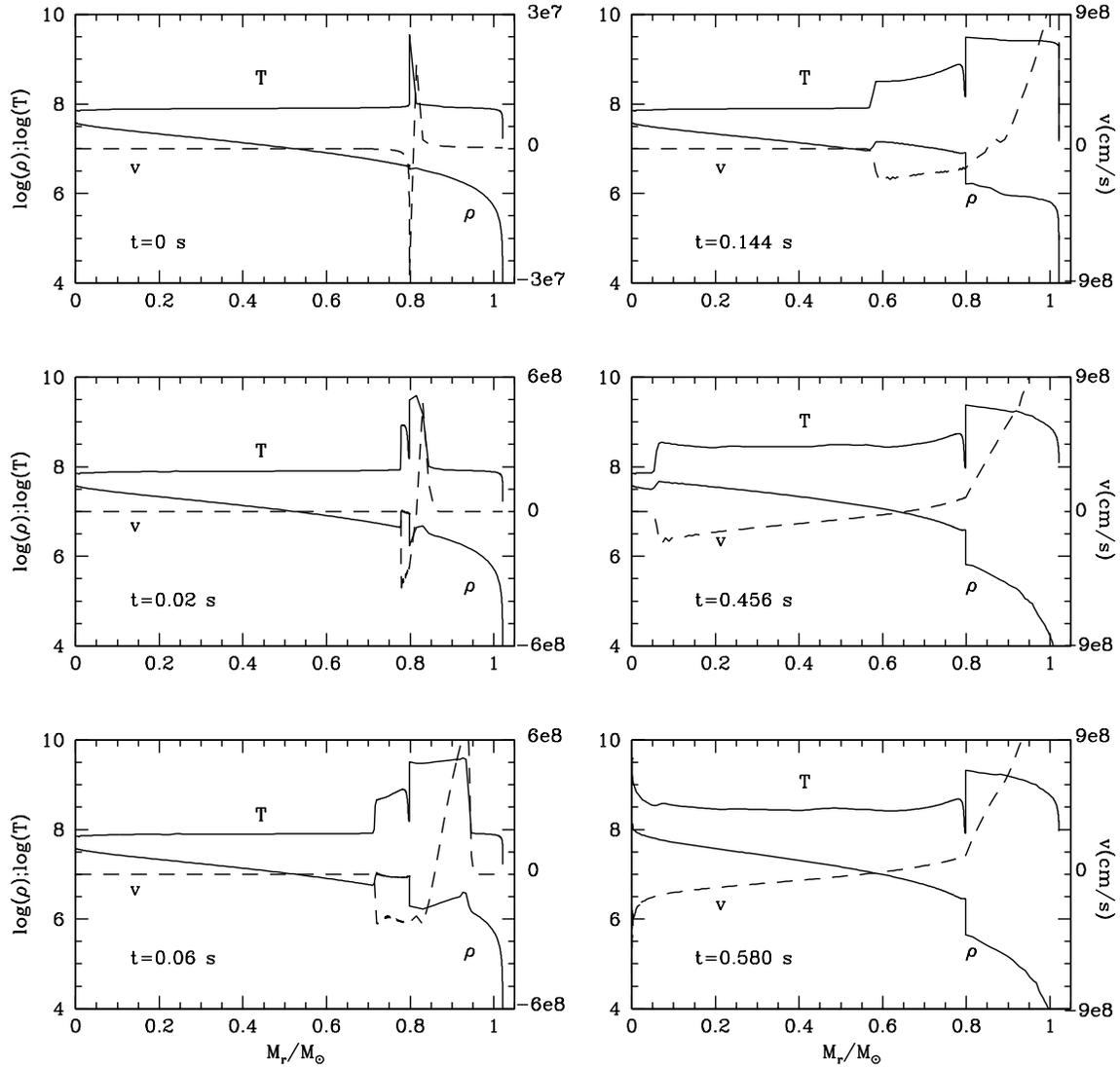}
\caption{Temperature, density and velocity profiles resulting from
He-rich accretion onto a 0.8 M$_\odot$ CO white dwarf, 
at a rate of $\dot M = 3.5 \times 10^{-8}$ M$_\odot$ yr$^{-1}$,
following the progress of the detonation
front originating at the envelope's base. The horizontal axis
corresponds to the (internal) mass coordinate, ranging from 0 
(center of the star) to 1 (surface). Notice how the 
ingoing compressional wave that propagates inward
 provokes carbon ignition near the center (last panel).
 The
simulation has been performed with the 1-D hydrodynamic code SHIVA 
\cite{Jos96}. 
\label{SubchaJJ}}
\end{figure}

Although the canonical scenario is based on the explosion of a 
Chandrasekhar mass CO white dwarf,  He-rich accretion onto low-mass 
white dwarfs has also received a particular interest, mostly in 
connection with subluminous supernovae. 
In these models, helium ignites at the core-envelope interface, 
driving a double shock front (Fig. \ref{SubchaJJ}): a detonation wave 
that moves outward, reaching the stellar 
 surface after incinerating most of the envelope to $^{56}$Ni, 
 and an ingoing compressional wave that propagates toward the
 center, where it provokes carbon ignition \cite{WW94}.
 This scenario has been suggested as a likely site for 
a number of nuclear processes, including {\it $\alpha$p-process}, 
 {\it pn-process} (i.e., a variant of the {\it rp-process}; Sec. \ref{rpp}) and, more 
importantly, {\it p-process} (Sec. \ref{pproc}), capable of overproducing almost 
all p-nuclei in solar proportions, including the elusive 
Mo and Ru species (Refs. \cite{Gor02,Gor05}; see also Ref. \cite{Kus11}
for recent work on the synthesis of p-nuclei in type Ia supernovae). 
 The crucial issue turns out to 
be the precise He ignition density that determines the 
strength of the explosion, and therefore, the range of 
temperatures expected in the envelope. The overall amount of $^{56}$Ni
produced in these {\it sub-Chandrasekhar} models is often found 
to be smaller than in standard C-deflagration or detonation models, 
leading to fainter events and hence providing  
a possible origin for dim supernovae, like the sub-luminous 
SN 1991bg \cite{WW94,Hof95}. 
From the nucleosynthesis viewpoint, these sub-Chandrasekhar models
are characterized by a significant overproduction of a number of nuclear 
species, including $^{44}$Ca (from the $\beta$-decay of $^{44}$Ti) 
and $^{47}$Ti, which are systematically underproduced in type II and 
type Ia models. 
The main problem associated with this scenario is
that only a very narrow range of mass-accretion rates
($\sim 10^{-9} \leq \dot M_{acc}$(M$_\odot$ yr$^{-1}$) $\leq 
5 \times 10^{-8}$) 
gives rise to a central C-detonation and thus the models require 
some fine tuning. It is also worth noting that most of these studies have been 
performed in spherical symmetry and claims have been made that the quasi-central 
C-detonation is in fact  imposed by the adopted geometry. 
Nevertheless, 3-D simulations of sub-Chandrasekhar models with 
asynchronous ignition (i.e., ignition takes place in a suite of
different envelope points) produced successful supernova-like explosions
\cite{GB05}. Nevertheless, the main arguments against sub-Chandrasekhar
models (at least, as the likely scenario for the majority of type Ia 
supernova explosions) result from a number of spectroscopic anomalies, 
namely the prediction of a fast expanding layer consisting of $^{56}$Ni and 
$^{4}$He, which is not observed in the spectra of SN Ia, and shorter than observed rise
times to peak luminosity \cite{HK96}.

More recently, other supernova-like explosions, driven by He-mass 
transfer in AM CVn stellar binaries\footnote{AM CVn stars
are a class of cataclysmic variables named after AM Canum Venaticorum.
In these variables, a white dwarf accretes H-deficient matter from a compact 
stellar companion. They have extremely short orbital periods ($<$1 hr), 
and are predicted to be strong sources of gravitational waves.} 
have received some attention. Bildsten et al. \cite{Bil07}
conclude that He-detonations producing primarily 
$^{48}$Cr, $^{52}$Fe, and $^{56}$Ni are likely expected in some of these 
systems. They also suggest that these faint ``.Ia" supernovae 
may represent between 1.6\% and 4\% of the local type 
Ia supernova rate. 

\subsection{Type Ia supernovae as standard candles}
The increasing number of supernovae discovered has revealed a
diversity among SN Ia, raising serious doubts on the uniqueness of the 
progenitor. 
About 85\% of the observed SN Ia are believed to belong to a remarkably
homogeneous class of events \cite{Bra93}, with a dispersion of only 
 $\Delta M \le 0.3^{mag}$ \cite{Cad85,Ham96} when 
 normalized to peak luminosity. These SN Ia are known as ``Branch-normals",
with canonical examples such as SN 1972E, 1981B, 1989B or 1994D. But
the number of SN Ia deviating from this homogeneous
class may be as high as $\sim$30\% \cite{Li00,Rop11}. 
Already in the 1970s \cite{Bar73}, it was evident that two
speed classes existed, a fast and a slow one, 
characterized by different rates of decline 
in their light curves. 
This diversity was soon interpreted in terms of 
explosion strength and envelope opacity: 
weaker explosions result in less luminous (and redder)
events, showing a faster declining light curve and somewhat smaller 
velocities in the ejecta compared to more energetic supernovae \cite{Bra98}.
Whereas SN 1991T is considered one of the most energetic supernovae ever 
observed, SN 1991bg and 1992K are the reddest, fastest and most subluminous 
SN Ia known to date \cite{Fil92,Ham94}. 
The difference in brightness between both groups amounts to $\sim 2^{mag}$. 
The existence of these different SN Ia classes was firmly established by
Phillips \cite{Phi93} on the basis of an empirical relation between 
maximum brightness and rate of decline during the first 15 days after 
peak. This correlation is actually used to renormalize the observed
peak magnitudes, thereby substantially reducing the dispersion in absolute 
magnitudes, and thus making type Ia supernova one of the most precise 
cosmological distance indicators.

The variations in light curve (shape, peak luminosity, characteristic
timescales, and so on) are interpreted as driven by a 
different amount of $^{56}$Ni synthesized in the explosion,
for the same overall ejected mass. 
Since peak luminosity is proportional to the $^{56}$Ni mass, brighter events 
are found when more $^{56}$Ni is produced. This is accompanied by larger 
expansion velocities and broader peaks because of the increased opacity 
(deeply influenced by the presence of Fe-peak elements). 
Models based on light curve analysis of the dim SN 1991bg
suggest that the amount of $^ {56}$Ni present in the ejecta was only 
$0.07$ M$_\odot$ \cite{Maz97}, much smaller than the canonical 
value of $\sim0.6$ M$_\odot$ associated
with normal SN Ia \cite{HK96}. 

\subsection{Ignition regimes}
Carbon ignition in the degenerate interior of a white dwarf close to the 
Chandrasekhar limit can result in the propagation of a supersonic wave (a {\it detonation}
driving nuclear burning by a shock wave) 
or a subsonic wave (a {\it deflagration} inducing burning by
thermal conduction  of the degenerate electrons). The actual outcome depends on the specific 
density, temperature, chemical composition and velocity profiles at the time 
ignition sets in (with low densities favoring the onset of detonations). 
In this section, we will briefly summarize the main characteristics of the 
different regimes proposed so far. More details can be found in Refs. 
\cite{Ise11,HN00,Rop08}. 

\subsubsection{The prompt detonation model.} It assumes a pure detonation
at the white dwarf center that propagates outward \cite{Arn69}. 
Although energetically
feasible, it fails to account for the distribution of intermediate-mass
elements inferred from the spectra of type Ia supernovae. Indeed, 
a detonation wave prevents the expansion of the layers ahead of the burning
front and hence the fuel is almost totally incinerated into Fe-peak elements.
The tiny amount of intermediate-mass elements produced is insufficient to provide the characteristic strong Si II feature, thus ruling out
this model as the likely mechanism powering SN Ia. 

\subsubsection{The deflagration model.} When assuming a subsonic rather than a
supersonic flame propagation, fuel layers can react to the
advancing front with an expansion \cite{Nom76}. 
Thus, a fraction of the star's matter can in principle 
be processed into intermediate-mass elements. 
The feasibility of a deflagration model depends on its ability to 
avoid unburnt pockets of carbon and oxygen behind the front \cite{NiWo97}.
This has been explored in the framework of high-resolution, multipoint 
ignition simulations \cite{Rop06}, which show the inability to
reach $^{56}$Ni masses of $>$0.7 M$_\odot$ and kinetic energies exceeding $0.7\times 
10^{51}$ erg, and thus are unable to 
account for the bulk of {\it Branch-normal} type Ia supernovae. 
For the typical conditions that characterize white dwarf interiors, such
deflagration fronts are highly subsonic \cite{Tim92} and hence allow for an 
expansion of the outer white dwarf layers. The expansion ultimately quenches
the burning, at a time when the white dwarf is still gravitationally bound.
Moreover, large amounts of unburnt C and O 
($> 0.57$ M$_\odot$) are left behind \cite{Sch06}. Two additional problems associated
with deflagration
models include the lack of chemical stratification in the ejecta as well
as the presence of big clumps of radioactive 
$^{56}$Ni in the photosphere at peak luminosity \cite{Ise11}.

Many multidimensional turbulent deflagration models have been published
in recent years \cite{Rei02,Gam03,RH05,RH05b,SN06,Rop07}, examining the effects
of turbulence and instabilities on the propagation of the flame.

\subsubsection{The delayed detonation models.} In this model, a deflagration
front propagates and pre-expands the star. Subsequently, the
deflagration wave switches into a detonation
 ({\it deflagration--detonation transition}). 
 Several versions of this model exist, with \cite{Iva74} or without \cite{Kho91,Kho93}
 assuming a pulsation.
One-dimensional simulations of delayed detonations proved successful
in reproducing many observational features of type Ia supernovae.
Indeed, the combination of the two ignition regimes allows for the synthesis
 of intermediate-mass elements, while providing the required energy budget.
 Moreover, the model accounts for the expected 
 light curves and photospheric expansion velocities \cite{HK96}.
 The main uncertainty affecting these models deals with the mechanism
 driving the deflagration--detonation transition. Recent studies suggest
 that this naturally takes places once the flame
reaches the so-called {\it distributed burning regime} \cite{RoNi07},
at typical densities $\sim 10^7$ g cm$^{-3}$. 
This corresponds to the time when the laminar flame width equals the
Gibson scale, that is, the scale at which turbulent velocity fluctuations 
equal the laminar flame speed. When the Gibson scale
becomes smaller than the laminar flame width, turbulence affects
 the laminar flame structure. 

A somewhat related model, the {\it gravitationally confined detonation model,
has been recently proposed in the framework of multidimensional simulations
\cite{Ple07}.  It relies on asymmetric deflagration flame ignitions, pushing
burnt fuel towards the white dwarf surface. The collision of these ashes on the
far side may generate a detonation wave propagating inward and burning the
stellar core, potentially triggering very energetic supernovae 
\cite{Ple07b,Tow07}. This model also produces substantial amounts of
intermediate-mass elements. Notice, however, that the feasibility of this 
scenario has been questioned by other studies \cite{RWH07}.}

Alternative models based on pulsational delayed detonations have also 
been proposed to account for the bulk of type Ia supernovae. As in standard 
delayed detonation models, the initial deflagration wave pre-expands the star. 
Because of the slow (subsonic) flame velocity, the burning front is quenched 
and fails to unbind the star. But during recontraction,
compressional heating at the interface between burnt and unburnt material 
ultimately triggers a detonation.
Recent multidimensional deflagration models \cite{Rei02,Gam03}, with a more 
accurate treatment of flame instabilities, suggest that the star may get 
unbound rather than recontracting.
Finally, other models propose that when no detonation is triggered by the 
collision of ashes, a pulsational contraction
may trigger a detonation wave that revitalizes the explosion 
\cite{BGS06,BGS09a,BGS09b}.
The reader is also referred to Refs. \cite{Gam05,Gol05,RN07,BGS08,Mae10} 
for additional information on multidimensional studies of delayed detonations.

\subsection{Nucleosynthesis: type Ia supernovae as Galactic Fe factories}
About once per century in our Galaxy, but probably once per second in the 
observable Universe, a white dwarf is totally disrupted by a titanic 
supernova explosion.
The specific abundance pattern of the nuclear species freshly synthesized during
such explosions depends critically on the peak temperature
achieved, as well as on the neutron excess (Sec.~\ref{expnucsec}). The latter 
depends in 
turn on the metallicity of the
white dwarf progenitor, in particular the distribution of neutron-rich species (such as
$^{22}$Ne), as well as on the density at which the thermonuclear 
runaway occurs. In broad terms, the abundance pattern of the ejecta is the 
result of four different burning regimes: nuclear statistical equilibrium
in the inner regions, incomplete Si-burning, O-burning, and C/Ne-burning in the outermost layers (see Sec.~\ref{expnucsec} and Ref. \cite{Woo86b}).

\begin{figure}
\includegraphics[scale=.7]{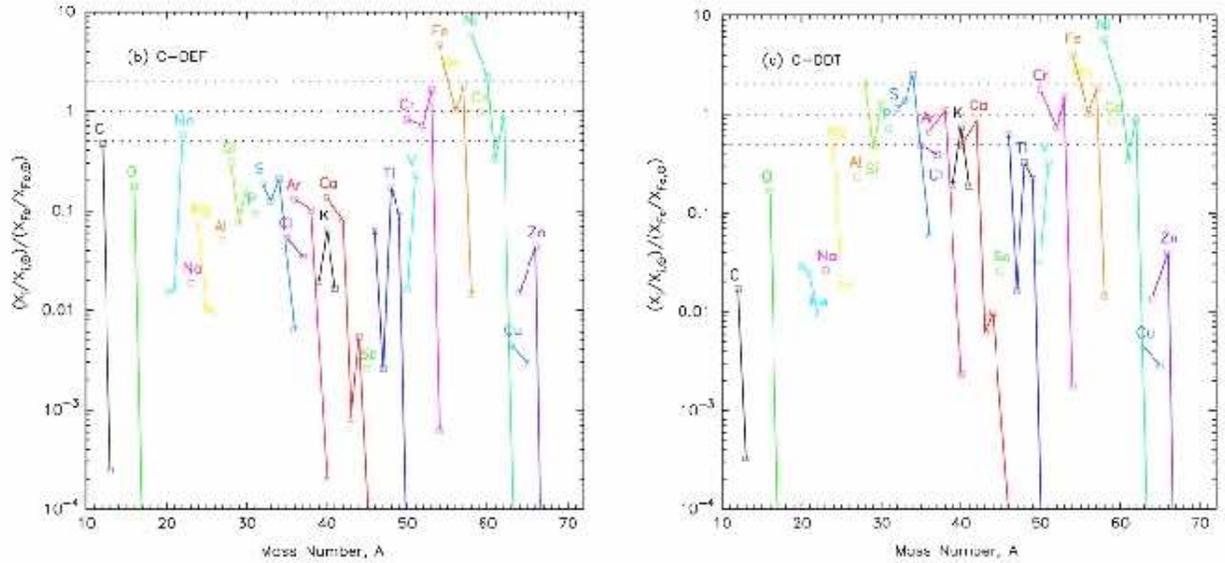}
\caption{Yields from 2-D models of type Ia supernova explosions: (left panel)
carbon deflagration; (right panel) delayed detonation.
Mass fractions are integrated over the whole ejecta, are 
shown relative to solar values, and are normalized to $^{56}$Fe.
Since delayed detonation models are initiated by a deflagration, the white
dwarf can expand prior to the onset of the detonation. Therefore,
the detonation moves through relatively low density layers producing only intermediate-mass elements 
(Fe-peak elements are only synthesized during the previous deflagration stage).
As a result, the distribution of Fe-peak elements is very similar in both C deflagration and delayed detonation models. 
Figure from Maeda et al. \cite{Mae10},
reproduced by permission of the American Astronomical Society.
\label{ropke}}
\end{figure}

The impact of nuclear uncertainties on the nucleosynthesis produced in type 
Ia supernovae has not been analyzed in the same detail as for other 
astrophysical scenarios, such as 
classical novae or X-ray bursts. Indeed, no sensitivity study aimed at 
identifying the key nuclear reactions whose uncertainties have the largest 
effect on the yield has been
performed to date. It is however worth mentioning the efforts made to 
accurately
determine the cross section of the $^{12}$C+$^{12}$C reaction, since it
triggers the explosion of these supernovae when temperature exceeds 
$\sim$ 700 MK. 
A new low-energy resonance, with a strong impact on the overall rate, has 
been reported in recent years \cite{Spi07}, but follow-up studies have not 
confirmed its presence.  
Further studies aimed at improving our knowledge of this rate are clearly 
required because of its relevant role as a trigger of type Ia supernova 
explosions, and hence, on the ignition density. 
Other properties of the explosion can depend as well on the specific 
composition
(more precisely on the $^{12}$C and $^{22}$Ne content). This has been 
analyzed in detail in a number of papers (see Refs. \cite{Cha07,Cha08,Tow09}, 
and references therein) that identified
systematic influences of composition on supernova explosions, including the 
ignition density, the specific density at which the initial deflagration
transforms into a detonation, the release of energy, or the flame speed, 
among others.
Hence, reactions involved in the synthesis and destruction of such nuclear species are
particularly relevant for our understanding of type Ia supernova explosions.

Estimates of the frequency of type Ia supernovae in our Galaxy 
yield a value of $\sim$0.2--0.3 events per century \cite{Cap03}. This is similar to the
expected occurrence of SN Ib/c ($\sim$0.65 per century), but smaller
than the inferred rate for core-collapse supernovae ($\sim$3.32 per century)
\cite{vdBT91}. Since the amount of $^{56}$Ni synthesized 
and ejected by each of the different supernova types
is about 0.6, 0.3 and 0.07 M$_\odot$, respectively,
it turns out that type Ia supernovae produce about half of the iron content in the Milky Way. Therefore, any 
model of type Ia supernovae must be able to 
reproduce the isotopic abundances of the  
Fe-peak elements determined in the Solar System (a task requiring 
Galactic chemical evolution models that take into account the different
contributions from all possible Fe-peak producers). For decades, 
models systematically overproduced neutron-rich species, 
such as $^{54}$Cr or $^{50}$Ti, with respect to the Solar System values \cite{Woo90b,Thi97,Nom97}.
This has greatly improved (but not yet solved) after the revision of the
stellar weak interaction rates \cite{LMP00}, a key
ingredient of all SN Ia nucleosynthesis models. 

For the different type Ia supernova ignition mechanisms discussed above,
somewhat different thermal histories and thus
different nucleosynthesis imprints are expected. This has been investigated
in detail in the framework of spherically symmetric, hydrodynamic models,
but only partially through multidimensional simulations. An early 
attempt, the famous carbon deflagration model W7 of Nomoto et al. \cite{Nom84}, was
successful in providing the abundance pattern required by 
Galactic chemical evolution models, as well as in reproducing the basic 
features of the observed spectra and light curves of some ``Branch-normal" SN Ia.
The extension of this model to the delayed detonation regime
improved the agreement with observations. 

Recent efforts to model the nucleosynthesis accompanying the explosions
in a truly multidimensional framework have been carried out by a handful
of groups \cite{RoNi07,Gam05,BGS08,Mae10}. An example of the expected 
nucleosynthesis has been recently reported in the
framework of 2-D simulations by Maeda et al. \cite{Mae10} (Fig. \ref{ropke}). 
We will briefly describe the differences reported between
 a pure turbulent deflagration model and a delayed detonation model.
At the first stages of the deflagration, nuclear statistical equilibrium ensues.
Electron captures become very important at the densities
achieved. The most abundant species in this fully 
incinerated region
are $^{54,56}$Fe, and $^{56}$Ni (which later decays into $^{56}$Fe).
Because of the subsonic nature of deflagrations, the white dwarf reacts
by expanding moderately, and hence the deflagration wave propagates through
progressively lower density layers. Hence, the front reaches layers where 
nuclear statistical equilibrium
still holds, but where electron captures are no longer relevant. 
The most abundant species in these layers are $^{56,58}$Ni.
 The front continues to advance through 
 successively lower temperature and density regions, where matter undergoes O-burning or incomplete 
Si-burning,
producing intermediate-mass elements, such as $^{28}$Si or
 $^{32}$S. Carbon- or Ne-burning regions are encountered subsequently, 
 resulting in large amounts of $^{16}$O and $^{24}$Mg. At some point, the low temperatures prevent
further thermonuclear reactions.

\begin{figure}
\includegraphics[scale=.8]{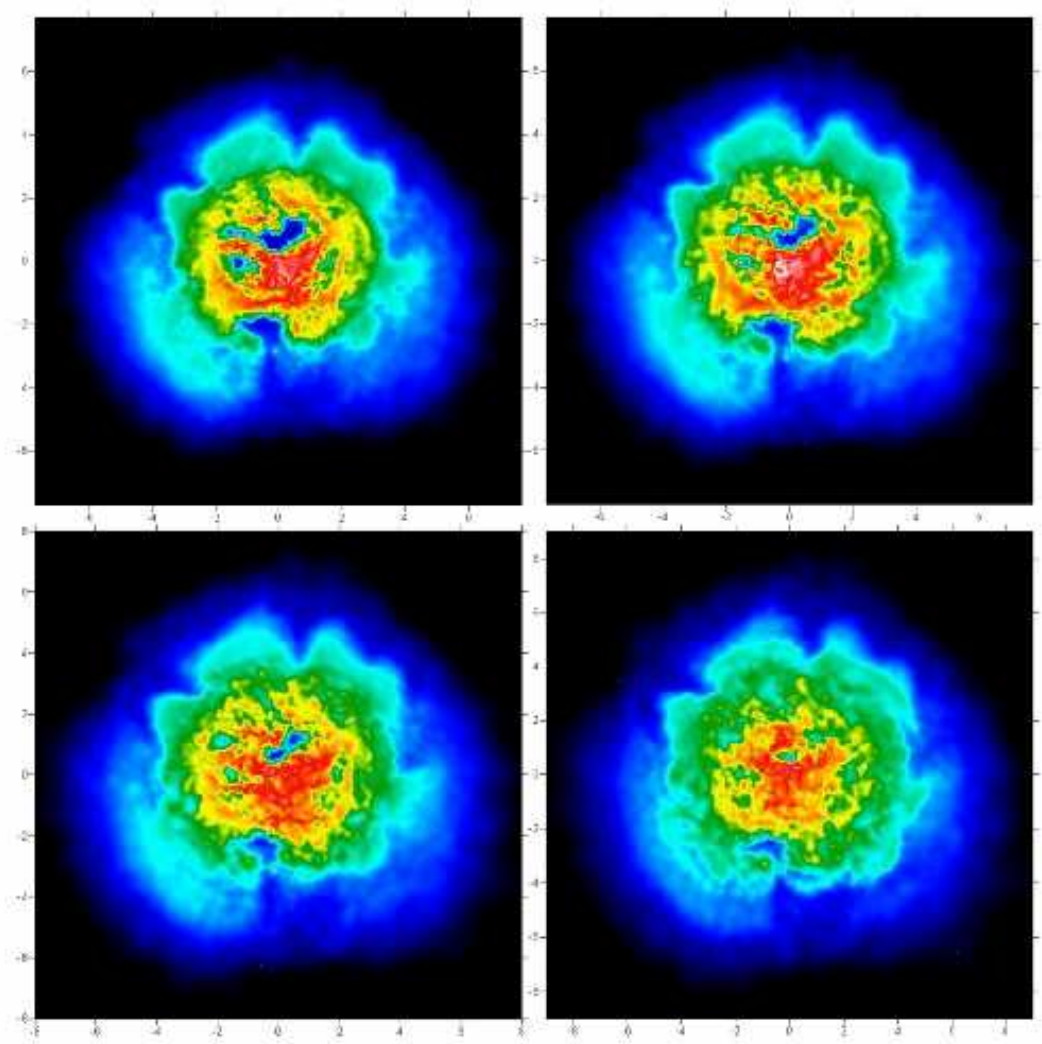}
\caption{Snapshots of the evolution of a type Ia supernova explosion,
in the framework of a 3-D delayed detonation model. Panels show a temperature
map during the detonation regime, at times t = 0.8, 0.9, 1.0, and
1.1 s. Regions in red are characterized by temperatures of about 5 GK, while
blue regions achieve at most 0.5 GK.
Credit: E. Bravo and D. Garc\'\i a--Senz, A\&A, vol. 478, p. 849, 2008 
\cite{BGS08}, reproduced with permission \copyright \, ESO.
\label{perro}}
\end{figure}

Similarly, in the delayed detonation
framework, the detonation is ignited after significant expansion
of the white dwarf has taken place, thus reducing the impact of electron captures. 
As the wave propagates outward, where the density decreases, the resulting nucleosynthesis is driven
by oxygen burning, carbon burning, and eventually burning
ceases. 

In terms of chemical stratification, the C deflagration model exhibits 
a broad region in which electron capture products ($^{54,56}$Fe, $^{58}$Ni),
complete-Si burning products ($^{56}$Ni), intermediate-mass elements
($^{32}$S, $^{28}$Si, $^{24}$Mg), and unburnt elements ($^{16}$O, $^{12}$C) 
co-exist.  This is surrounded by a shell with unburnt material in regions that are never reached 
by the flame. The delayed detonation model
results in a similar structure,  but in the innermost regions unburnt C
and O are efficiently processed into intermediate-mass elements by the delayed detonation wave. 
In the particular model reported by Maeda et al. \cite{Mae10}, C is almost completely burned, but O remains with a mean mass fraction of $\sim 0.1$. 
Surrounding the deflagration region, there is a layer containing products of O burning, namely
$^{28}$Si, $^{32}$S, and $^{24}$Mg. This, in turn, is surrounded by a shell containing a mixture of 
C-burning products and unburnt material. To summarize,  
the delayed detonation model reproduces the characteristic layered structure, in contrast to
what is found in the C deflagration model. 

It is finally worth noting that a layered structure, with a surface dominated by intermediate-mass
elements in agreement with observations, has also been obtained in other delayed detonation
models, such as the 3-D simulations of R\"opke \& Niemeyer \cite{RoNi07}. In contrast,  
Bravo \& Garc\'\i a--Senz \cite{BGS08} (Fig. \ref{perro}) found 
large amounts of Fe-peak elements near the surface regions
and a structure that is not clearly layered. This abundance
stratification has also been obtained in the 3-D delayed detonation models
of Gamezo et al. \cite{Gam05}. The reasons for the reported differences between
Bravo \& Garc\'\i a--Senz and the other groups are yet unclear, 
although they have been
attributed to differences in the modeling of the detonation, which is
poorly understood in multidimensional frameworks. 
 Extensive multidimensional simulations are needed to clarify these issues.
 The interested reader is referred to some of the existing monographs
 and reviews on type Ia supernovae, for further information 
 \cite{Ise11,HN00,Rop08,Bra01,RCI97,BMZ94,Pet90}.

\section{X-ray bursts. The $\alpha p-$ and $rp-$processes\label{rpp}}
Since the launch of NASA's X-ray satellite Uhuru,
in December 1970, astronomers have opened a new window to
observe violent phenomena that occur in the cosmos. 
Indeed, a suite of different astrophysical
sources have revealed an intense X-ray emission, including AGNs (active galactic
nuclei powered by black holes), supernova remnants, and binary stellar systems hosting compact objects,
such as white dwarfs (cataclysmic variables; 
super soft X-ray sources), neutron stars (X-ray bursters) or black holes
(X-ray novae). Among these objects, X-ray bursters (XRBs), 
serendipitously discovered in the 1970s, have deserved particular attention.

XRBs exhibit brief, recurrent bursts. 
Their light curves show a large variety of shapes (with single,
double or triple-peaked bursts) and are characterized by a fast rise
($\sim 1 - 10$ s) toward a peak luminosity of $\sim 10^{38}$ erg s$^{-1}$,
followed by a slower (sometimes exponential-like) decline. Recurrence
periods range from hours to days \cite{Lew93,SR06,SB06}. 

Even though the bursting episodes detected in the X-ray source 3U 1820-30
by the Astronomical Netherlands Satellite (1974-1976) 
\cite{Gri76} are often referred to as the first X-ray bursts ever observed,
 similar events were previously recorded by the Vela-5b satellite  
(1969-1970) \cite{Bel72,Bel76} 
during a survey in the constellation Norma. 
These early discoveries were soon followed by the identification of
additional bursting sources, one of which, the enigmatic {\it Rapid Burster} 
(MXB 1730-335 \cite{Lew76}), is characterized by puzzling recurrence
times that are as short as $\sim$10 s. 
To date, $92$ Galactic
X-ray bursting sources have been identified \cite{intZ10}. 
The first extragalactic X-ray burst sources
have been discovered in two globular cluster candidate sources of the Andromeda
galaxy \cite{PH05}. Plans to extend the survey to other galaxies within
the Local Group are currently underway. 

The thermonuclear nature of X-ray bursts was suggested shortly after their discovery. 
Maraschi \& Cavaliere \cite{MC77}, and independently 
Woosley \& Taam \cite{WT76}, were the first to propose that X-ray bursts are
driven by thermonuclear runaways that occur on the
surface of accreting neutron stars. Since the fast succession of bursts 
exhibited by the Rapid Burster did not match the general behavior of most X-ray bursts,
two different categories, type I and type II bursts, were established.
The former are accretion-powered thermonuclear runaways, while the 
latter are most likely associated with  
instabilities in the accretion disk.
In the following, we will restrict the discussion to type I X-ray bursts, 
the most frequent 
kind of thermonuclear stellar explosion in the Galaxy, and the third most energetic
(after supernovae and classical novae). 

The spatial distribution of type I X-ray bursts shows a clear concentration toward the 
Galactic center, and the sources are frequently located inside globular clusters. 
This observational pattern points toward old progenitors
 \cite{Lew93}. Indeed, the donors that transfer 
material onto the neutron star surface in X-ray bursts are known to be faint, low-mass stars 
(with M $\lesssim 1$ M$_\odot$), corresponding either to main sequence or red giant
stars.  
The nature of the compact object hosting the explosion has been particularly 
controversial. 
Grindlay \& Gursky \cite{GG76} proposed a pioneering model of X-ray bursts that 
assumed accretion onto 
massive black holes ($\gtrsim 100$ M$_\odot$). But shortly after, observations  
of bursting sources belonging to globular clusters \cite{Swa77}, from which accurate 
distance determinations could be inferred, suggested
instead a much smaller object, either a neutron star or a stellar mass
black hole. A number of spectroscopic features, together with 
estimates of masses inferred for those systems, pointed clearly toward a 
neutron star \cite{vPM95}. 
It has been suggested that time-resolved spectroscopy may allow 
measurements of the surface gravitational redshift \cite{Dam90,Sma01},
from which constraints on the mass-radius relationship could be derived.

The orbital periods determined in X-ray burst sources range typically between 
1 - 15 hours \cite{WNP95}.
Hence, these binary systems are so close that mass transfer episodes, driven
by Roche-lobe overflows, ensue. This 
leads to the formation of an accretion disk surrounding the compact star. 
The maximum mass-accretion
rate is set by the Eddington limit (\.M$_{Edd} \sim 2 \times 10^{-8} \, 
M_\odot$ yr$^{-1}$, for H-rich accretion
onto a 1.4 $M_\odot$ neutron star).  
It was soon realized that the ratio of the gravitational potential 
energy released by the matter falling onto the neutron star during
the accretion stage
(G M$_{NS}$/R$_{NS} \sim$ 200 MeV per nucleon) 
 and the nuclear energy released during the burst 
($\sim$ 5 MeV per nucleon, for a fuel of 
solar composition), matches the observed ratio of time-integrated persistent and burst fluxes 
($\sim$ 40--100). 

Progress in the understanding of the physics of X-ray bursts has been achieved 
through multi-wavelength observations beyond the X-ray domain. In 1978, 
the first simultaneous optical/X-ray emission
was detected from the X-ray burst source 1735-444 \cite{Gri78}. 
The flux in the optical was interpreted as 
 the low-energy tail of the blackbody X-ray burst
emission \cite{Lew93}. But more importantly, the optical burst was delayed by 
$\sim$3 seconds with respect to the burst observed in X-rays \cite{McC79}. A similar delay has been
observed in many other X-ray burst sources. It has been suggested
that this optical emission is caused by 
reprocessing of X-rays in material within a few light-seconds from the source, 
either at the accretion disk that surrounds the neutron 
star or at the hemisphere of the secondary, which is directly illuminated by the 
source. In other words, the delay in the optical emission could be explained 
by travel-time differences between the X-rays directly emitted by the source, 
and those that hit the disk or the secondary and subsequently are scattered
down in energy and become optical photons.
Although efforts have been invested at other wavelengths, neither infrared nor radio emission has been 
unambiguously detected simultaneously with X-rays from a bursting
source. 

\subsection{Models of type I X-ray bursts}
Van Horn \& Hansen \cite{vHH74,HvH75} first pointed out that 
nuclear burning on the surface of neutron stars can be unstable. The 
different regimes for unstable burning have been extensively discussed
elsewhere \cite{SB06,FHM81,FL87}. For a chemical mixture with a CNO mass fraction of
$\sim 0.01$, for instance, mixed H/He-burning is expected 
for mass accretion rates of \.M$<2 \times 10^{-10}$ M$_\odot$ yr$^{-1}$, which is triggered by thermally 
unstable H ignition.
Pure He-shell ignition results for mass accretion rates in the range from
$2 \times 10^{-10}$ M$_\odot$ yr$^{-1}$ to $4.4 \times 
10^{-10}$ M$_\odot$ yr$^{-1}$,
following completion of H-burning. Mixed H/He-burning is expected 
for \.M$>4.4 \times 10^{-10}$  M$_\odot$ yr$^{-1}$, 
this time triggered by thermally unstable He ignition. 

Since the pioneering work  of Maraschi \& Cavaliere \cite{MC77}, 
Woosley \& Taam \cite{WT76}, and Joss \cite{Jos77}, 
many groups have performed simulations of the dynamics of X-ray bursts and their
associated nucleosynthesis. 
With a neutron star as the underlying compact object hosting the explosion, 
temperatures and densities in the accreted envelope 
reach rather high values ($T_{peak}\gtrsim 10^9$ K, $\rho_{max}\sim10^6$ g cm$^{-3}$).  
As a result, detailed nucleosynthesis 
studies require the use of hundreds of nuclides, up to the SnSbTe-mass region
\cite{Sch01} or beyond \cite{Koi04}, 
and thousands of nuclear interactions. 
Indeed, the endpoint of the nucleosynthesis in X-ray bursts is 
still a matter of debate. For example, recent experimental work suggests
that it will be more difficult to reach the SnSbTe-mass region
\cite{Elo09}. 
Because of computational limitations, studies of X-ray burst nucleosynthesis
have usually been performed with limited nuclear reaction networks that are truncated 
near Ni \cite{Woo84,Taa93,Taa96}, Kr \cite{Han83,Koi99},  
Cd \cite{Wal84}, or Y \cite{Wal81}.  More recently, Schatz et al. 
\cite{Sch01,Sch99}
carried out detailed nucleosynthesis calculations with a network 
containing more than 600 nuclides (up to Xe), 
but assuming a single burning zone only. Koike et al. \cite{Koi04} also performed 
one-zone nucleosynthesis calculations, with temperature and density profiles 
obtained from a spherically symmetric evolutionary code,
linked to a 1270-nuclide network extending up to $^{198}$Bi. 
Extensive 1-D hydrodynamic calculations with detailed nuclear reaction networks (up to Te) 
have been published in recent years \cite{Fis08,Jos10,Woo04}.
To date, no multidimensional calculation for realistic X-ray burst conditions has been 
successfully performed, although several attempts are currently underway with hydro codes such as MAESTRO \cite{Mal11} or FLASH. 
Notice that the 2-D study reported in Zingale et al. 
\cite{Zin01} analyzes He-detonations on neutron stars, whereas thermonuclear runaways are
expected to propagate deflagratively in X-ray bursts.

\begin{figure}
\includegraphics[scale=.60]{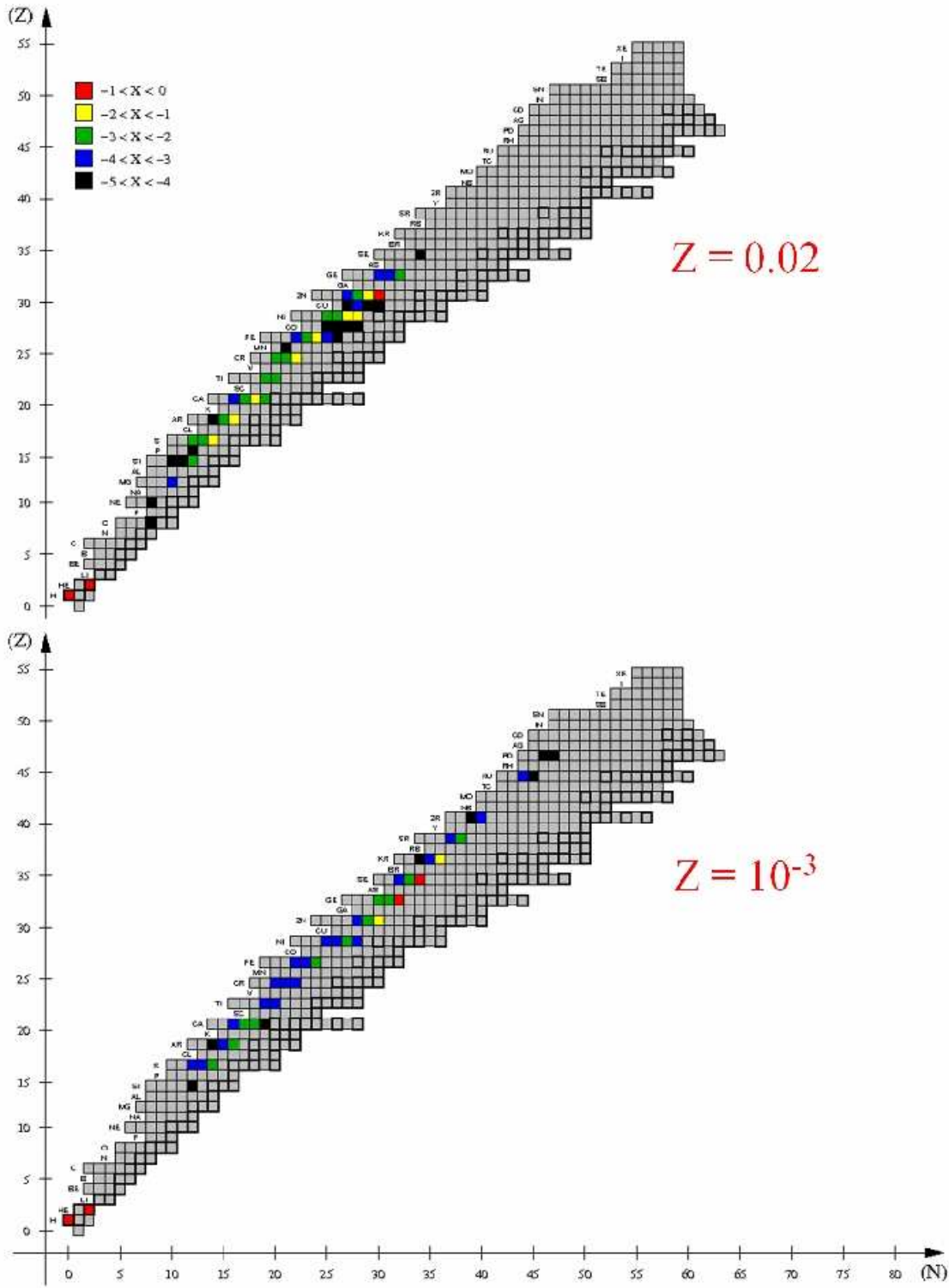}
\caption{Main nuclear activity shown in terms of the most abundant species
at the innermost envelope shell, for two different XRB models. 
Color legends indicate different ranges of mass fractions (in 
logarithmic scale).  Calculations have been performed with the code SHIVA 
\cite{Jos96}, and rely on 1.4 M$_\odot$ neutron stars (L$_{ini} = 1.6 \times 
10^{34}$ erg s$^{-1}$ = 4.14 L$_\odot$), accreting mass a rate $\dot M_{acc} =
1.8 \times 10^{-9}$ M$_\odot$ yr$^{-1}$ (0.08 $\dot M_{Edd}$).
The composition of the accreted material is assumed
to be solar-like (X = 0.7048, Y = 0.2752, Z = 0.02) [upper panel],
or metal-deficient (X = 0.759, Y = 0.240, and Z = 0.001 = Z$_\odot$/20)
[lower panel]. Both panels display the nuclear activity at T$_{peak}$, for
the first computed burst.  
\label{XRBnuc}}
\end{figure}

Detailed accounts of the main nuclear paths during type I X-ray bursts,
both in the framework of one-zone models and hydrodynamic (1-D) simulations,
have been given elsewhere 
\cite{Ili07,SR06,Wal81,Fis08,Jos10,CW92,Sch98}. 
To summarize, X-ray bursts are powered by a suite of different nuclear processes, 
including the {\it rp-process} (rapid proton-captures and $\beta^+$-decays), 
the $3\alpha$-reaction (Sec~\ref{heburnsec}), and the {\it $\alpha$p-process} (a sequence of 
($\alpha$,p) and (p,$\gamma$) reactions). The main nuclear flow  
proceeds far away from the valley of
stability (Fig. \ref{XRBnuc}), reaching the proton drip-line beyond A = 38 \cite{Sch99}.
Contrary to the thermonuclear runaways characteristic of other scenarios (e.g., classical novae),
X-ray bursts are halted by fuel consumption instead of expansion when degeneracy is lifted. In fact, 
breakout from the CNO region via $^{15}$O($\alpha$,$\gamma$)$^{19}$Ne 
leads the path toward the synthesis of
intermediate-mass and heavier species.
During the final stages of the burst, the main nuclear
flow is dominated by a multitude of successive $\beta$-decays that power the
decline of the light curve. 
The final composition of the envelope
depends on a number of parameters in the simulations (mainly, the 
mass-accretion rate and the adopted initial metallicity), but tends
to favor species in the mass range A=60-70 (mainly
$^{64}$Zn  and $^{68}$Ge \cite{Woo04}). Other studies \cite{Jos10} find production of additional species
with masses near A$\sim$105.

A major inconvenience in the modeling of X-ray bursts comes from the lack of 
clear observational nucleosynthesis constraints. The
potential impact of X-ray burst nucleosynthesis on the Galactic
abundances is indeed a matter of debate. Although ejection of matter from a neutron star 
is very unlikely because of the extremely large escape velocities required, 
it has been proposed that late radiation-driven winds during photospheric radius
expansion may lead to the ejection of a tiny fraction of envelope matter that 
contains processed material \cite{Wei06a,MacA07}. Although
it has been
claimed \cite{Sch98} that X-ray bursts may help to explain the Galactic 
abundances of the elusive light {\it p-nuclei} (Sec.~\ref{pproc}),
recent hydrodynamic calculations \cite{Jos10} have shown that the concentration
of these p-nuclei in the outermost layers of a neutron star envelope is orders
of magnitude below the required levels.
It has also been claimed that the nuclear ashes produced by X-ray bursts may
provide characteristic signatures, such as gravitationally redshifted atomic 
absorption lines that could be identified through high-resolution
X-ray spectra, representing a valuable tool to constrain X-ray burst models
\cite{Wei06a,Bil03,Cha05,Cha06}. However, this issue is controversial at present. 
Cottam et al. \cite{Cot02} reported features in the burst spectra 
of 28 X-ray bursts detected from the source EXO 0748-676, which were
interpreted as gravitationally redshifted absorption lines of Fe XXVI 
(during the early phase of the bursts), Fe XXV, and perhaps O VIII (at 
late stages). But no evidence for such
spectral features was found either during the 16 X-ray bursts observed from  
GS 1826-24 \cite{Kon07} or from the bursting episodes detected from the original source  
\cite{Cot08,Rau08}.

Reliable estimates of the neutron star surface composition
are nevertheless important since they determine a suite of different 
 thermal \cite{Sch99,MPH90}, radiative \cite{Pac83}, electrical 
\cite{Sch99,BB98}, and mechanical properties \cite{BC95,BC98} of the neutron star.  
Also, the diversity of shapes of X-ray burst light curves is probably driven by
different nuclear histories \cite{Han83,Heg07}.

Finally, it is important to stress that 
most of the reaction rates adopted in these extensive nucleosynthesis
calculations rely on theoretical
estimates using statistical models of nuclear reactions and, therefore, may be subject to significant
uncertainties. 
Efforts to quantify the impact of such nuclear physics
uncertainties on the overall abundance pattern for X-ray burst nucleosynthesis
have been performed by a number of groups
\cite{Koi04,Ili99,Wal81,Sch98,Thi01,Woo04,Amt06,Fis08,Cyb10}.
The most extensive works to date were performed by Parikh et al. \cite{Par08,Par09}, 
who identified the key nuclear processes (about 50 nuclear reactions
out of 3500, including $^{65}$As(p,$\gamma$)$^{66}$Se,  $^{61}$Ga(p,$\gamma$)$^{62}$Ge, $^{12}$C($\alpha$,$\gamma$)$^{16}$O, and $^{96}$Ag(p,$\gamma$)$^{97}$Cd), as well as key nuclear masses (in particular of $^{62}$Ge, $^{65}$As, $^{66}$Se, $^{69}$Br, 
$^{70}$Kr, $^{84}$Nb, $^{85}$Mo, $^{86,87}$Tc, $^{96}$Ag, $^{97}$Cd, 
$^{103}$Sn and $^{106}$Sb),  
whose uncertainties deeply influence the final yields.

\subsection{Superbursts: the next frontier}
A few extremely energetic bursts have been reported in recent years as a result of
 better performances in monitoring achieved with
X-ray satellites, such as Chandra or XMM-Newton. These rare and rather violent
events are known as {\it superbursts} \cite{Kuu04,Cum05}.
They are characterized by long durations 
with typical exponential-like decay time ranging from 1 to 3 hours 
(including an extreme case, KS 1731--260, that lasted for more than 10 hours 
\cite{Kuu02}). With a released energy of about $\sim$1000 times that of 
a typical X-ray burst ($\sim10^{42}$ erg), superbursts exhibit much longer
recurrence periods. For example, the period amounts to 4.7 yr for the system 4U 1636--53, for which two
superbursts have been observed to date \cite{Wij01}. Although
superburst sources also exhibit type I X-ray bursts, their occurrence is 
quenched for about a month after each superburst.
The first superburst was reported by Cornelisse et al. \cite{Cor00}
in the framework of a ``common" type I bursting source, the BeppoSAX source 
4U1735--44.  To date, 15 superbursts from 10 different bursting sources
 have been discovered, including GX 17+2, for which
4 superbursts have been identified \cite{intZ04}. 

The duration and energetics of 
superbursts suggest that they result from thermonuclear
flashes occurring in fuel layers at greater depths than what is typically encountered for 
X-ray bursts (corresponding to densities exceeding $10^9$ g cm$^{-3}$ \cite{CB01}). Thus they are more likely  to involve
 C-rich ashes resulting from prior type I X-ray bursts 
\cite{BB98,CB01,TP78,SBC03,Wei06b,WB07}, 
as first proposed by Woosley \& Taam \cite{WT76}.
Unfortunately, it is not clear whether enough 
 carbon is left after a type I burst \cite{Sch01,Sch99,Woo04,Jos10}.
 Nevertheless, Cumming \& Bildsten \cite{CB01} suggested
that low concentrations of carbon are sufficient to power a superburst, especially
in neutron star ``oceans" enriched in heavy nuclei. Moreover, 
recent studies indicate that both stable and unstable burning of the accreted 
H/He mixture may be required to power a superburst \cite{intZ03}. 

\section{Stellar mergers\label{mergsec}}
A small fraction of all
stars, perhaps 10\%, are expected to merge. Such merging episodes
have been invoked to explain the peculiar properties of a suite of different objects, such
as {\it blue straggler} stars (presumably products of two old, red stars merging
in a dense stellar environment), 
 R stars (suggested to result from the merger of a He white dwarf with a H-burning red giant branch star \cite{Izz07}; but see also
Ref.  \cite{Pie10}), 
HdC,  R Cr B, and AM CVn stars (presumably produced in double white dwarf mergers \cite{Hin06,Cla07,GarH09,Sta11,Bro11}),
 V838 Mon-type stars (mergers of a main sequence and a pre-main sequence star \cite{Tyl06}),
and short $\gamma$-ray bursts (from double neutron star or neutron star-black hole mergers; 
Sect. 6.10)\footnote{Blue stragglers are main sequence stars that belong to open or globular clusters, and are more luminous and bluer than stars 
at the main sequence turnoff point for the cluster; R stars are K-type giants, enriched in $^{12,13}$C and $^{14}$N; all of these are single stars
and standard stellar evolution theory cannot explain why they are C-rich;  
HdC stars are a type of H-deficient, supergiant C-rich star; 
R Cr B stars are H-poor, C- and He-rich, high-luminosity objects resembling simultaneously erupting and pulsating variables. 
Finally, V838 Monocerotis is a red variable star that experienced a major outburst in
2002. Originally believed to be a classical nova, V838 Mon exhibited an unusual light curve. The nature of the outburst is still not well understood.}. 
Because of their importance as potential sources of
gravitational waves, particular attention has been focused on the merging of compact 
objects \cite{Ace08,Gro08,Sig08}. Galactic compact binaries, such as double neutron stars, neutron star-white
dwarf binaries, or close double white dwarfs, are prime targets for the planned space-borne gravitational wave interferometer LISA.
In fact, the emission from Galactic (close) white dwarf binary systems is expected to be the dominant source of 
the background noise in the low frequency range.
Simulations of the merging of a white dwarf and a neutron star \cite{Gar07}, 
and of double white dwarfs \cite{Gue04,Lor05,Gui10}, have received little  attention so far compared to the coalescence of double neutron stars 
(Ref. \cite{Ros11}, and references therein).

\subsection{Double white dwarf mergers\label{MergWDWD}}
The merging of two white dwarfs by emission of gravitational radiation can be considered
the final fate of a significant fraction of this type of  systems, with an estimated
frequency of $\sim 8.3 \times 10^{-3}$ yr$^{-1}$.

\begin{figure}
\includegraphics[scale=.7]{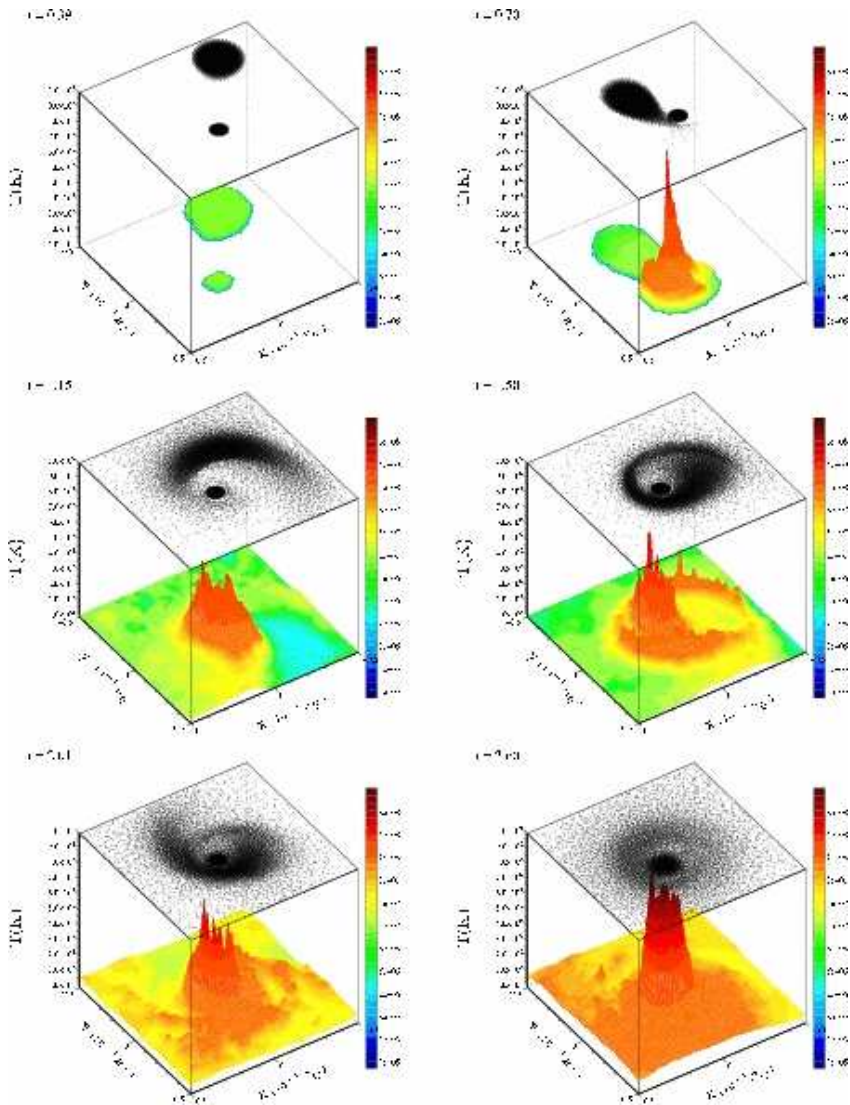}
\caption{Snapshots of the merging episode of two white
dwarfs of masses 1.2 and 0.4 M$_\odot$. 
Calculations have been performed with a {\it smoothed particle hydrodynamics}
(SPH) code.  Each panel shows the equatorial
temperature contours as well as the location of the SPH particles.
When the first particles of the secondary star hit the surface of the primary (top right panel),
the shocked regions reach temperatures up to $5 \times 10^8$ K. 
The colliding particles are reflected almost instantaneously  by the hard
boundary of the primary. As mass transfer proceeds (central panels) the accreted
matter is shocked and heated on top of the surface of the primary. This
matter is actually ejected from the surface of the primary and a hot
toroidal structure forms (right middle panel). Temperatures
exceed $10^9$  K and nuclear reactions transform
the accreted He into C and O.
Credit: J. Guerrero, E. Garc\'\i a--Berro, and J. Isern,
A\&A, vol. 413, p. 268, 2004 
\cite{Gue04}, reproduced with permission \copyright \, ESO.
\label{merge}}
\end{figure}

Several hydrodynamic simulations of the coalescence of double white
dwarf systems have been performed \cite{Gue04,Gui10,Dan11,Lor08} for different white dwarf masses and 
compositions\footnote{The nucleosynthesis in double white dwarf systems undergoing
collisions or close encounters has also been modeled recently \cite{Lor10}.}. 
Mass transfer ensues  as soon as the less massive white dwarf fills its Roche lobe (Fig. \ref{merge}). For certain combinations of masses of the double white dwarf system, 
mass transfer is stable \cite{Mar04}. In those cases,
 since the white dwarf radius scales as $M^{-1/3}$, 
when the secondary loses
 mass its radius increases, and hence, its mass-loss rate would increase, 
 reinforcing the overall mass transfer episode. 
 An accretion ``arm" forms, extending all the way from the secondary 
 to the surface of the more massive white dwarf. In fact, the arm becomes entangled because
of the orbital motion of the 
 coalescing white dwarfs, giving rise to a spiral shape. Finally, the secondary becomes fully disrupted 
 and a Keplerian disk surrounding the primary star forms after some orbital periods, depending
on the initial conditions.

 Models involving a He white dwarf undergo partial thermonuclear fusion
 as the temperatures achieved during the merger exceed $10^8$ K. This translates into
 noticeable abundances of Ca, Mg, S, Si and Fe. In fact, the abundances of some of these 
species are even larger in the hot corona 
 surrounding the primary star. This has been claimed as a feasible explanation for the 
 origin of the recently discovered metal-rich DAZd white dwarfs, a subclass with strong H lines and dusty disks \cite{Gar07}. Indeed, the reason for such a metallicity enhancement 
 could be related to the accretion of a minor planet that condensed in the metal-rich disks resulting from a merger episode. 
 
The expected gravitational wave emission from these systems
would be characterized by an initial, almost
 sinusoidal, pattern of increasing frequency (the so-called {\it chirping phase}), followed
 by a sudden disappearance of the gravitational wave signal, on a timescale of the order of the
orbital period as the merger proceeds. 
 All such merging episodes could be detectable by LISA with different signal-to-noise ratios.

\subsection{Neutron star - white dwarf mergers\label{MergNSWD}}
Another interesting case of compact star mergers involves systems composed of
a neutron star and a white dwarf.  About 40 binaries of this type have been
discovered to date \cite{Ker05}, although the
number of systems expected to merge in less than the age of the Universe within a distance of
$<3$ kpc from the Sun 
could be as high as $\sim$850 \cite{Edw01}. This results from the relatively high expected number
of these binaries, with an estimated neutron star--white dwarf formation rate that is about
10--20 times larger than what is predicted for double neutron stars \cite{Tau00}. 

A neutron star--white dwarf binary is expected \cite{Sta04} to form after an explosion
of a massive primary star in a core-collapse 
supernova (Sec.~\ref{sn}). For an appropriate range of primary star masses, the explosion would 
be accompanied by the formation of a neutron star.
In its evolution toward a neutron star--white dwarf binary, the system must first
avoid disruption by the supernova blast. This
depends critically on a number of parameters, including the mass ratio between
the stellar components. The final evolution is governed hereafter by
the secondary, which undergoes severe mass-loss episodes during its red
giant phase, and ends up as a white dwarf. 

Two classes of neutron star-white dwarf binaries can be distinguished \cite{Sta04,Cam01}: 
{\it intermediate-mass binary pulsars}, 
characterized by very eccentric
orbits and masses of the secondary stars ranging between 0.5--1.1 M$_\odot$,
and {\it low-mass binary pulsars}, with almost circular orbits 
and  masses of the secondaries between 0.15 and 0.45 M$_\odot$.
Both classes are expected to merge because of the emission
of gravitational waves with an estimated frequency
of $1.4 \times 10^{-4}$ yr$^{-1}$
\cite{Nel01}. Neutron star--white dwarf
 binaries expected to merge in less than the age of the Universe include the sources
PSR 30751+1807, PSR 31757-5322, and PSR 31141-6545.

Simulations of these compact star mergers are scarce 
and despite their interest as potential nucleosynthesis sites, most of the
efforts have focused on their role as future laboratories for 
gravitational wave studies. After coalescence, theses systems evolve into a
configuration clearly resembling that discussed for a double white dwarf merger,
with a a central compact object (a neutron star, in this case) surrounded
by a heavy Keplerian, self-supported accretion disk \cite{Gar07b}.
The expected gravitational wave emission from those systems
bears also a clear resemblance to that of a double white dwarf merger, 
with a short chirping phase detectable by the future LISA interferometer up to an estimated distance of about 250 kpc.
Even though the simulations performed to date face severe limitations 
(e.g., no post-Newtonian corrections, and very
limited nuclear reaction networks), some
interesting results have  already been obtained. In particular, it has been estimated
that the mass ejected from these systems amounts to about $0.01 \%$ of the overall
system mass assuming a merger of a 1.4 M$_\odot$ neutron star and a 0.6 M$_\odot$
white dwarf \cite{Gar07b}. 
Further studies aimed at determining the nucleosynthesis impact of
these mergers on the Galactic abundances are clearly needed.

\subsection{Double neutron star and neutron star--black hole mergers.
            Another r-process site?\label{MergNSNS}}
Mergers of stellar binaries containing double neutron stars, or a neutron star and a black hole,
have been extensively studied as r-process nucleosynthesis sites, as potential sources of gravitational waves,
and as the central engines for powering short $\gamma$-ray bursts.

To date about ten stellar binaries that are likely containing double neutron stars have been identified \cite{Lori08}. Of these,
about half are expected to merge over the age of the Universe. This is roughly in 
agreement with the predicted
frequency, between 40 and 700 Myr$^{-1}$ \cite{Kal04,Bel07}, for a galaxy like ours.
The frequency of neutron star--black hole mergers is highly uncertain, with estimates varying by orders
of magnitude 
\cite{Bel07,Bet98}. Nevertheless, with one of the stars being a black hole, these systems are expected
to play a minor role in the chemical evolution of the Galaxy.

Pioneering simulations of double neutron star and neutron star--black hole mergers were performed by 
Oohara \& Nakamura \cite{Ooh89}, assuming Newtonian gravity and polytropic equations of state
(see also Refs. \cite{Ras94,Zhu94,Lee99}). Several improvements, aimed at providing a more realistic description compatible
with general relativity include the use of pseudo-relativistic potentials \cite{Lee99b,Ros05},
 as well as post-Newtonian corrections \cite{Fab00,Aya01}.
Another approximation adopted, halfway between Newtonian gravity and general relativity, is the  
{\it conformal flatness approximation (CFA)}, in which the spatial part of the metric is assumed to be conformally flat
\cite{Wil95,Oec02,Fab06,Ise08}. The first fully general relativistic calculations of such mergers were performed by Shibata and coworkers \cite{Shi00}.
It is worth noting that two complementary general relativistic formulations are presently used, the {\it Baumgarte-Shapiro-Shibata-Nakamura
(BSSN)} formulation,
 and the {\it generalized harmonic} formalism (see Refs. \cite{Bau99,Shi95,Pre05}, and references therein).
More details on the different methods adopted in the modeling of double neutron star and neutron star--black hole mergers
can be found in Ref. \cite{Ros11}.

It has been suggested that the contribution of these mergers to the Galactic inventory of chemical species can proceed through
three different processes: (i) direct ejection, (ii) neutrino-driven winds, and (iii) disk disintegration. 
Lattimer \& Schramm \cite{Lat74} first provided estimates of the fraction of a neutron star that
becomes unbound after a neutron star--black hole merger. Their calculations yield a mean ejected mass per
merger of about 0.05 M$_{NS}$, or equivalently, 
an accumulated value of $2 \times 10^{-4}$ M$_\odot$ pc$^{-2}$, which is
approximately equal to the mass fraction of r-process elements in the Solar System. 
More realistic simulations \cite{Ros05,Ros99,Oec07} yield  a mean ejected mass of $3 \times 10^{-3}$ M$_\odot$,
with a mean electron mole fraction in the range of 0.01 $< Y_e <$ 0.5.
Neutrino-driven winds emitted by the
merger remnant are considered a promising alternative to the more traditional core-collapse supernovae as a production
site for r-process nuclei (see Refs. \cite{Kiz10,Qia96,Sur08}, and Sect. \ref{GRBSN}). 
Simulations suggest that such neutrino-driven winds drive ejection of about $\sim 10^{-4}$ M$_\odot$ of matter with $Y_e \sim 0.1 - 0.2$
per event \cite{Des09}.
Finally, the disk resulting after the merging episode may give rise to the ejection of neutron-rich material,
either through viscous heating in an advective disk or through the energy produced during recombination
of nucleons into nuclei \cite{Met08,Che07}. It has been claimed that about 0.03 M$_\odot$ 
of moderately neutron-rich material (0.1 $< Y_e <$ 0.5) may  be ejected from the disk \cite{Met09}.

First estimates of the impact of such double neutron star mergers on the Galactic nucleosynthesis \cite{Fre99} 
yielded a promising agreement with the observed r-process pattern beyond Ba. However, 
this was questioned by detailed inhomogeneous chemical evolution studies \cite{Arg04} on the basis of the 
low expected frequency of such mergers. The reasons included
a lack of consistency between the theoretical predictions and observations  
at very low metallicities, and the fact that
 injection of r-process material produced by a single merger would lead to a 
 scatter in the ratios of r-process and Fe abundances
 at times too long compared to observations. 
While core-collapse supernovae seem to be favored as the main r-process site (Sec.~\ref{rprocsec}), a contribution from neutron star mergers
and the existence of multiple r-process sites still remain open questions.

\section{Non-stellar processes. Cosmic ray nucleosynthesis}
\subsection{Quest for the origin of the light nuclides}
We have addressed so far the origin of most nuclides, except of the 
light species $^6$Li, $^9$Be, $^{10}$B and $^{11}$B. Their solar 
abundances are smaller by about 6 orders of magnitude compared to 
other light nuclides (although their abundances are much larger compared 
to almost all s-, r-, or p-nuclides). As was the case for $^2$H and $^7$Li, 
their cross sections for proton-induced reactions are so large, because 
of the small Coulomb barriers, that they are destroyed during the 
hydrogen burning phase in stellar interiors already at temperatures 
below a few million kelvin. In fact, under such conditions their mean 
lifetimes amount to less than a few Gyr. On the other hand, standard 
big bang nucleosynthesis (Sec. \ref{BBN}) produces only negligible 
amounts of $^6$Li, $^9$Be, $^{10}$B and $^{11}$B, with predicted number 
abundances less than 4 orders of magnitude compared to $^7$Li \cite{Van99}.
 Their origin seemed so obscure that they were attributed in the seminal 
work of Burbidge, Burbidge, Fowler and Hoyle \cite{Bur57} to some unknown 
``x process". Among the suggestions made in that work was the production 
by spallation reactions involving high-energy protons, neutrons or 
$\alpha$-particles of energies in excess of 100 MeV per nucleon on abundant
 CNO nuclei in stellar atmospheres. While it was convincingly demonstrated 
that the energy available during the T Tauri phase of young stars is 
insufficient to produce the light nuclides in stellar atmospheres 
\cite{Ryt70}, the association of lithium, beryllium and boron  synthesis 
with spallation reactions proved correct.

\begin{figure}
\includegraphics[scale=.60]{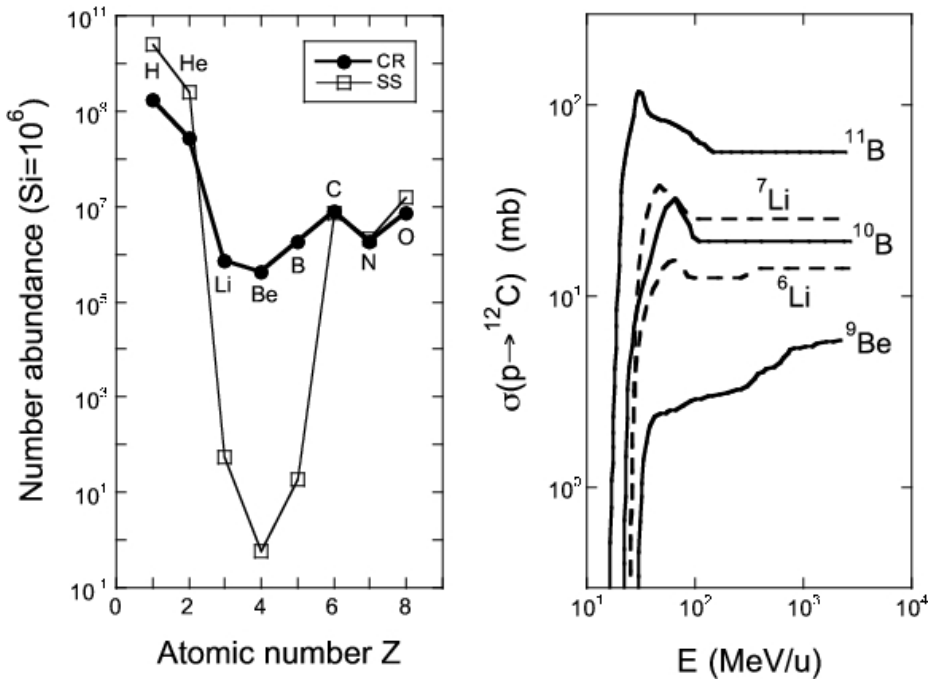}
\caption{(Left panel) Number abundances of light elements in Galactic 
cosmic Rays (solid circles) and the Solar System (open squares), 
normalized to silicon. The cosmic ray values for H and He are adopted 
from measurements by the balloon-borne BESS instrument and the GSFC 
instrument on the IMP-8 spacecraft, while those for the other elements 
shown are from CRIS measurements. The cosmic ray abundances were 
measured during solar minimum at 170 MeV per nucleon \cite{Cal97}. 
The Solar System abundances are adopted from Lodders \cite{Lod09}. 
Note that in cosmic rays, B is more abundant than Li, which in turn 
is more abundant than Be. However, in the Solar System, Li is more 
abundant than B, which is more abundant than Be, supporting the 
conjecture that a significant fraction of the Solar System $^{7}$Li 
must have been produced by some stellar source (classical novae or 
AGB stars). (Right panel) Cross sections for the production of Li, 
Be and B by the spallation of protons incident on $^{12}$C as a 
function of energy. Data are adopted from Ramaty et al. \cite{Ram97}. 
Note the decreasing sequence of B, Li and Be.
\label{figCR}}
\end{figure}

An important piece of evidence was uncovered around 1970, when it was 
pointed out that in the Solar System the ratio of number abundances for 
Li, Be and B compared to C, N and O amounts to $\sim$10$^{-6}$, while 
the value in Galactic cosmic rays is near $\sim$0.2 \cite{Rev70}. 
Consequently, the idea was born that the source of Li, Be and B are 
spallation reactions between Galactic cosmic rays and nuclei in the 
interstellar medium. The conjecture is supported by the observation 
that the spallation cross sections of protons and $\alpha$-particles 
on C, N and O nuclei at energies in excess of 100 MeV per nucleon favor 
the production, in decreasing order, of B, Li and Be \cite{Ram97}. 
The observed Galactic cosmic ray abundances of these light species 
exhibit the same exact ordering. More information on abundances and 
associated spallation cross sections is provided in Fig.~\ref{figCR}. 
First quantitative models \cite{Men71}, based on measured spallation 
cross sections, showed that this scenario could reasonably account 
for the abundances of $^6$Li, $^9$Be and $^{10}$B observed in the 
Solar System and in stars after 10 Gyr of Galactic chemical evolution. 

On the other hand, $^{7}$Li and $^{11}$B were considerably 
underproduced in the standard cosmic ray spallation scenario. It is 
now believed that additional mechanisms contribute to their 
production. During the first billion years, Big Bang nucleosynthesis 
dominates the synthesis of $^7$Li (Sec. \ref{BBN}), while later a 
stellar source, perhaps classical novae or AGB stars, contributed 
significantly to its production (Fig.~\ref{figCR}). Thus $^7$Li is 
exceptional among all naturally occurring nuclides in that its origin 
can be attributed to three distinct sites: the Big Bang, Galactic 
cosmic rays, and stars. In the case of $^{11}$B it was found that 
neutrino-induced interactions with $^{12}$C during core-collapse 
supernovae may also produce significant amounts of this species 
(Sec.~\ref{sn}). However, the predicted yields for the stellar production 
of $^{7}$Li and for the neutrino-induced synthesis of $^{11}$B
 presently have large associated uncertainties. 

\subsection{Galactic cosmic rays and spallation}
Galactic cosmic rays consist predominantly of energetic protons, 
$\alpha$-particles and heavier nuclei, with kinetic energies up to 
the several 100 TeV per nucleon range and beyond. Their energy density 
amounts to about 2 eV cm$^{-3}$ \cite{Web97} and thus exceeds the energy
 density of star light ($\sim0.3$ eV cm$^{-3}$). Near an energy of 10
 GeV per nucleon, for example, the number fractions of hydrogen, helium,
 CNO and LiBeB nuclei amount to 95\%, 4.5\%, 0.4\% and 0.07\%, 
respectively \cite{Eng90,San00}. An energetic cosmic ray particle 
propagating through the Galaxy faces a number of possibilities. First, 
it may undergo a high-energy nuclear collision (spallation) with 
interstellar nuclei. Second, an energetic cosmic ray particle may slow 
down significantly because of many collisions with interstellar 
electrons. Such ionization losses are responsible for the eventual 
deposition of cosmic ray matter into the interstellar medium. Third, 
the cosmic ray particle may escape into intergalactic space, depending
 on the value of the escape length, that is, the amount of matter the
 cosmic ray particle traverses between its source and the Galactic 
boundary. The escape length, which can be inferred from the abundance
 ratio of spallation products (for example, Li, Be and B) and target 
nuclei (mostly C, N and O), amounts to $\sim$10 g cm$^{-2}$ for Galactic
 cosmic rays near an energy of 1 GeV per nucleon \cite{Str07}. During
 propagation an energetic charged particle is strongly deflected by 
Galactic magnetic fields, such that most information regarding its 
original direction of motion is lost. 

Spallation may involve a proton or $\alpha$-particle with energies in 
excess of several MeV per nucleon impinging on a heavy nucleus in the 
interstellar medium, of which C, N and O are most abundant.  
Alternatively, an energetic heavy nucleus, again mainly C, N or O, may 
interact with interstellar H and He. A third possibility, thought to
 be especially important for the production of the lithium isotopes in 
the early Galaxy when the CNO abundance was small, is the spallation 
of $\alpha$-particles with energies in the tens of MeV per nucleon range 
with other $^4$He nuclei of the interstellar medium. The production 
rate of a given LiBeB species, $i$, expressed in terms of the number 
abundance ratio with respect to hydrogen, $Y_i=N_i/N_H$, is given by an
 expression of the form $dY_i/dt=\Sigma_{j,k}~F_j\sigma_{jk}Y_k P_i$, 
with $F$ the average cosmic ray flux (number of protons per cm$^2$ and s),
 $\sigma$ the average spallation cross section (in cm$^2$), and $P$ the 
probability that after production species $i$ will be thermalized and
 retained in the interstellar medium; the indices $j$ and $k$ denote 
the projectile and target, respectively, that is, protons, 
$\alpha$-particles or CNO nuclei. It is instructive to consider a 
simple example: a typical cosmic ray proton flux of 
$F_p\sim$10~cm$^{-2}$~s$^{-1}$, an average spallation cross section 
of $\sigma^{Be}_{p\rightarrow CO}\sim 5\times10^{-27}$~cm$^2$ 
\cite{Ram97} for the production of $^9$Be, assuming energetic protons 
impinging on C or O nuclei near an energy of 200 MeV per nucleon (i.e.,
 the maximum of the spectrum), an abundance of $Y_{CNO}\sim 10^{-3}$ 
in the interstellar medium, a probability of $P_{Be}\sim$1, and 
integrating over a time period of $\sim 10^{10}$~yr, yields a value 
of $Y_{Be}\sim 1.6\times10^{-11}$. The measured Solar System $^9$Be 
abundance from solar photosphere and meteoritic data amounts to 
$Y^{obs}_{Be}=(2.1\pm0.1)\times10^{-11}$ \cite{Lod09}, in good 
overall agreement with the predicted result. Of course, this simple
 estimate does not account for the change of the $^9$Be abundance over 
time.

\subsection{Evolution of Li, Be, B abundances. 
Origin of Galactic cosmic rays}
The study of the Li, Be and B abundances, and in particular their 
evolution over the past several billion years, yields important clues 
for testing ideas regarding the nucleosynthesis in stars, for the 
physics of stellar outer layers, and for models of the early Galaxy. 
Moreover, the Li, Be and B abundance evolutions may hint at the 
origin of Galactic cosmic rays. Although it can be safely assumed 
that the sources of the cosmic ray particles, except perhaps those 
of the very highest energies, are located within the Galaxy, their
 origin and acceleration mechanism have not found a satisfactory 
explanation yet. In fact, the mystery of cosmic ray origin represents
 a long-standing problem of foremost importance. Energy arguments 
severely constrain the kind of candidate sources: for an assumed 
energy density of $\epsilon=2$ eV cm$^{-3}$, maintained on average over
 the volume of our Galaxy, $V\sim 7\times 10^{66}$~cm$^3$, and a
 cosmic ray lifetime of $\tau\sim$20 Myr near 1 GeV per nucleon,
 as derived from the cosmic ray abundances of the radioactive 
``clock" isotopes $^{10}$Be, $^{26}$Al, $^{36}$Cl and $^{54}$Mn 
\cite{Mew01}, one finds a total power of 
$P_{CR}=\epsilon V / \tau\sim 4\times 10^{40}$ erg s$^{-1}$. Thus, Galactic 
cosmic rays are expected to be associated with the most energetic
 astronomical objects. For example, a typical core-collapse  
supernova releases a kinetic energy of $(1-2)\times 10^{51}$~erg 
(Sec.~\ref{sn}). Assuming a frequency of about two such events per 
century gives a power of $P_{SN}\sim 10^{42}$ erg s$^{-1}$, that is, 
significantly larger than what is needed for accelerating Galactic 
cosmic rays. Indeed, there is now substantial observational evidence
 that the origin of Galactic cosmic rays is related to supernova 
remnants \cite{Acc09,Tav10}.

Consider now the abundance evolution of $^9$Be and suppose that 
supernovae are the source of Galactic cosmic rays. One possibility 
for producing this species is energetic protons or $\alpha$-particles
 colliding with interstellar C, N and O nuclei. The production 
{\it rate} of $^9$Be will then depend on the cosmic ray flux and 
the CNO abundance. The former factor is proportional to the supernova 
rate, $dn_{SN}/dt$, while the latter is proportional to the total 
number of supernovae up to that time, $n_{SN}$. Consequently, the 
observed $^8$Be abundance at any given instant should be proportional
 to $n_{SN}^2$ or, in other words, to the square of the metallicity
 (as measured, for example, by the Fe abundance). On the other hand, 
$^9$Be may be produced by energetic CNO nuclei colliding with 
interstellar protons or $\alpha$-particles. In this case, assuming a
 constant abundance of CNO nuclei in Galactic cosmic rays, the 
cosmic ray flux is again proportional to the supernova rate, but the
 H or He abundance in the interstellar medium is independent of 
$n_{SN}$. Thus one would expect that the $^9$Be abundance at any 
given time is directly proportional to the metallicity 
\cite{Dun92,Van00}. Alternatively, a linear relationship between 
the $^9$Be abundance and metallicity could naturally arise if 
supernova-produced protons or $\alpha$-particles undergo spallation 
reactions with CNO nuclei in the same supernova's ejecta \cite{Gil92}. 
Similar arguments hold for the boron abundance evolution. 
       
Observations in the 1990s of beryllium and boron in metal-poor stars 
were of paramount importance in this regard \cite{Dun92,Rya92,Boe93}.
Unlike lithium, which exhibits the Spite plateau at low metallicities 
(Sec.~\ref{BBN}), the beryllium and boron abundances showed no such 
plateau and, furthermore, revealed a linear dependence on metallicity. 
Therefore, it is unlikely that Galactic cosmic rays originate from 
supernova-induced energetic protons or $\alpha$-particles colliding 
with CNO nuclei of the interstellar medium (which would give rise 
to a beryllium or boron abundance proportional to the metallicity 
{\it squared}). Neither can the observations be explained by a CNO 
abundance in Galactic cosmic rays that, similar to the CNO abundance 
in the interstellar medium, would increase with time (since then the
 relationship between the beryllium or boron abundance and metallicity 
could not be {\it linear}). In addition, the CRIS instrument onboard 
the Advanced Composition Explorer (ACE) detected significant amounts
 of $^{59}$Co in Galactic cosmic rays, but only very small amounts 
of the radioactive precursor $^{59}$Ni \cite{Mew01}. The latter 
isotope decays by electron-capture (with a half-life of 
$T_{1/2}=7.6\times 10^4$~yr in the laboratory), but is stable once 
fully stripped of its electrons and accelerated to high energies.
 From the observed abundance ratio of $^{59}$Ni and $^{59}$Co it
 can be estimated that the time delay between explosive 
nucleosynthesis and acceleration of radioactive $^{59}$Ni must 
exceed $\sim 10^5$~yr, when the ejecta of a particular supernova 
are presumably diluted and mixed in the interstellar medium.
 Consequently, it seems unlikely that individual supernovae 
accelerate their own ejecta.

Several ideas have been proposed. A popular model is the production 
of Galactic cosmic rays in superbubbles, which result from the 
evolution of massive star clusters \cite{Mac88,Hig98,Ali02}. First, 
through their strong winds, the massive stars create cavities of 
hot, low-density, gas in the interstellar medium. The cavities 
increase in size with each supernova explosion, and eventually 
merge to create superbubbles. Assuming that, in the simplest case,
 the composition of the metal-rich, low-density, material within 
the superbubble does not change with time, Galactic cosmic rays 
could be launched from this reservoir by subsequent supernova 
shocks. The superbubble origin has been criticized recently and 
other ideas have been proposed \cite{Pra10}. Future developments 
in this field at the intersection of nucleosynthesis, astroparticle
 physics and Galactic chemical evolution promise to be exciting.

\section{Conclusions and future perspectives}
The last fifty years have witnessed
extraordinary progress in the modeling of stellar evolution, as well as in
our understanding of the origin of the elements. First,
supercomputers have allowed astrophysicists to perform complex 
hydrodynamic simulations coupled with extensive nuclear reaction networks at 
astounding levels of resolution. Indeed, last decades have seen 
the emergence of multidimensional simulations in many areas of stellar astrophysics research. 
Second, better observational constraints have been obtained through
combined efforts in spectroscopy, the dawn of high-energy astrophysics 
with space-borne observatories, and laboratory measurements of presolar
meteoritic grains that condensed in a suite of stellar environments. 
And third, nuclear physics has witnessed the birth of 
radioactive ion beam facilities, as well as direct measurements at astrophysically important energies using stable ion beam facilities.

Further progress eagerly awaits future developments aimed at solving key 
remaining questions
in nuclear astrophysics, chief among them:
\begin{enumerate}
\item Why do predictions of helioseismology disagree with those of the standard solar model?
\item What is the solution to the lithium problem in Big Bang nucleosynthesis? 
\item What do the observed light-nuclide and 
      s-process abundances tell us about convection and dredge-up in massive stars and AGB stars? 
\item What are the production sites 
      of the $\gamma$-ray emitting radioisotopes $^{26}$Al, $^{44}$Ti and $^{60}$Fe? 
\item What is the origin of about 30 rare and neutron deficient nuclides beyond the iron peak (p-nuclides)?
\item What causes core-collapse supernovae to explode? 
\item What is the extend of neutrino-induced nucleosynthesis ($\nu$-process)? 
\item What is the extend of the nucleosynthesis
      in proton-rich outflows in the early ejecta of core-collapse supernovae ($\nu p$-process)? 
\item What are the sites of the r-process? 
\item What causes the discrepancy between models and observations regarding the mass ejected during classical nova outbursts?
\item Which are the physical mechanisms driving convective mixing in novae?
\item What are the progenitors of type Ia supernovae?
\item What is the nucleosynthesis endpoint in type I X-ray bursts? Is there any matter ejected from those systems?
\item What is the impact of stellar mergers on Galactic chemical abundances?
\item What are the production and acceleration sites of Galactic cosmic rays? 
\end{enumerate}

\noindent These questions will challenge young and courageous researchers for years to come, thereby keeping nuclear astrophysics in the spot light of science.

\section*{Acknowledgements}
The authors would like to thank Alessandro Chieffi, Alain Coc, 
Carla Fr\"ohlich, Enrique Garc\'\i a--Berro, Domingo Garc\'\i a--Senz, 
Margarita Hernanz, Jordi Isern, Daid Kahl, Karl-Ludwig Kratz, Marco Limongi, Richard Longland, Maria Lugaro, Brad Meyer, Larry Nittler, 
Anuj Parikh, Nikos Prantzos, Aldo Serenelli, Steve Shore, and 
Frank Timmes, for helpful discussions. Larry Nittler kindly 
provided us with Fig. 2. 
This work was supported in part by the Spanish MICINN grants AYA2010-
15685 and EUI2009-04167, by the E.U. FEDER funds, 
by the ESF EUROCORES Program EuroGENESIS, by the National Science Foundation under Grant No. AST-1008355,
and by the U.S. Department of Energy under Contract No. DE-FG02-97ER41041.

\end{document}